\documentclass[%
reprint,
superscriptaddress,
 amsmath,amssymb,
 aps,
]{revtex4-2}

\usepackage{changes}
\usepackage{graphicx}
\usepackage{dcolumn}
\usepackage{bm}

\usepackage{xcolor}
\usepackage[normalem]{ulem}

\usepackage{braket}

\usepackage[pagebackref=false]{hyperref}
\usepackage[capitalize]{cleveref}
\begin{document}

\author{A.~Rothstein}
\email{alexander.rothstein@rwth-aachen.de}
\affiliation{JARA-FIT and 2nd Institute of Physics, RWTH Aachen University, 52074 Aachen, Germany,~EU}%
\affiliation{Peter Gr\"unberg Institute  (PGI-9), Forschungszentrum J\"ulich, 52425 J\"ulich,~Germany,~EU}

\author{R. J.~Dolleman}
\affiliation{JARA-FIT and 2nd Institute of Physics, RWTH Aachen University, 52074 Aachen, Germany,~EU}%

\author{L. Klebl}
\affiliation{I. Institute for Theoretical Physics, University of Hamburg, Notkestraße 9-11, 22607 Hamburg, Germany, EU}
\affiliation{Institute for Theoretical Physics and Astrophysics, University of Würzburg, Am Hubland, 97074 Würzburg, Germany,~EU}

\author{A. Achtermann}
\affiliation{JARA-FIT and 2nd Institute of Physics, RWTH Aachen University, 52074 Aachen, Germany,~EU}%

\author{F. Volmer}
\affiliation{JARA-FIT and 2nd Institute of Physics, RWTH Aachen University, 52074 Aachen, Germany,~EU}%

\author{K.~Watanabe}
\affiliation{Research Center for Electronic and Optical Materials, National Institute for Materials Science, 1-1 Namiki, Tsukuba 305-0044, Japan}

\author{T.~Taniguchi}
\affiliation{Research Center for Materials Nanoarchitectonics, National Institute for Materials Science,  1-1 Namiki, Tsukuba 305-0044, Japan}%

\author{F.~Hassler}
\affiliation{Institute for Quantum Information, RWTH Aachen University, 52056 Aachen, Germany, EU}

\author{L. Banszerus}
\affiliation{Faculty of Physics, University of Vienna, Boltzmanngasse 5, 1090 Vienna, Austria, EU}

\author{B.~Beschoten}
\affiliation{JARA-FIT and 2nd Institute of Physics, RWTH Aachen University, 52074 Aachen, Germany,~EU}%

\author{C.~Stampfer}
\email{stampfer@physik.rwth-aachen.de}
\affiliation{JARA-FIT and 2nd Institute of Physics, RWTH Aachen University, 52074 Aachen, Germany,~EU}%
\affiliation{Peter Gr\"unberg Institute  (PGI-9), Forschungszentrum J\"ulich, 52425 J\"ulich,~Germany,~EU}%

\title{Gate-tunable Josephson diodes in magic-angle twisted bilayer graphene}

\date{\today}


\begin{abstract} 
We report low-temperature measurements of two adjacent, gate-defined Josephson junctions (JJs) in magic-angle twisted bilayer graphene (MATBG) at a moiré filling factor near $\nu = -2$.
We show that both junctions exhibit a prominent, gate-tunable Josephson diode effect, which we explain by a combination of large kinetic inductance and non-uniform supercurrent distribution.
Despite their proximity, the JJs display differences in their interference patterns and different diode behavior, underscoring that microscopic inhomogeneities such as twist angle variations shape the non-uniform supercurrent and drive the diode behavior.
As a result, the nonreciprocal supercurrent can be tuned by gate voltage, enabling tuning of the diode efficiency and even reversing the polarity at fixed magnetic fields.
Our findings offer potential routes for tailoring Josephson diode performance in superconducting quantum circuits. 
\end{abstract}

\maketitle

\begin{figure*}[!tbh]
\centering
\includegraphics[draft=false,keepaspectratio=true,clip,width=1\linewidth]{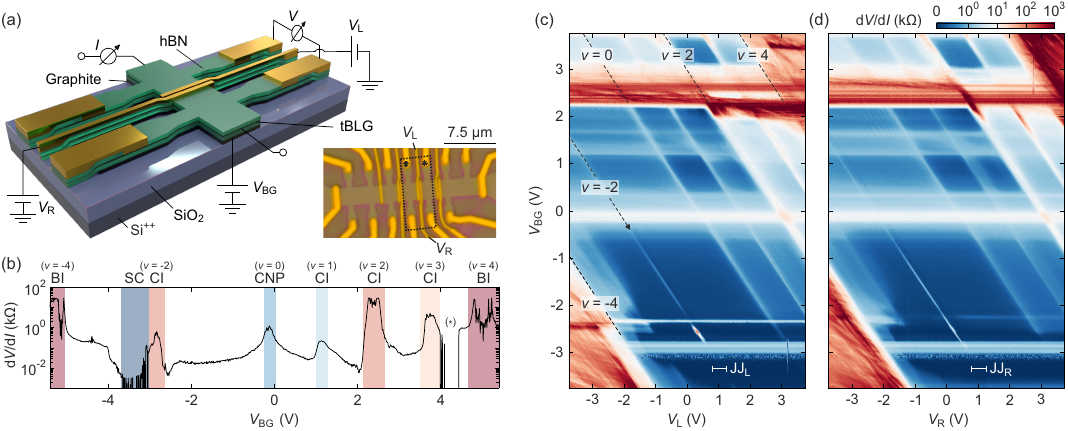}
\caption{(a) Schematic of the device and the measurement setup. 
Inset shows an optical image. 
The region under investigation is highlighted in the optical image.
(b) Differential resistance as a function of $V_\mathrm{BG}$.
Highlighted are the correlated and band insulating states (CI/BI), the charge neutrality point (CNP) and the superconducting (SC) area.
The vanishing differential resistance in the area marked by the asterisk is a measurement artifact.
(c) Differential resistance $\mathrm d V/\mathrm d I$ as a function of $V_\mathrm{BG}$ and $V_\mathrm{L}$. 
The JJ is formed at the gate configurations where the $\nu = -2$ correlated insulator intersects the superconducting area at fixed $V_\mathrm{BG} = -3.34 \, \mathrm{V}$ (see white bar and label).
The used colorscale is linear between $0$ and $10^0\, \mathrm{k \Omega}$ and logarithmic between $10^0$ and $10^3 \, \mathrm{k \Omega}$.
(d) Same as in panel (c) but as a function of $V_\mathrm{BG}$ and $V_\mathrm{R}$.
}
\label{f1}
\end{figure*}
The superconducting diode effect -- where a supercurrent flows without dissipation in one direction but a dissipative current flows in the opposite direction -- offers a promising route to realizing nonreciprocal elements in quantum circuits~\cite{Nadeem2023Sep}.
Superconducting diodes have been realized in different platforms, including superconducting thin films~\cite{Ando2020Aug, Miyasaka2021Jun, Kawarazaki2022Nov, Itahashi2020Mar, Schumann2020Mar}, two-dimensional materials~\cite{Wakatsuki2017Apr, Lin2022Oct, Scammell2022Mar}, superconductor/ferromagnet hybrids~\cite{Narita2022Aug}, and Josephson junctions (JJs) -- commonly referred to as Josephson diodes (JDs) when engineered to exhibit this behavior \cite{Baumgartner2022Jan, Jeon2022Sep, Wu2022Apr, Pal2022Oct, Bauriedl2022Jul, Diez-Merida2023Apr, Chen2024Mar,Kudriashov2025Jun}.
Among these, twisted multilayer graphene stands out as a particularly promising platform for implementing JDs~\cite{Cao2018Apr2, Khalaf2019Aug, Park2021Feb, Park2022Aug, Zhang2022Sep, Hao2021Mar}.
Its appeal lies in the coexistence of unconventional superconductivity~\cite{Cao2018Apr2, Lu2019Oct, Yankowitz2019Mar, Oh2021Dec}, correlated insulating states~\cite{Cao2018Apr, Stepanov2020Jul}, orbital magnetism~\cite{Sharpe2019Jul, Serlin2020Feb, Stepanov2021Nov}, and topological Chern bands~\cite{Das2021Jun}; all of which are highly tunable via electrostatic gating.
This gate tunability allows independent control of the weak link and superconducting leads with atomically clean interfaces, thereby enabling systematic exploration of the interplay among different correlated phases in a monolithic platform~\cite{Rodan-Legrain2021Jul, deVries2021Jul,Portoles2022Nov, Diez-Merida2023Apr,Perego2024Oct,Portoles2025May}. 
To obtain a JD effect, both time-reversal symmetry and inversion symmetry must be broken.
In magic-angle twisted bilayer graphene (MATBG) JDs, several microscopic mechanisms have been proposed to achieve this symmetry breaking, including finite-momentum Cooper pairing~\cite{Davydova2022Jun}, orbital magnetism~\cite{Diez-Merida2023Apr}, valley polarization~\cite{Wei2022Oct, Hu2023Jun, Xie2023Apr}, and chiral pairing combined with a topological band structure~\cite{Alvarado2023Sep}. 
However, the precise mechanism underlying the observation of the diode effect in MATBG JJs remains unclear.

To address this, we study monolithically integrated JJs formed in MATBG, each 200~nm long and spaced 180~nm apart, with independent gate control.
Magnetotransport measurements reveal strongly skewed interference patterns which are a hallmark of the JD effect.
At the same time, both junctions show pronounced differences in their interference patterns, indicating that local disorder -- likely arising from twist angle variations over tens to hundreds of nanometers -- plays a key role in shaping the supercurrent distribution.
This spatially non-uniform current distribution breaks the inversion symmetry, making the kinetic inductance of the superconducting phase a critical factor in the JJ dynamics~\cite{Annunziata2010Oct, Barone1982Jul}.
In MATBG, the low carrier density and large effective mass of the charge carriers enhance the effects of kinetic inductance compared to other materials, resulting in the JD effect~\cite{Lopez-Nunez2023Nov,Portoles2025May, Banszerus2025Feb, Jha2024Mar}.
From our measurements, we estimate the kinetic inductance to be on the order of $10 \, \mathrm{nH}$.
Taken together, our results demonstrate that the interplay between kinetic inductance and local disorder breaks inversion symmetry, whereas time-reversal symmetry is broken only by the externally applied magnetic field, with no evidence for intrinsic time-reversal symmetry breaking.
This combined symmetry breaking naturally accounts for the emergence of the JD effect in MATBG.
Moreover, we find that the JD is highly tunable by gate voltage, enabling a reversal of the diode polarity at fixed magnetic fields.
A schematic illustration and an optical image of the device are shown in \cref{f1}(a).
The device consists of MATBG encapsulated in hexagonal boron nitride (hBN) with a global graphite back gate. 
The MATBG is fabricated into a multi-probe Hall bar with $\approx 200 \, \mathrm{nm}$ long metal gates placed on top of the structure, to locally tune the charge carrier density (for details, see Methods).
This allows us to create JJs at different spatial locations on the device. 
We focus on one region within two longitudinal voltage probes, highlighted by the dotted rectangle in the optical image.
The twist angle within this region is estimated to be $ \approx 1.09^\circ \pm 0.01^\circ$ (see Supporting Information). 
Within the studied region two top gates are present, separated by $\approx 180\, \mathrm{nm}$, which allows us to define and compare the transport behavior of two separate, closely spaced JJs.  

\cref{f1}(b) shows a resistance trace as a function of $V_\mathrm{BG}$. 
This trace highlights several important features, such as the charge neutrality point ($\nu = 0$, where $\nu$ is the filling factor denoting the number of charge carriers per moiré superlattice unit cell), and the single-particle band gaps ($\nu = \pm 4$).
Furthermore, correlated insulating states are visible at $\nu = -2, 1, 2$ and $3$. 
Adjacent to the correlated insulator at $\nu = -2$, a superconducting regime is found at $\nu < -2$. 
To find the gate configurations required for the JJ regime, we measure the differential resistance as a function of the back gate voltage $V_\mathrm{BG}$ and the voltage applied to either the left or the right top gate [see \cref{f1}(c,d)], $V_\mathrm{L}$ or $V_\mathrm{R}$ respectively, with the unused gate grounded. 
Here, the visible horizontal features emerge from the area not covered by the top gate, as they are not tuned by $V_\mathrm{L}$ or $V_\mathrm{R}$.
In contrast, the diagonal features indicate a tuning by both $V_\mathrm{BG}$ and $V_\mathrm{L,R}$, and therefore emerge from the area below the top gate. 
Following the diagonal feature corresponding to $\nu = -2$ in \cref{f1}(c,d), we reach a region near $V_\mathrm{L,R} \approx 1~\mathrm{V}$ (white bar and label), where the correlated insulator intersects the superconducting regime. 
Note, that the diagonal feature disappears in the superconducting region, as expected for the transmission of a supercurrent across the weak link between the superconducting leads formed by the correlated insulator. 

\begin{figure*}[!tbh]
\centering
\includegraphics[draft=false,keepaspectratio=true,clip,width=1\linewidth]{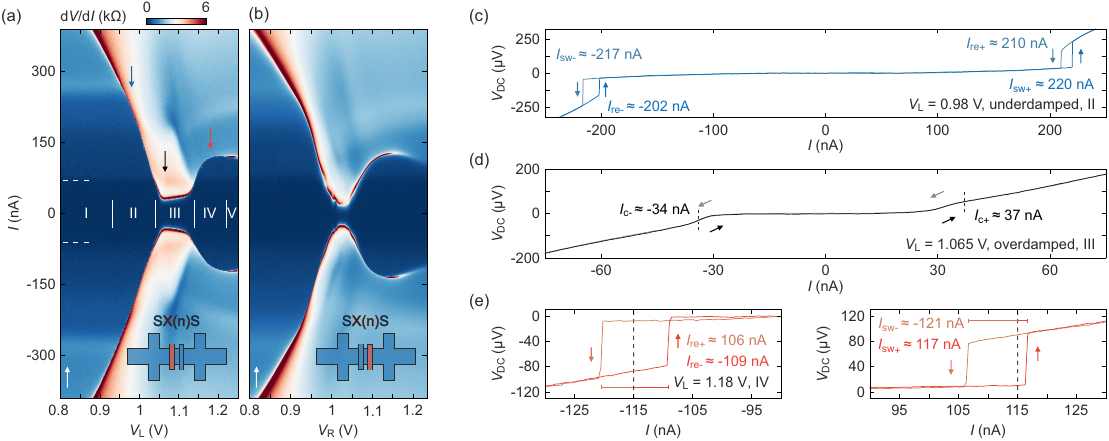}
\caption{(a) Differential resistance $\mathrm d V/\mathrm d I$ as a function of bias current $I$ and $V_\mathrm{L}$ measured at fixed $V_\mathrm{BG}$. 
Inset shows the formation of the JJ by weakly coupling the superconducting leads (S) via the correlated insulator at $\nu = -2$ below $V_\mathrm{L}$ resulting in the formation of a SX($n$)S junction, where $\mathrm{X}(n)$ denotes the gate-tunability of the weak link defined by the correlated insulator. 
White arrow indicates the sweep direction of the bias current when taking the map.
(b) An analogous measurement to panel (a), but using the right top gate to form the JJ.
(c-e) Current-voltage characteristics for both bias current sweep directions measured at the positions indicated by the colored arrows in panel (a).
}
\label{f2}
\end{figure*}

In \cref{f2}(a,b) we show the differential resistance $\mathrm{d}V/\mathrm{d}I$ as a function of bias current $I$ and the top gate voltages $V_\mathrm L$ or $V_\mathrm R$ in the area, where the correlated insulator intersects the superconducting regime in \cref{f1}(c,d).
The measurement reveals sharp maxima in differential resistance, roughly corresponding to the critical current $I_\mathrm{c+}$ or $I_\mathrm{c-}$ (where $I_\mathrm{c+}$, $I_\mathrm{c-}$ are the critical currents at positive and negative bias, respectively), which tunes with $V_\mathrm{L,R}$.
For both top gates $|I_\mathrm{c\pm}|$ exhibits a minimum at a top gate voltage which corresponds to the value where the diagonal feature of the $\nu = -2$ correlated insulator in \cref{f1}(c,d) intersects the superconducting region at $V_\mathrm{BG} = -3.34 \, \mathrm V$.
At this particular back gate voltage, measurements as a function of bias current (see Supporting Information)  reveal the largest critical current, therefore we fix this voltage in all further experiments.
Line cuts taken at different gate voltages indeed show the characteristic current-voltage curve of a JJ and reveal a continuous tuning between underdamped and overdamped junction regimes \cite{tinkham1975}.
We observe overdamped behavior of the junction dynamics in three distinct regimes, labeled as I, III and V in \cref{f2}(a), while we observe underdamped behavior evidenced by hysteresis in regimes II and IV. 
In \cref{f2}(c-e) we show three exemplary line cuts taken at the positions indicated by the arrows in \cref{f2}(a) (see Supporting Information for line cuts taken in all regions).
Independent of the presence of the JD effect, a residual resistance is observed at finite $I$, independent of $V_\mathrm{L}$ or $V_\mathrm{R}$, as indicated by the transition from dark to lighter blue regions in \cref{f2}(a,b) [see white dashed lines in \cref{f2}(a)]. 
This residual resistance might be explained by the presence of a soft superconducting gap or pseudogap \cite{Oh2021Dec}, or a disorder-induced loss of the phase stiffness \cite{Sacepe2020Jul}. 

Focusing on the critical current, an inspection of \cref{f2}(c) and \cref{f2}(e) reveals that the switching (sw) and retrapping (re) currents for positive and negative bias are unequal within our measurement resolution, i.e. $I_\mathrm{sw+} \neq |I_\mathrm{sw-}|$ and $I_\mathrm{re+} \neq |I_\mathrm{re-}|$.
Since in an underdamped junction the switching current corresponds to the critical current $I_\mathrm{c}$, this observation implies that a JD behavior is present.
The efficiency of such a JD can be quantified by the fraction
\begin{align}
    \eta_j = \frac{I_{j+} - |I_{j-}|}{I_{j+} + |I_{j-}|}, \label{diode_eff}
\end{align}
where $j$ denotes either the switching or retrapping current for the underdamped regions or the critical current (c) for the overdamped regions. 
For the underdamped regions II and IV, we can directly read off the switching (and retrapping current) [see labels in \cref{f2}(c,e)] yielding diode efficiencies of $\eta_\mathrm{sw} = 0.7\, \%$  ($\eta_\mathrm{re} = 1.9 \, \%$) and $\eta_\mathrm{sw} = -1.7 \,\%$  ($\eta_\mathrm{re} = -1.4 \,\%$), respectively.
Moreover, the observation of a JD behavior is not restricted to the hysteretic regions II and IV. 
From the non-hysteretic region III [\cref{f2}(d)], we extract the critical current by fitting the Ivanchenko-Zil'berman model \cite{Glick2017Oct,Ivanchenko1969Jun,Ambegaokar1969Jun} (see Supporting Information for details) to the current-voltage curves which results in a diode efficiency of $\eta_\mathrm{c} \approx 4.2 \%$.

Despite the similar appearance of the junction regimes in \cref{f2}(a,b), we observe several differences between the two junctions.
For example, the minimum critical current is lower for the left junction compared to the right junction.
The gate voltages corresponding to the minimum critical current also differ, with $V_\mathrm{L} \approx 1.06~\mathrm{V}$ and $V_\mathrm{R} \approx 1.03~\mathrm{V}$.
Due to the identical geometries of both top gates and their close spatial distance this observation hints towards the presence and influence of disorder to the transport characteristics of Josephson diodes in MATBG. 
\begin{figure*}[!tbh]
\centering
\includegraphics[draft=false,keepaspectratio=true,clip,width=1\linewidth]{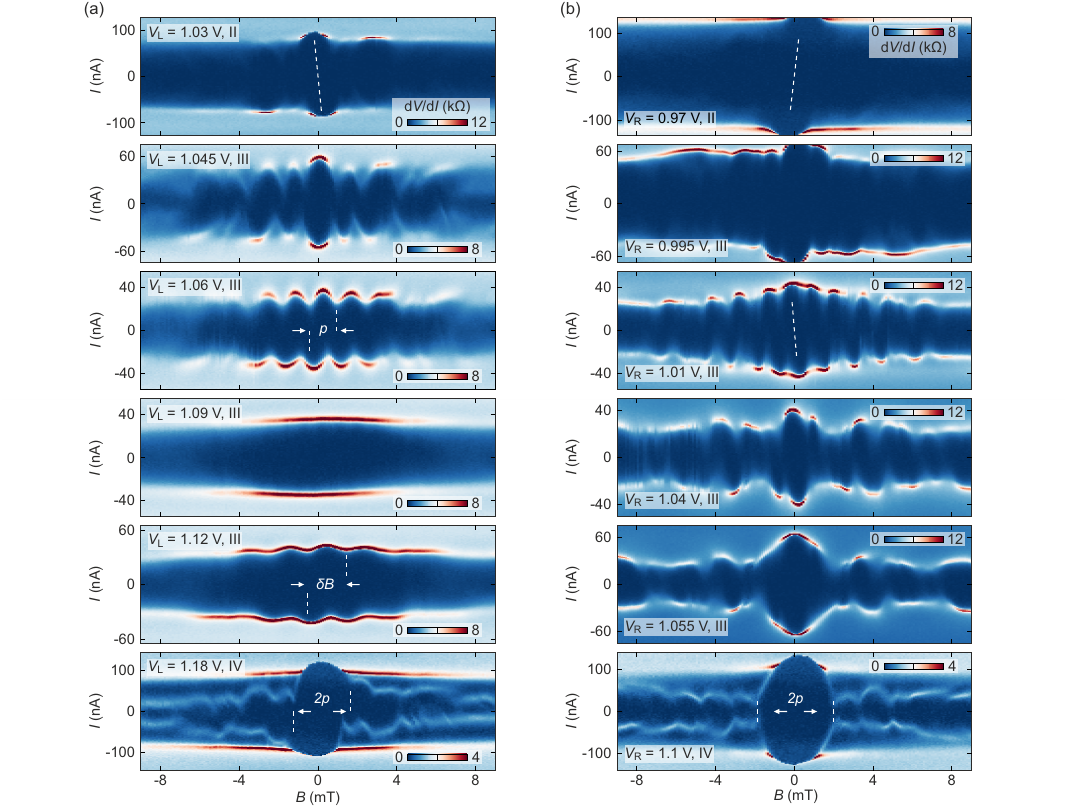}
\caption{(a) Magnetospectroscopy measurements taken in the regions II-IV of the junction regime shown in \cref{f2}(a) for different fixed $V_\mathrm{L}$.
The interference pattern exhibit a complex dependence on the charge carrier density of the weak link.
The bias current is swept on the fast axis from negative to positive values, while the magnetic field is stepped on the slow axis from negative to positive values. 
Before each map is taken, we initialize the magnet at $B = -0.75 \, \mathrm{T}$.
(b) Same as in panel (a) but measured for the junction tuned by $V_\mathrm{R}$ as shown in \cref{f2}(b). 
}
\label{f3}
\end{figure*}
To study the JD behavior in more detail, we study the interference patterns in out-of-plane magnetic field for various fixed gate voltages as shown in \cref{f3}(a,b). 
Several interesting observations can be made in these patterns. 
First, they are not symmetric with respect to an inversion of the current axis ($I \rightarrow -I$), giving rise to the JD effect.
The same is true for an isolated inversion of the out-of-plane magnetic field axis ($B \rightarrow -B$).
However, all patterns exhibit an almost perfect point symmetry with respect to a combined inversion of these two axes ($I \rightarrow -I$ and $B \rightarrow -B$), suggesting that time-reversal symmetry is only externally broken due to the applied out-of-plane magnetic field.
We investigated bi-directional magnetic field sweeps and found no evidence of hysteresis, indicating no intrinsically broken time reversal symmetry in our device (see Methods and Supporting Information).

Second, the skewness (quantified by the offset $\delta B$ of the center of the central oscillation) of the interference pattern tuned by $V_\mathrm{L}$ in \cref{f3}(a) changes sign as the applied voltage $V_\mathrm{L}$ increases from $1.03 \, \mathrm V$ to $1.06 \, \mathrm V$ (see dashed lines).
This is associated with an inversion of the diode polarity as already seen in \cref{f2}(c,e).
Third, the periodicity of the oscillations change with gate voltage. 
For $V_\mathrm{L} = 1.045\, \mathrm V$, multiple frequency components emerge, while only one dominant frequency component is visible in other patterns (e.g. $V_\mathrm{L} = 1.06~$V). 
Furthermore, the oscillations disappear completely for $V_\mathrm{L} = 1.09\, \mathrm{V}$ and reappear for higher gate voltages. 
Finally, the prominence of the central oscillation changes between the interference patterns.
For example, at $V_\mathrm{L} = 1.06 \,\mathrm{V}$, the interference pattern resembles that of an asymmetric superconducting quantum interference device (SQUID)~\cite{Portoles2022Nov} since the supercurrent extrema of the central oscillation are similar to that of the surrounding.
On the other hand, for $V_\mathrm{L} = 1.18 \  \mathrm V$, the prominence of the central oscillation increases.

The interference patterns for the junction tuned by $V_\mathrm{R}$ are generally more complex and exhibit multiple frequency components.
Nevertheless, similarities are also evident. 
For instance, the interference patterns are skewed left at certain voltages ($V_\mathrm{R} = 1.01\, \mathrm V$, $V_\mathrm{R} = 1.04\, \mathrm V$) and right at others ($V_\mathrm{R} = 0.97\, \mathrm V$, $V_\mathrm{R} = 1.1\, \mathrm V$). 
Furthermore, the prominence of the central oscillation increases at $V_\mathrm{R} = 1.055\, \mathrm V$ and $V_\mathrm{R} = 1.1\, \mathrm V$.

\begin{figure}[!tbh]
\centering
\includegraphics[draft=false,keepaspectratio=true,clip,width=1\linewidth]{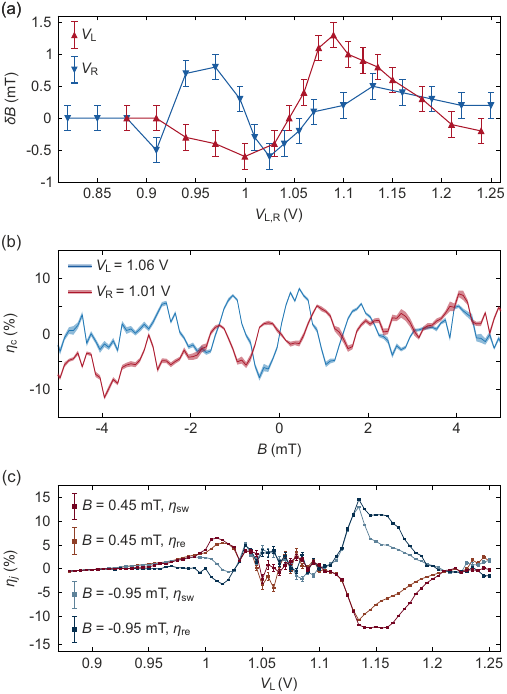}
\caption{(a) Extracted shift in magnetic field $\delta B$ between the supercurrent maxima of the central interference for positive and negative bias currents as a function of $V_\mathrm{L}$ (red trace) and $V_\mathrm{R}$ (blue trace).
The shifts are extracted from magnetospectroscopy measurements, partially shown in \cref{f3}(a,b).
The error bars correspond to an estimated reading uncertainty of $0.2 \, \mathrm{mT}$.
(b) Diode efficiency as a function of magnetic field extracted from the interference pattern taken at $V_\mathrm L = 1.06 \, \mathrm V$ and $V_\mathrm R = 1.01 \, \mathrm V$ in \cref{f3}.
The shaded areas correspond to the errors arising from the fitting procedure.
(c) Diode efficiency $\eta_j$ extracted from measurements at constant magnetic field values of $B = 0.45 \, \mathrm{mT}$ and $B = -0.95 \, \mathrm{mT}$ for the junction defined by $V_\mathrm{L}$.
The diode efficiency is calculated via \cref{diode_eff} for the critical currents $I_\mathrm{sw \pm}$ and analogously for the retrapping currents $I_\mathrm{re \pm}$.
The error bars arise from the fitting method used to extract the critical currents (see Supporting Information).
}
\label{f4}
\end{figure}

We quantify the JD effect by first extracting the skewness of the interference patterns ($\delta B$, defined as the difference in supercurrent maxima between positive and negative bias within the central oscillation) for all investigated gate voltages.
\cref{f4}(a) shows $\delta B$ as a function of $V_\mathrm{L}$ and $V_\mathrm{R}$, showing that for both junctions $\delta B$ changes sign, albeit at different gate voltages. 
Next, we extract the diode efficiency, $\eta_\mathrm{c}$ (calculated with the extracted critical currents, see Eq.~\ref{diode_eff}) for $V_\mathrm{L} = 1.06~$V and $V_\mathrm{R} = 1.01$~V, where pronounced oscillations are visible [see \cref{f4}(b)].
Thus, a strong tuning of $\eta_\mathrm{c}$ is observed as a function of magnetic field.
For large $|B|$ the point symmetry is not perfectly reflected, which is probably due to uncertainty from the fitting procedure to extract the critical current (see Supporting Information).
Lastly, we extract the diode efficiencies $\eta_\mathrm{sw}$ and $\eta_\mathrm{re}$ at fixed magnetic fields for $V_\mathrm{L}$, as shown in \cref{f4}(c).
This demonstrates a tuning of the diode efficiency and reversal of its polarity by changing the gate voltage. 
Similar control over diode behavior was only recently demonstrated in InAs devices with Al electrodes~\cite{Schiela2025Apr}.
Moreover, when comparing \cref{f4}(a) and \cref{f4}(c), the analysis reveals the connection of the inversion of the diode polarity with the sign change of the skewness of the interference pattern.
From \cref{f4}(c), we extract maximal values of the diode efficiency for $B = -0.95$~mT, of $\eta_\mathrm{sw} \approx 14.6~\%$ at $V_\mathrm{L} = 1.135$~V which can be reversed to $\eta_\mathrm{sw} \approx -3.2~\%$ by sweeping the gate to $V_\mathrm{L} = 1.015 \, \mathrm V$. 

\begin{figure*}[!tbh]
\centering
\includegraphics[draft=false,keepaspectratio=true,clip,width=1\linewidth]{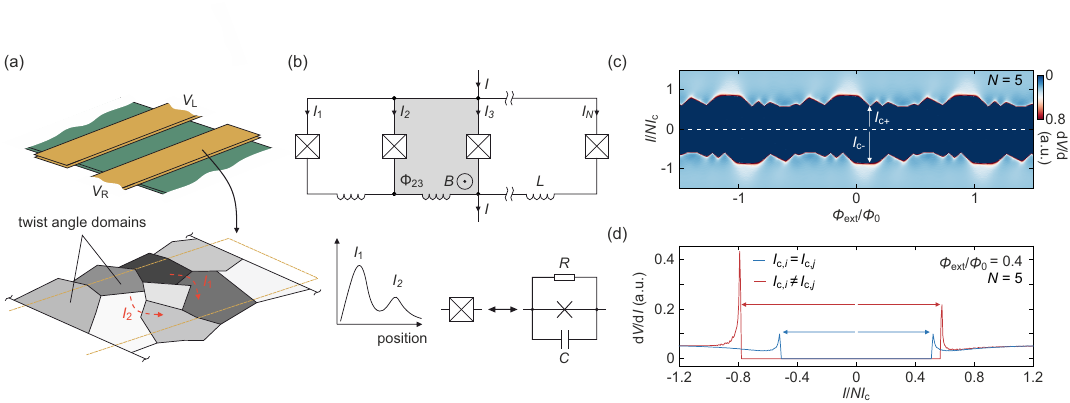}
\caption{(a) Schematic illustration of different twist angle domains below a top gate (projection into the plane). 
The different twist angle domains give rise to multiple current paths.
(b) Circuit diagram schematic showing the resulting JJ network defined by $N$ parallel current paths.
Each current path is described within the RCSJ model where we additionally incorporate the kinetic inductance of the superconductor. 
Gray shading indicates a finite area enclosed by two adjacent current paths which is penetrated by the externally applied magnetic field.
(c) Magnetotransport simulation of the differential resistance as a function of the total current through the junction for $N = 5$ current paths.
The model produces a skewed interference pattern.
(d) Line-cut extracted from panel (c) at fixed $\Phi_\mathrm{ext}/\Phi_0 = 0.4$. 
For equal critical currents of the five individual current paths (blue curve) we do receive an overall symmetric critical current of the JJ for positive and negative bias current. 
For (slightly) different critical currents of the five current paths (red curve) we observe the onset of a JD effect, i.e. the critical current of the overall JJ is not equal for positive and negative bias current.}
\label{f5}
\end{figure*}

In short, we observe that two closely spaced JDs in MATBG exhibit striking differences in supercurrent transmission and interference patterns.
Since both JJs feature the same geometric dimensions, we attribute these differences to disorder within the MATBG.
While hBN encapsulation and a graphite back gate minimize potential disorder, twist angle inhomogeneities remain a known source of disorder in MATBG. 
These inhomogeneities form domains with relatively uniform but distinct twist angles separated by sharp boundaries, occurring on length scales of tens to hundreds of nanometers~\cite{McGilly2020Jul, Schapers2022Jul, Uri2020May, Choi2021Jan, Dolleman2024Apr,Lau2022Feb}. 
For our geometry, we thus expect that domains with different twist angles are covered by a single top gate as illustrated in \cref{f5}(a). 
The different twist angles shift the energy gap of the correlated insulator at $\nu = -2$ between the single domains relative to the electrochemical potential, resulting in a non-uniform supercurrent distribution across the junction. 
Since these domains are covered by a single top gate, changing the gate voltage of this top gate can tune the current distribution in a complex manner by selectively enhancing or reducing the current path contribution of a particular domain, which may account for the observed changes in the shape and frequency of the interference patterns, as shown in \cref{f3}.
Furthermore, due to the spatial distance of the two top gates and their geometric dimension, it is likely that both top gates cover different domains, giving a potential explanation of the observed differences between the two JJs observed in \cref{f2}(a,b) and \cref{f3}. 

The inversion symmetry breaking required to observe a JD effect can, in principle, be induced by spin-orbit coupling~\cite{Reinhardt2024May} or by an unequal current distribution across the width of the junction in the presence of finite kinetic inductance~\cite{Kudriashov2025Jun}. Since spin-orbit coupling is generally weak in graphene, we rule out this mechanism and focus our discussion on the effect of non-uniform current distributions.
Such non-uniformity can naturally arise from twist-angle domains, which generate distinct current paths enclosing finite areas [\cref{f5}(a)], leading to the exemplary current distribution shown in the lower left of \cref{f5}(b). 
A current distribution breaking inversion symmetry produces a circulating component to the supercurrent near the junction, which induces a magnetic (self-)field proportional to both the current and the inductance of the superconductor~\cite{Barone1982Jul}.
This inductance is set by the effective mass of the Cooper pairs~\cite{tinkham1975} and is therefore termed the {\it kinetic inductance}. It gives rise to a phase shift between distinct current paths, which depends on both the current direction (see Supporting Information) and the external magnetic flux threading the circulating loop. As a result, the kinetic inductance skews the interference pattern characteristic of the JD effect~\cite{Nadeem2023Sep}.

To investigate the influence of the kinetic inductance, we can estimate its value from the experimental interference patterns in \cref{f3}.
First, we estimate an effective junction area from the oscillation period $p$ using $A_\mathrm{eff} = \Phi_0^*/p$, where $\Phi_0^*$ is the superconducting flux quantum. 
We start with the interference patterns that most closely resemble a Fraunhofer pattern, for example at \(V_\mathrm{L} = 1.18 \, \mathrm{V}\) and \(V_\mathrm{R} = 1.10 \, \mathrm{V}\), where the width of the central lobe corresponds to \(2p\). From this, we obtain \(A_\mathrm{eff} \approx 1.5 \, \mu\mathrm{m}^2\) and \(A_\mathrm{eff} \approx 1.1 \, \mu\mathrm{m}^2\), respectively.
In addition, for the interference patterns resembling a SQUID-like behavior at $V_\mathrm{L} = 1.06 \, \mathrm V$ and $V_\mathrm{L} = 1.12 \, \mathrm{V}$ we take the periodicity and find $A_\mathrm{eff} = 1.5 \, \mathrm{\mu m^2}$ and $A_\mathrm{eff} \approx 1 \, \mathrm{\mu m^2}$.
These values exceed the gate-defined area of $0.2 \, \mathrm{\mu m} \times 2.5 \, \mathrm{\mu m} = 0.5 \, \mathrm{\mu m^2}$, which can be attributed to the fact that the film thickness is significantly smaller than the London penetration depth~\cite{Randle2023Sep}.
In the second step, we use the skewness of the exemplary interference patterns to extract the flux difference between positive and negative currents, \(\Delta \Phi\).
Combined with the critical current, this analysis enables us to estimate the maximum possible kinetic inductance, assuming a maximally imbalanced SQUID model.
In that case, the maximum kinetic inductance is given by
$L_\mathrm{K,max} = |\Delta \Phi|/(2I_{c+})$ \cite{Portoles2022Nov}.
Using this approach, we find for $V_\mathrm{L} = 1.06 \, \mathrm V$, $V_\mathrm{L} = 1.12 \, \mathrm{V}$, $V_\mathrm{L} = 1.18$~V and $V_\mathrm R = 1.1 \, \mathrm{V}$ kinetic inductance values of $L_\mathrm{K,max} \approx 8.7 \, \mathrm{nH}$, $L_\mathrm{K,max} \approx 11.2 \, \mathrm{nH}$, $L_\mathrm{K,max} \approx 1.9 \, \mathrm{nH}$ and $L_\mathrm{K,max} \approx 0.9 \, \mathrm{nH}$, respectively.

For comparison with literature, we use the thin-film expression of \citeauthor{Annunziata2010Oct} in the limit \(T \ll T_\mathrm{c}\),  
\(L_\mathrm{K} \approx L R_\square h /(2 \pi^2 W \Delta \tanh(\Delta/2 k_B T))\)~\cite{Annunziata2010Oct}.
Here, $T_\mathrm{c} \approx 0.7~$K is the critical temperature, $L, W$ are the length and width of the superconducting region, respectively, $h$ is Planck's constant, $k_\mathrm{B}$ is the Boltzmann constant, $R_\square$ is the sheet resistance and $\Delta \approx 1.76k_BT_\mathrm{c}$ is the superconducting gap.
In our case, $L$ and $W$ roughly correspond to the length of a twist angle domain and to the penetration depth, respectively, which we both estimate to be around $100 \, \mathrm{nm}$.
Consequently, we find $L_\mathrm{K} \approx 4~$nH which is in the correct order of magnitude. 
This estimate shows that kinetic inductance can indeed explain the observed diode effect.
Note that in MATBG the kinetic inductance is exceptionally large, arising from the high effective electron mass and elevated normal-state resistance compared to other materials~\cite{Lopez-Nunez2023Nov,Portoles2025May, Banszerus2025Feb, Jha2024Mar}.

Next, we consider a more comprehensive model of the junction, treating it as a multi-path interferometer composed of a network of \(N\) parallel current paths.
Each path is represented as an ideal Josephson junction within the resistively and capacitively shunted junction (RCSJ) method (see Supporting Information), as illustrated in Fig.~\ref{f5}(b).
This provides more insight on the effect of the non-uniform current distribution and the kinetic inductance, and how it leads to the gate-tunability of the diode efficiency and polarity. 
To reproduce the situation in our experimental junction, we randomly vary the area between each current path, and vary the critical current value between each path.
As an example, simulations with \(N=5\) current paths [Fig.~\ref{f5}(c)] reproduce both the skewness and the point symmetry observed in the experimental interference pattern [Fig.~\ref{f3}].
For finite magnetic flux, a Josephson diode effect emerges, with $I_\mathrm{c+} \neq -|I_\mathrm{c-}|$, as illustrated by the red curve in \cref{f5}(d).
Importantly, when all current paths have identical critical currents, the interference pattern remains symmetric ($I_\mathrm{c+} = -|I_\mathrm{c-}|$), as shown by the blue curve in \cref{f5}(d), highlighting the importance of the imbalance between the different current paths \cite{Hooper2004Jul, Gupta2023May}.
Note that the polarity of the JD crucially depends on the current distribution across the junction. For example, inverting the current distribution shown in the lower left panel of Fig.~\ref{f5}(b) would reverse the diode polarity.
The exact shape of the simulated interference pattern — and the associated diode effect — depends on the number of contributing paths and the variance of their critical currents, which we systematically study in the Supporting Information. 
Based on these simulation results, we argue that the transmittance through the junction strongly depends on the twist angle disorder, leading to a high sensitivity to the applied gate voltage controlling the junction. 
As a consequence, even small changes in gate voltage strongly affect the distribution of current paths and their effective areas, leading to pronounced modifications of the interference pattern, as observed in Fig.~\ref{f3}.

In conclusion, we have demonstrated a gate-tunable Josephson diode effect in MATBG JJs near filling factor $\nu =-2$. By carefully studying two neighboring junctions defined within the same device, we identify two key factors underlying this behavior: (i) a non-uniform supercurrent distribution arising from microscopic twist-angle variations, and (ii) the intrinsically large kinetic inductance of MATBG, rooted in its large effective Cooper pair mass. Twist-angle disorder, which strongly affects current distributions across the junction, allows the local top gate that defines the junction to substantially reshape the supercurrent profile. As a direct consequence, the diode efficiency becomes gate-tunable, with the possibility of reversing its polarity at fixed magnetic field.
Josephson network simulations reproduce the central features of the experimental data and provide strong support for this interpretation. Importantly, our results show that neither orbital magnetism nor other intrinsic mechanisms of time-reversal symmetry breaking are required to account for the Josephson diode effect in MATBG.
More broadly, this work highlights the crucial role of microscopic disorder and kinetic inductance in shaping nonreciprocal superconducting transport in moiré materials. The ability to achieve gate-controlled and reversible diode polarity establishes MATBG JJs as a versatile platform for programmable superconducting electronics. Beyond their fundamental significance, these findings might be interesting for the design of energy-efficient superconducting rectifiers, reconfigurable superconducting circuits, and the controlled exploration of nonreciprocal phenomena in correlated 2D materials.

\textbf{Acknowledgements} 
This project has received funding from the European Research Council (ERC) under grant agreement No. 820254, the Deutsche Forschungsgemeinschaft (DFG, German Research Foundation) through SPP 2244 (Project No. 535377524) and under Germany’s Excellence Strategy - Cluster of Excellence Matter and Light for Quantum Computing (ML4Q) EXC 2004/1 - 390534769, the FLAG-ERA grants No. 437214324 TATTOOS, No. 471733165 PhotoTBG, No. 534269806 ThinQ by the DFG, and by the Helmholtz Nano Facility~\cite{Albrecht2017May}. K.W.~and T.T.~acknowledge support from the JSPS KAKENHI (Grant Numbers 21H05233 and 23H02052), the CREST (JPMJCR24A5) and World Premier International Research Center Initiative (WPI), MEXT, Japan. L.K.~acknowledges support by the DFG through the Würzburg-Dresden Cluster of Excellence on Complexity and Topology in Quantum Matter -- ct.qmat, Project-ID 390858490 -- EXC 2147, and through the Research Unit QUAST, Project-ID 449872909 -- FOR5249. 
The authors thank Sebastian Staacks for the preparation of the rendered device schematic.\\

\textbf{Author contributions} 
C.S. and A.R. conceived the experiment. 
A.R. built the device with help from A.A..
K.W. and T.T. provided the hBN crystals.
A.R. and R.J.D. performed the measurements.
F.V. provided assistence with the measurement setup.
A.R. and R.J.D. performed the data analysis.
A.R., R.J.D., L.K., F.H., L.B., B.B. and C.S. discussed the experimental data. 
L.B. and F.H. provided insights into the underlying theoretical model. 
L.K. implemented the model and provided the theoretical simulations. 
A.R., and R.J.D. wrote the manuscript with input from L.K., F.H., L.B., B.B. and C.S.. 

\textbf{Data availability} 
The data supporting the findings of this study are available in a Zenodo repository under \url{https://doi.org/10.5281/zenodo.17376229}.

\section*{Methods}
\subsection*{Device Fabrication}
For the exfoliation of graphene, graphite (source: Naturgraphit GmbH) and hBN we use commercially available $\mathrm{SiO_2/Si}$ wafer with a $90$~nm oxide thickness.
Suitable flakes are then identified via optical microscopy. The fabrication of the van-der-Waals heterostructure is performed with the "laser cut and stack" \cite{Park2021Apr} technique using a  polycarbonate/polydimethylsiloxane (PC/PDMS) droplet \cite{Kim2016Mar,Bisswanger2022Jun}. 
During the assembly of the heterostructure we adjust a twist angle of $1.2^\circ$, slightly overshooting the aimed magic angle of $1.1^\circ$ to compensate for relaxation effects.
We pick up the top and bottom hBN flakes (thicknesses for both flakes are approximately $35$~nm) at a temperature of $90\, ^\circ \mathrm{C}$, while the two graphene pick-ups are performed at $40\, \mathrm{^\circ C}$.
The final pick-up of the graphite flake is performed at $120\, \mathrm{^\circ C}$.
The finished stacked is placed on a marked $\mathrm{SiO_2/Si}$ chip at $165 \, \mathrm{^\circ C}$.
We remove the PC in chloroform and clean the heterostructure subsequently in isopropyl alcohol.
We first fabricate ohmic contacts to the MATBG by means of electron beam lithography, reactive ion etching ($\mathrm{CF_4/O_2}$) and metal evaporation ($5/50$~nm Cr/Au) \cite{Wang2013Nov, Uwanno2018Aug}.
In the next step, we evaporate the leads to the ohmic contacts together with the top gates ($5/70$~nm Cr/Au) ontop of our heterostructure. 
The final Hall bar shape is defined after the fabrication of a hardmask ($85$~nm Al) by means of atomic layer etching ($\mathrm{SF_6/Ar/O_2}$).
Finally, we remove the Al hardmask with tetramethylammonium hydroxide (TMAH; $2.38\%$ in DI water).

\subsection*{Measurement Setup}
All measurements presented in this manuscript were performed in a dilution ${}^3\mathrm{He}/{}^4\mathrm{He}$ dilution refrigerator at a base temperature of $50 \, \mathrm{mK}$.
We bias our sample via a $10 \, \mathrm{M \Omega}$ resistor using standard low-frequency AC (Stanford Research SR830) and DC (Yokogawa GS200) measurement techniques. 
The recorded signal were amplified with a low-noise differential amplifier (Stanford Research SR560).
All measurements were performed in a four-terminal setup (see \cref{f1}(a)). 

\subsection*{Twist angle extraction}
The twist angle of the device is extracted from the Landau fan given in the Supporting Information. 
The charge carrier density corresponding to full filling of the moiré superlattice unit cell is extracted to be $n_s = (2.75 \pm 0.05) \times 10^{12} \, \mathrm{cm^{-2}}$.
By using the relation $\theta = \sqrt{3n_sa^2/8}$ \cite{Cao2018Apr2}, where $a = 0.246 \, \mathrm{nm}$ is the lattice constant of graphene, we extract a twist angle of $\theta = 1.09^\circ \pm 0.01^\circ$.\\

\subsection*{Magnetic hysteresis and residual field correction}
In the Supporting Information, we find a hysteresis in $I_\mathrm{c}$ that is similar to that reported by \citeauthor{Diez-Merida2023Apr}~\cite{Diez-Merida2023Apr}.
However, control experiments where $V_{\mathrm{BG}}$ was tuned away from superconductivity and $V_{\mathrm{TG}}$ was tuned away from the correlated insulator at $\nu = -2$, revealed an identical hysteresis loop. 
Based on this observation, we conclude that any magnetic hysteresis can be attributed to an artifact in our experimental setup, and we find no evidence of intrinsic time reversal symmetry breaking. 
To ensure a consistent magnetic field, we initialized our superconducting magnet to $B = -750$~mT before each measurement presented in \cref{f3,f4}.
This field exceeds the saturation field of the magnetic hysteresis (see Supporting Information). 
We find that each of the interference patterns is shifted by a constant magnetic field of $\approx 0.55 \, \mathrm{mT}$ and thus the point symmetry is present with respect to $(I,B) \approx (0 \, \mathrm{nA}, 0.55 \, \mathrm{mT})$ (not regarding the effect of switching and retrapping current) in the raw data. 
We correct for this constant shift induced by the residual field by defining $B = B_\mathrm{raw} - 0.55 \, \mathrm{mT}$, where $B_\mathrm{raw}$ and $B$ are the nominal and effective out-of-plane magnetic field of our magnet.


\begin{thebibliography}{70}%
\makeatletter
\providecommand \@ifxundefined [1]{%
 \@ifx{#1\undefined}
}%
\providecommand \@ifnum [1]{%
 \ifnum #1\expandafter \@firstoftwo
 \else \expandafter \@secondoftwo
 \fi
}%
\providecommand \@ifx [1]{%
 \ifx #1\expandafter \@firstoftwo
 \else \expandafter \@secondoftwo
 \fi
}%
\providecommand \natexlab [1]{#1}%
\providecommand \enquote  [1]{``#1''}%
\providecommand \bibnamefont  [1]{#1}%
\providecommand \bibfnamefont [1]{#1}%
\providecommand \citenamefont [1]{#1}%
\providecommand \href@noop [0]{\@secondoftwo}%
\providecommand \href [0]{\begingroup \@sanitize@url \@href}%
\providecommand \@href[1]{\@@startlink{#1}\@@href}%
\providecommand \@@href[1]{\endgroup#1\@@endlink}%
\providecommand \@sanitize@url [0]{\catcode `\\12\catcode `\$12\catcode
  `\&12\catcode `\#12\catcode `\^12\catcode `\_12\catcode `\%12\relax}%
\providecommand \@@startlink[1]{}%
\providecommand \@@endlink[0]{}%
\providecommand \url  [0]{\begingroup\@sanitize@url \@url }%
\providecommand \@url [1]{\endgroup\@href {#1}{\urlprefix }}%
\providecommand \urlprefix  [0]{URL }%
\providecommand \Eprint [0]{\href }%
\providecommand \doibase [0]{https://doi.org/}%
\providecommand \selectlanguage [0]{\@gobble}%
\providecommand \bibinfo  [0]{\@secondoftwo}%
\providecommand \bibfield  [0]{\@secondoftwo}%
\providecommand \translation [1]{[#1]}%
\providecommand \BibitemOpen [0]{}%
\providecommand \bibitemStop [0]{}%
\providecommand \bibitemNoStop [0]{.\EOS\space}%
\providecommand \EOS [0]{\spacefactor3000\relax}%
\providecommand \BibitemShut  [1]{\csname bibitem#1\endcsname}%
\let\auto@bib@innerbib\@empty
\bibitem [{\citenamefont {Nadeem}\ \emph {et~al.}(2023)\citenamefont {Nadeem},
  \citenamefont {Fuhrer},\ and\ \citenamefont {Wang}}]{Nadeem2023Sep}%
  \BibitemOpen
  \bibfield  {author} {\bibinfo {author} {\bibfnamefont {M.}~\bibnamefont
  {Nadeem}}, \bibinfo {author} {\bibfnamefont {M.~S.}\ \bibnamefont {Fuhrer}},\
  and\ \bibinfo {author} {\bibfnamefont {X.}~\bibnamefont {Wang}},\ }\bibfield
  {title} {\bibinfo {title} {{The superconducting diode effect}},\ }\href
  {https://doi.org/10.1038/s42254-023-00632-w} {\bibfield  {journal} {\bibinfo
  {journal} {Nat. Rev. Phys.}\ }\textbf {\bibinfo {volume} {5}},\ \bibinfo
  {pages} {558} (\bibinfo {year} {2023})}\BibitemShut {NoStop}%
\bibitem [{\citenamefont {Ando}\ \emph {et~al.}(2020)\citenamefont {Ando},
  \citenamefont {Miyasaka}, \citenamefont {Li}, \citenamefont {Ishizuka},
  \citenamefont {Arakawa}, \citenamefont {Shiota}, \citenamefont {Moriyama},
  \citenamefont {Yanase},\ and\ \citenamefont {Ono}}]{Ando2020Aug}%
  \BibitemOpen
  \bibfield  {author} {\bibinfo {author} {\bibfnamefont {F.}~\bibnamefont
  {Ando}}, \bibinfo {author} {\bibfnamefont {Y.}~\bibnamefont {Miyasaka}},
  \bibinfo {author} {\bibfnamefont {T.}~\bibnamefont {Li}}, \bibinfo {author}
  {\bibfnamefont {J.}~\bibnamefont {Ishizuka}}, \bibinfo {author}
  {\bibfnamefont {T.}~\bibnamefont {Arakawa}}, \bibinfo {author} {\bibfnamefont
  {Y.}~\bibnamefont {Shiota}}, \bibinfo {author} {\bibfnamefont
  {T.}~\bibnamefont {Moriyama}}, \bibinfo {author} {\bibfnamefont
  {Y.}~\bibnamefont {Yanase}},\ and\ \bibinfo {author} {\bibfnamefont
  {T.}~\bibnamefont {Ono}},\ }\bibfield  {title} {\bibinfo {title}
  {{Observation of superconducting diode effect}},\ }\href
  {https://doi.org/10.1038/s41586-020-2590-4} {\bibfield  {journal} {\bibinfo
  {journal} {Nature}\ }\textbf {\bibinfo {volume} {584}},\ \bibinfo {pages}
  {373} (\bibinfo {year} {2020})}\BibitemShut {NoStop}%
\bibitem [{\citenamefont {Miyasaka}\ \emph {et~al.}(2021)\citenamefont
  {Miyasaka}, \citenamefont {Kawarazaki}, \citenamefont {Narita}, \citenamefont
  {Ando}, \citenamefont {Ikeda}, \citenamefont {Hisatomi}, \citenamefont
  {Daido}, \citenamefont {Shiota}, \citenamefont {Moriyama}, \citenamefont
  {Yanase},\ and\ \citenamefont {Ono}}]{Miyasaka2021Jun}%
  \BibitemOpen
  \bibfield  {author} {\bibinfo {author} {\bibfnamefont {Y.}~\bibnamefont
  {Miyasaka}}, \bibinfo {author} {\bibfnamefont {R.}~\bibnamefont
  {Kawarazaki}}, \bibinfo {author} {\bibfnamefont {H.}~\bibnamefont {Narita}},
  \bibinfo {author} {\bibfnamefont {F.}~\bibnamefont {Ando}}, \bibinfo {author}
  {\bibfnamefont {Y.}~\bibnamefont {Ikeda}}, \bibinfo {author} {\bibfnamefont
  {R.}~\bibnamefont {Hisatomi}}, \bibinfo {author} {\bibfnamefont
  {A.}~\bibnamefont {Daido}}, \bibinfo {author} {\bibfnamefont
  {Y.}~\bibnamefont {Shiota}}, \bibinfo {author} {\bibfnamefont
  {T.}~\bibnamefont {Moriyama}}, \bibinfo {author} {\bibfnamefont
  {Y.}~\bibnamefont {Yanase}},\ and\ \bibinfo {author} {\bibfnamefont
  {T.}~\bibnamefont {Ono}},\ }\bibfield  {title} {\bibinfo {title}
  {{Observation of nonreciprocal superconducting critical field}},\ }\href
  {https://doi.org/10.35848/1882-0786/ac03c0} {\bibfield  {journal} {\bibinfo
  {journal} {Appl. Phys. Express}\ }\textbf {\bibinfo {volume} {14}},\ \bibinfo
  {pages} {073003} (\bibinfo {year} {2021})}\BibitemShut {NoStop}%
\bibitem [{\citenamefont {Kawarazaki}\ \emph {et~al.}(2022)\citenamefont
  {Kawarazaki}, \citenamefont {Narita}, \citenamefont {Miyasaka}, \citenamefont
  {Ikeda}, \citenamefont {Hisatomi}, \citenamefont {Daido}, \citenamefont
  {Shiota}, \citenamefont {Moriyama}, \citenamefont {Yanase}, \citenamefont
  {Ognev}, \citenamefont {Samardak},\ and\ \citenamefont
  {Ono}}]{Kawarazaki2022Nov}%
  \BibitemOpen
  \bibfield  {author} {\bibinfo {author} {\bibfnamefont {R.}~\bibnamefont
  {Kawarazaki}}, \bibinfo {author} {\bibfnamefont {H.}~\bibnamefont {Narita}},
  \bibinfo {author} {\bibfnamefont {Y.}~\bibnamefont {Miyasaka}}, \bibinfo
  {author} {\bibfnamefont {Y.}~\bibnamefont {Ikeda}}, \bibinfo {author}
  {\bibfnamefont {R.}~\bibnamefont {Hisatomi}}, \bibinfo {author}
  {\bibfnamefont {A.}~\bibnamefont {Daido}}, \bibinfo {author} {\bibfnamefont
  {Y.}~\bibnamefont {Shiota}}, \bibinfo {author} {\bibfnamefont
  {T.}~\bibnamefont {Moriyama}}, \bibinfo {author} {\bibfnamefont
  {Y.}~\bibnamefont {Yanase}}, \bibinfo {author} {\bibfnamefont {A.~V.}\
  \bibnamefont {Ognev}}, \bibinfo {author} {\bibfnamefont {A.~S.}\ \bibnamefont
  {Samardak}},\ and\ \bibinfo {author} {\bibfnamefont {T.}~\bibnamefont
  {Ono}},\ }\bibfield  {title} {\bibinfo {title} {{Magnetic-field-induced
  polarity oscillation of superconducting diode effect}},\ }\href
  {https://doi.org/10.35848/1882-0786/ac99b9} {\bibfield  {journal} {\bibinfo
  {journal} {Appl. Phys. Express}\ }\textbf {\bibinfo {volume} {15}},\ \bibinfo
  {pages} {113001} (\bibinfo {year} {2022})}\BibitemShut {NoStop}%
\bibitem [{\citenamefont {Itahashi}\ \emph {et~al.}(2020)\citenamefont
  {Itahashi}, \citenamefont {Ideue}, \citenamefont {Saito}, \citenamefont
  {Shimizu}, \citenamefont {Ouchi}, \citenamefont {Nojima},\ and\ \citenamefont
  {Iwasa}}]{Itahashi2020Mar}%
  \BibitemOpen
  \bibfield  {author} {\bibinfo {author} {\bibfnamefont {Y.~M.}\ \bibnamefont
  {Itahashi}}, \bibinfo {author} {\bibfnamefont {T.}~\bibnamefont {Ideue}},
  \bibinfo {author} {\bibfnamefont {Y.}~\bibnamefont {Saito}}, \bibinfo
  {author} {\bibfnamefont {S.}~\bibnamefont {Shimizu}}, \bibinfo {author}
  {\bibfnamefont {T.}~\bibnamefont {Ouchi}}, \bibinfo {author} {\bibfnamefont
  {T.}~\bibnamefont {Nojima}},\ and\ \bibinfo {author} {\bibfnamefont
  {Y.}~\bibnamefont {Iwasa}},\ }\bibfield  {title} {\bibinfo {title}
  {{Nonreciprocal transport in gate-induced polar superconductor SrTiO$_3$}},\
  }\href {https://doi.org/10.1126/sciadv.aay9120} {\bibfield  {journal}
  {\bibinfo  {journal} {Sci. Adv.}\ }\textbf {\bibinfo {volume} {6}},\ \bibinfo
  {pages} {eaay9120} (\bibinfo {year} {2020})}\BibitemShut {NoStop}%
\bibitem [{\citenamefont {Schumann}\ \emph {et~al.}(2020)\citenamefont
  {Schumann}, \citenamefont {Galletti}, \citenamefont {Jeong}, \citenamefont
  {Ahadi}, \citenamefont {Strickland}, \citenamefont {Salmani-Rezaie},\ and\
  \citenamefont {Stemmer}}]{Schumann2020Mar}%
  \BibitemOpen
  \bibfield  {author} {\bibinfo {author} {\bibfnamefont {T.}~\bibnamefont
  {Schumann}}, \bibinfo {author} {\bibfnamefont {L.}~\bibnamefont {Galletti}},
  \bibinfo {author} {\bibfnamefont {H.}~\bibnamefont {Jeong}}, \bibinfo
  {author} {\bibfnamefont {K.}~\bibnamefont {Ahadi}}, \bibinfo {author}
  {\bibfnamefont {W.~M.}\ \bibnamefont {Strickland}}, \bibinfo {author}
  {\bibfnamefont {S.}~\bibnamefont {Salmani-Rezaie}},\ and\ \bibinfo {author}
  {\bibfnamefont {S.}~\bibnamefont {Stemmer}},\ }\bibfield  {title} {\bibinfo
  {title} {{Possible signatures of mixed-parity superconductivity in doped
  polar $\mathrm{SrTi}{\mathrm{O}}_{3}$ films}},\ }\href
  {https://doi.org/10.1103/PhysRevB.101.100503} {\bibfield  {journal} {\bibinfo
   {journal} {Phys. Rev. B}\ }\textbf {\bibinfo {volume} {101}},\ \bibinfo
  {pages} {100503} (\bibinfo {year} {2020})}\BibitemShut {NoStop}%
\bibitem [{\citenamefont {Wakatsuki}\ \emph {et~al.}(2017)\citenamefont
  {Wakatsuki}, \citenamefont {Saito}, \citenamefont {Hoshino}, \citenamefont
  {Itahashi}, \citenamefont {Ideue}, \citenamefont {Ezawa}, \citenamefont
  {Iwasa},\ and\ \citenamefont {Nagaosa}}]{Wakatsuki2017Apr}%
  \BibitemOpen
  \bibfield  {author} {\bibinfo {author} {\bibfnamefont {R.}~\bibnamefont
  {Wakatsuki}}, \bibinfo {author} {\bibfnamefont {Y.}~\bibnamefont {Saito}},
  \bibinfo {author} {\bibfnamefont {S.}~\bibnamefont {Hoshino}}, \bibinfo
  {author} {\bibfnamefont {Y.~M.}\ \bibnamefont {Itahashi}}, \bibinfo {author}
  {\bibfnamefont {T.}~\bibnamefont {Ideue}}, \bibinfo {author} {\bibfnamefont
  {M.}~\bibnamefont {Ezawa}}, \bibinfo {author} {\bibfnamefont
  {Y.}~\bibnamefont {Iwasa}},\ and\ \bibinfo {author} {\bibfnamefont
  {N.}~\bibnamefont {Nagaosa}},\ }\bibfield  {title} {\bibinfo {title}
  {{Nonreciprocal charge transport in noncentrosymmetric superconductors}},\
  }\href {https://doi.org/10.1126/sciadv.1602390} {\bibfield  {journal}
  {\bibinfo  {journal} {Sci. Adv.}\ }\textbf {\bibinfo {volume} {3}},\ \bibinfo
  {pages} {e1602390} (\bibinfo {year} {2017})}\BibitemShut {NoStop}%
\bibitem [{\citenamefont {Lin}\ \emph {et~al.}(2022)\citenamefont {Lin},
  \citenamefont {Siriviboon}, \citenamefont {Scammell}, \citenamefont {Liu},
  \citenamefont {Rhodes}, \citenamefont {Watanabe}, \citenamefont {Taniguchi},
  \citenamefont {Hone}, \citenamefont {Scheurer},\ and\ \citenamefont
  {Li}}]{Lin2022Oct}%
  \BibitemOpen
  \bibfield  {author} {\bibinfo {author} {\bibfnamefont {J.-X.}\ \bibnamefont
  {Lin}}, \bibinfo {author} {\bibfnamefont {P.}~\bibnamefont {Siriviboon}},
  \bibinfo {author} {\bibfnamefont {H.~D.}\ \bibnamefont {Scammell}}, \bibinfo
  {author} {\bibfnamefont {S.}~\bibnamefont {Liu}}, \bibinfo {author}
  {\bibfnamefont {D.}~\bibnamefont {Rhodes}}, \bibinfo {author} {\bibfnamefont
  {K.}~\bibnamefont {Watanabe}}, \bibinfo {author} {\bibfnamefont
  {T.}~\bibnamefont {Taniguchi}}, \bibinfo {author} {\bibfnamefont
  {J.}~\bibnamefont {Hone}}, \bibinfo {author} {\bibfnamefont {M.~S.}\
  \bibnamefont {Scheurer}},\ and\ \bibinfo {author} {\bibfnamefont {J.~I.~A.}\
  \bibnamefont {Li}},\ }\bibfield  {title} {\bibinfo {title} {{Zero-field
  superconducting diode effect in small-twist-angle trilayer graphene}},\
  }\href {https://doi.org/10.1038/s41567-022-01700-1} {\bibfield  {journal}
  {\bibinfo  {journal} {Nat. Phys.}\ }\textbf {\bibinfo {volume} {18}},\
  \bibinfo {pages} {1221} (\bibinfo {year} {2022})}\BibitemShut {NoStop}%
\bibitem [{\citenamefont {Scammell}\ \emph {et~al.}(2022)\citenamefont
  {Scammell}, \citenamefont {Li},\ and\ \citenamefont
  {Scheurer}}]{Scammell2022Mar}%
  \BibitemOpen
  \bibfield  {author} {\bibinfo {author} {\bibfnamefont {H.~D.}\ \bibnamefont
  {Scammell}}, \bibinfo {author} {\bibfnamefont {J.~I.~A.}\ \bibnamefont
  {Li}},\ and\ \bibinfo {author} {\bibfnamefont {M.~S.}\ \bibnamefont
  {Scheurer}},\ }\bibfield  {title} {\bibinfo {title} {{Theory of zero-field
  superconducting diode effect in twisted trilayer graphene}},\ }\href
  {https://doi.org/10.1088/2053-1583/ac5b16} {\bibfield  {journal} {\bibinfo
  {journal} {2D Mater.}\ }\textbf {\bibinfo {volume} {9}},\ \bibinfo {pages}
  {025027} (\bibinfo {year} {2022})}\BibitemShut {NoStop}%
\bibitem [{\citenamefont {Narita}\ \emph {et~al.}(2022)\citenamefont {Narita},
  \citenamefont {Ishizuka}, \citenamefont {Kawarazaki}, \citenamefont {Kan},
  \citenamefont {Shiota}, \citenamefont {Moriyama}, \citenamefont {Shimakawa},
  \citenamefont {Ognev}, \citenamefont {Samardak}, \citenamefont {Yanase},\
  and\ \citenamefont {Ono}}]{Narita2022Aug}%
  \BibitemOpen
  \bibfield  {author} {\bibinfo {author} {\bibfnamefont {H.}~\bibnamefont
  {Narita}}, \bibinfo {author} {\bibfnamefont {J.}~\bibnamefont {Ishizuka}},
  \bibinfo {author} {\bibfnamefont {R.}~\bibnamefont {Kawarazaki}}, \bibinfo
  {author} {\bibfnamefont {D.}~\bibnamefont {Kan}}, \bibinfo {author}
  {\bibfnamefont {Y.}~\bibnamefont {Shiota}}, \bibinfo {author} {\bibfnamefont
  {T.}~\bibnamefont {Moriyama}}, \bibinfo {author} {\bibfnamefont
  {Y.}~\bibnamefont {Shimakawa}}, \bibinfo {author} {\bibfnamefont {A.~V.}\
  \bibnamefont {Ognev}}, \bibinfo {author} {\bibfnamefont {A.~S.}\ \bibnamefont
  {Samardak}}, \bibinfo {author} {\bibfnamefont {Y.}~\bibnamefont {Yanase}},\
  and\ \bibinfo {author} {\bibfnamefont {T.}~\bibnamefont {Ono}},\ }\bibfield
  {title} {\bibinfo {title} {{Field-free superconducting diode effect in
  noncentrosymmetric superconductor/ferromagnet multilayers}},\ }\href
  {https://doi.org/10.1038/s41565-022-01159-4} {\bibfield  {journal} {\bibinfo
  {journal} {Nat. Nanotechnol.}\ }\textbf {\bibinfo {volume} {17}},\ \bibinfo
  {pages} {823} (\bibinfo {year} {2022})}\BibitemShut {NoStop}%
\bibitem [{\citenamefont {Baumgartner}\ \emph {et~al.}(2022)\citenamefont
  {Baumgartner}, \citenamefont {Fuchs}, \citenamefont {Costa}, \citenamefont
  {Reinhardt}, \citenamefont {Gronin}, \citenamefont {Gardner}, \citenamefont
  {Lindemann}, \citenamefont {Manfra}, \citenamefont {Faria~Junior},
  \citenamefont {Kochan}, \citenamefont {Fabian}, \citenamefont {Paradiso},\
  and\ \citenamefont {Strunk}}]{Baumgartner2022Jan}%
  \BibitemOpen
  \bibfield  {author} {\bibinfo {author} {\bibfnamefont {C.}~\bibnamefont
  {Baumgartner}}, \bibinfo {author} {\bibfnamefont {L.}~\bibnamefont {Fuchs}},
  \bibinfo {author} {\bibfnamefont {A.}~\bibnamefont {Costa}}, \bibinfo
  {author} {\bibfnamefont {S.}~\bibnamefont {Reinhardt}}, \bibinfo {author}
  {\bibfnamefont {S.}~\bibnamefont {Gronin}}, \bibinfo {author} {\bibfnamefont
  {G.~C.}\ \bibnamefont {Gardner}}, \bibinfo {author} {\bibfnamefont
  {T.}~\bibnamefont {Lindemann}}, \bibinfo {author} {\bibfnamefont {M.~J.}\
  \bibnamefont {Manfra}}, \bibinfo {author} {\bibfnamefont {P.~E.}\
  \bibnamefont {Faria~Junior}}, \bibinfo {author} {\bibfnamefont
  {D.}~\bibnamefont {Kochan}}, \bibinfo {author} {\bibfnamefont
  {J.}~\bibnamefont {Fabian}}, \bibinfo {author} {\bibfnamefont
  {N.}~\bibnamefont {Paradiso}},\ and\ \bibinfo {author} {\bibfnamefont
  {C.}~\bibnamefont {Strunk}},\ }\bibfield  {title} {\bibinfo {title}
  {{Supercurrent rectification and magnetochiral effects in symmetric Josephson
  junctions}},\ }\href {https://doi.org/10.1038/s41565-021-01009-9} {\bibfield
  {journal} {\bibinfo  {journal} {Nat. Nanotechnol.}\ }\textbf {\bibinfo
  {volume} {17}},\ \bibinfo {pages} {39} (\bibinfo {year} {2022})}\BibitemShut
  {NoStop}%
\bibitem [{\citenamefont {Jeon}\ \emph {et~al.}(2022)\citenamefont {Jeon},
  \citenamefont {Kim}, \citenamefont {Yoon}, \citenamefont {Jeon},
  \citenamefont {Han}, \citenamefont {Cottet}, \citenamefont {Kontos},\ and\
  \citenamefont {Parkin}}]{Jeon2022Sep}%
  \BibitemOpen
  \bibfield  {author} {\bibinfo {author} {\bibfnamefont {K.-R.}\ \bibnamefont
  {Jeon}}, \bibinfo {author} {\bibfnamefont {J.-K.}\ \bibnamefont {Kim}},
  \bibinfo {author} {\bibfnamefont {J.}~\bibnamefont {Yoon}}, \bibinfo {author}
  {\bibfnamefont {J.-C.}\ \bibnamefont {Jeon}}, \bibinfo {author}
  {\bibfnamefont {H.}~\bibnamefont {Han}}, \bibinfo {author} {\bibfnamefont
  {A.}~\bibnamefont {Cottet}}, \bibinfo {author} {\bibfnamefont
  {T.}~\bibnamefont {Kontos}},\ and\ \bibinfo {author} {\bibfnamefont
  {S.~S.~P.}\ \bibnamefont {Parkin}},\ }\bibfield  {title} {\bibinfo {title}
  {{Zero-field polarity-reversible Josephson supercurrent diodes enabled by a
  proximity-magnetized Pt barrier}},\ }\href
  {https://doi.org/10.1038/s41563-022-01300-7} {\bibfield  {journal} {\bibinfo
  {journal} {Nat. Mater.}\ }\textbf {\bibinfo {volume} {21}},\ \bibinfo {pages}
  {1008} (\bibinfo {year} {2022})}\BibitemShut {NoStop}%
\bibitem [{\citenamefont {Wu}\ \emph {et~al.}(2022)\citenamefont {Wu},
  \citenamefont {Wang}, \citenamefont {Xu}, \citenamefont {Sivakumar},
  \citenamefont {Pasco}, \citenamefont {Filippozzi}, \citenamefont {Parkin},
  \citenamefont {Zeng}, \citenamefont {McQueen},\ and\ \citenamefont
  {Ali}}]{Wu2022Apr}%
  \BibitemOpen
  \bibfield  {author} {\bibinfo {author} {\bibfnamefont {H.}~\bibnamefont
  {Wu}}, \bibinfo {author} {\bibfnamefont {Y.}~\bibnamefont {Wang}}, \bibinfo
  {author} {\bibfnamefont {Y.}~\bibnamefont {Xu}}, \bibinfo {author}
  {\bibfnamefont {P.~K.}\ \bibnamefont {Sivakumar}}, \bibinfo {author}
  {\bibfnamefont {C.}~\bibnamefont {Pasco}}, \bibinfo {author} {\bibfnamefont
  {U.}~\bibnamefont {Filippozzi}}, \bibinfo {author} {\bibfnamefont {S.~S.~P.}\
  \bibnamefont {Parkin}}, \bibinfo {author} {\bibfnamefont {Y.-J.}\
  \bibnamefont {Zeng}}, \bibinfo {author} {\bibfnamefont {T.}~\bibnamefont
  {McQueen}},\ and\ \bibinfo {author} {\bibfnamefont {M.~N.}\ \bibnamefont
  {Ali}},\ }\bibfield  {title} {\bibinfo {title} {{The field-free Josephson
  diode in a van der Waals heterostructure}},\ }\href
  {https://doi.org/10.1038/s41586-022-04504-8} {\bibfield  {journal} {\bibinfo
  {journal} {Nature}\ }\textbf {\bibinfo {volume} {604}},\ \bibinfo {pages}
  {653} (\bibinfo {year} {2022})}\BibitemShut {NoStop}%
\bibitem [{\citenamefont {Pal}\ \emph {et~al.}(2022)\citenamefont {Pal},
  \citenamefont {Chakraborty}, \citenamefont {Sivakumar}, \citenamefont
  {Davydova}, \citenamefont {Gopi}, \citenamefont {Pandeya}, \citenamefont
  {Krieger}, \citenamefont {Zhang}, \citenamefont {Date}, \citenamefont {Ju},
  \citenamefont {Yuan}, \citenamefont {Schr{\ifmmode\ddot{o}\else\"{o}\fi}ter},
  \citenamefont {Fu},\ and\ \citenamefont {Parkin}}]{Pal2022Oct}%
  \BibitemOpen
  \bibfield  {author} {\bibinfo {author} {\bibfnamefont {B.}~\bibnamefont
  {Pal}}, \bibinfo {author} {\bibfnamefont {A.}~\bibnamefont {Chakraborty}},
  \bibinfo {author} {\bibfnamefont {P.~K.}\ \bibnamefont {Sivakumar}}, \bibinfo
  {author} {\bibfnamefont {M.}~\bibnamefont {Davydova}}, \bibinfo {author}
  {\bibfnamefont {A.~K.}\ \bibnamefont {Gopi}}, \bibinfo {author}
  {\bibfnamefont {A.~K.}\ \bibnamefont {Pandeya}}, \bibinfo {author}
  {\bibfnamefont {J.~A.}\ \bibnamefont {Krieger}}, \bibinfo {author}
  {\bibfnamefont {Y.}~\bibnamefont {Zhang}}, \bibinfo {author} {\bibfnamefont
  {M.}~\bibnamefont {Date}}, \bibinfo {author} {\bibfnamefont {S.}~\bibnamefont
  {Ju}}, \bibinfo {author} {\bibfnamefont {N.}~\bibnamefont {Yuan}}, \bibinfo
  {author} {\bibfnamefont {N.~B.~M.}\ \bibnamefont
  {Schr{\ifmmode\ddot{o}\else\"{o}\fi}ter}}, \bibinfo {author} {\bibfnamefont
  {L.}~\bibnamefont {Fu}},\ and\ \bibinfo {author} {\bibfnamefont {S.~S.~P.}\
  \bibnamefont {Parkin}},\ }\bibfield  {title} {\bibinfo {title} {{Josephson
  diode effect from Cooper pair momentum in a topological semimetal}},\ }\href
  {https://doi.org/10.1038/s41567-022-01699-5} {\bibfield  {journal} {\bibinfo
  {journal} {Nat. Phys.}\ }\textbf {\bibinfo {volume} {18}},\ \bibinfo {pages}
  {1228} (\bibinfo {year} {2022})}\BibitemShut {NoStop}%
\bibitem [{\citenamefont {Bauriedl}\ \emph {et~al.}(2022)\citenamefont
  {Bauriedl}, \citenamefont {B{\ifmmode\ddot{a}\else\"{a}\fi}uml},
  \citenamefont {Fuchs}, \citenamefont {Baumgartner}, \citenamefont {Paulik},
  \citenamefont {Bauer}, \citenamefont {Lin}, \citenamefont {Lupton},
  \citenamefont {Taniguchi}, \citenamefont {Watanabe}, \citenamefont {Strunk},\
  and\ \citenamefont {Paradiso}}]{Bauriedl2022Jul}%
  \BibitemOpen
  \bibfield  {author} {\bibinfo {author} {\bibfnamefont {L.}~\bibnamefont
  {Bauriedl}}, \bibinfo {author} {\bibfnamefont {C.}~\bibnamefont
  {B{\ifmmode\ddot{a}\else\"{a}\fi}uml}}, \bibinfo {author} {\bibfnamefont
  {L.}~\bibnamefont {Fuchs}}, \bibinfo {author} {\bibfnamefont
  {C.}~\bibnamefont {Baumgartner}}, \bibinfo {author} {\bibfnamefont
  {N.}~\bibnamefont {Paulik}}, \bibinfo {author} {\bibfnamefont {J.~M.}\
  \bibnamefont {Bauer}}, \bibinfo {author} {\bibfnamefont {K.-Q.}\ \bibnamefont
  {Lin}}, \bibinfo {author} {\bibfnamefont {J.~M.}\ \bibnamefont {Lupton}},
  \bibinfo {author} {\bibfnamefont {T.}~\bibnamefont {Taniguchi}}, \bibinfo
  {author} {\bibfnamefont {K.}~\bibnamefont {Watanabe}}, \bibinfo {author}
  {\bibfnamefont {C.}~\bibnamefont {Strunk}},\ and\ \bibinfo {author}
  {\bibfnamefont {N.}~\bibnamefont {Paradiso}},\ }\bibfield  {title} {\bibinfo
  {title} {{Supercurrent diode effect and magnetochiral anisotropy in few-layer
  NbSe$_2$}},\ }\href {https://doi.org/10.1038/s41467-022-31954-5} {\bibfield
  {journal} {\bibinfo  {journal} {Nat. Commun.}\ }\textbf {\bibinfo {volume}
  {13}},\ \bibinfo {pages} {4266} (\bibinfo {year} {2022})}\BibitemShut
  {NoStop}%
\bibitem [{\citenamefont
  {D{\ifmmode\acute{\imath}\else\'{\i}\fi}ez-M{\ifmmode\acute{e}\else\'{e}\fi}rida}\
  \emph {et~al.}(2023)\citenamefont
  {D{\ifmmode\acute{\imath}\else\'{\i}\fi}ez-M{\ifmmode\acute{e}\else\'{e}\fi}rida},
  \citenamefont
  {D{\ifmmode\acute{\imath}\else\'{\i}\fi}ez-Carl{\ifmmode\acute{o}\else\'{o}\fi}n},
  \citenamefont {Yang}, \citenamefont {Xie}, \citenamefont {Gao}, \citenamefont
  {Senior}, \citenamefont {Watanabe}, \citenamefont {Taniguchi}, \citenamefont
  {Lu}, \citenamefont {Higginbotham}, \citenamefont {Law},\ and\ \citenamefont
  {Efetov}}]{Diez-Merida2023Apr}%
  \BibitemOpen
  \bibfield  {author} {\bibinfo {author} {\bibfnamefont {J.}~\bibnamefont
  {D{\ifmmode\acute{\imath}\else\'{\i}\fi}ez-M{\ifmmode\acute{e}\else\'{e}\fi}rida}},
  \bibinfo {author} {\bibfnamefont {A.}~\bibnamefont
  {D{\ifmmode\acute{\imath}\else\'{\i}\fi}ez-Carl{\ifmmode\acute{o}\else\'{o}\fi}n}},
  \bibinfo {author} {\bibfnamefont {S.~Y.}\ \bibnamefont {Yang}}, \bibinfo
  {author} {\bibfnamefont {Y.-M.}\ \bibnamefont {Xie}}, \bibinfo {author}
  {\bibfnamefont {X.-J.}\ \bibnamefont {Gao}}, \bibinfo {author} {\bibfnamefont
  {J.}~\bibnamefont {Senior}}, \bibinfo {author} {\bibfnamefont
  {K.}~\bibnamefont {Watanabe}}, \bibinfo {author} {\bibfnamefont
  {T.}~\bibnamefont {Taniguchi}}, \bibinfo {author} {\bibfnamefont
  {X.}~\bibnamefont {Lu}}, \bibinfo {author} {\bibfnamefont {A.~P.}\
  \bibnamefont {Higginbotham}}, \bibinfo {author} {\bibfnamefont {K.~T.}\
  \bibnamefont {Law}},\ and\ \bibinfo {author} {\bibfnamefont {D.~K.}\
  \bibnamefont {Efetov}},\ }\bibfield  {title} {\bibinfo {title}
  {{Symmetry-broken Josephson junctions and superconducting diodes in
  magic-angle twisted bilayer graphene}},\ }\href
  {https://doi.org/10.1038/s41467-023-38005-7} {\bibfield  {journal} {\bibinfo
  {journal} {Nat. Commun.}\ }\textbf {\bibinfo {volume} {14}},\ \bibinfo
  {pages} {2396} (\bibinfo {year} {2023})}\BibitemShut {NoStop}%
\bibitem [{\citenamefont {Chen}\ \emph {et~al.}(2024)\citenamefont {Chen},
  \citenamefont {Wang}, \citenamefont {Ye}, \citenamefont {Wang}, \citenamefont
  {Zhou}, \citenamefont {Tang}, \citenamefont {Wang}, \citenamefont {Wang},
  \citenamefont {Zhang}, \citenamefont {Mei}, \citenamefont {Chen},\ and\
  \citenamefont {He}}]{Chen2024Mar}%
  \BibitemOpen
  \bibfield  {author} {\bibinfo {author} {\bibfnamefont {P.}~\bibnamefont
  {Chen}}, \bibinfo {author} {\bibfnamefont {G.}~\bibnamefont {Wang}}, \bibinfo
  {author} {\bibfnamefont {B.}~\bibnamefont {Ye}}, \bibinfo {author}
  {\bibfnamefont {J.}~\bibnamefont {Wang}}, \bibinfo {author} {\bibfnamefont
  {L.}~\bibnamefont {Zhou}}, \bibinfo {author} {\bibfnamefont {Z.}~\bibnamefont
  {Tang}}, \bibinfo {author} {\bibfnamefont {L.}~\bibnamefont {Wang}}, \bibinfo
  {author} {\bibfnamefont {J.}~\bibnamefont {Wang}}, \bibinfo {author}
  {\bibfnamefont {W.}~\bibnamefont {Zhang}}, \bibinfo {author} {\bibfnamefont
  {J.}~\bibnamefont {Mei}}, \bibinfo {author} {\bibfnamefont {W.}~\bibnamefont
  {Chen}},\ and\ \bibinfo {author} {\bibfnamefont {H.}~\bibnamefont {He}},\
  }\bibfield  {title} {\bibinfo {title} {{Edelstein Effect Induced
  Superconducting Diode Effect in Inversion Symmetry Breaking MoTe$_2$
  Josephson Junctions}},\ }\href {https://doi.org/10.1002/adfm.202311229}
  {\bibfield  {journal} {\bibinfo  {journal} {Adv. Funct. Mater.}\ }\textbf
  {\bibinfo {volume} {34}},\ \bibinfo {pages} {2311229} (\bibinfo {year}
  {2024})}\BibitemShut {NoStop}%
\bibitem [{\citenamefont {Kudriashov}\ \emph {et~al.}(2025)\citenamefont
  {Kudriashov}, \citenamefont {Zhou}, \citenamefont {Hovhannisyan},
  \citenamefont {Frolov}, \citenamefont {Elesin}, \citenamefont {Wang},
  \citenamefont {Zharkova}, \citenamefont {Taniguchi}, \citenamefont
  {Watanabe}, \citenamefont {Liu}, \citenamefont {Novoselov}, \citenamefont
  {Yashina}, \citenamefont {Zhou},\ and\ \citenamefont
  {Bandurin}}]{Kudriashov2025Jun}%
  \BibitemOpen
  \bibfield  {author} {\bibinfo {author} {\bibfnamefont {A.}~\bibnamefont
  {Kudriashov}}, \bibinfo {author} {\bibfnamefont {X.}~\bibnamefont {Zhou}},
  \bibinfo {author} {\bibfnamefont {R.~A.}\ \bibnamefont {Hovhannisyan}},
  \bibinfo {author} {\bibfnamefont {A.~S.}\ \bibnamefont {Frolov}}, \bibinfo
  {author} {\bibfnamefont {L.}~\bibnamefont {Elesin}}, \bibinfo {author}
  {\bibfnamefont {Y.~B.}\ \bibnamefont {Wang}}, \bibinfo {author}
  {\bibfnamefont {E.~V.}\ \bibnamefont {Zharkova}}, \bibinfo {author}
  {\bibfnamefont {T.}~\bibnamefont {Taniguchi}}, \bibinfo {author}
  {\bibfnamefont {K.}~\bibnamefont {Watanabe}}, \bibinfo {author}
  {\bibfnamefont {Z.}~\bibnamefont {Liu}}, \bibinfo {author} {\bibfnamefont
  {K.~S.}\ \bibnamefont {Novoselov}}, \bibinfo {author} {\bibfnamefont {L.~V.}\
  \bibnamefont {Yashina}}, \bibinfo {author} {\bibfnamefont {X.}~\bibnamefont
  {Zhou}},\ and\ \bibinfo {author} {\bibfnamefont {D.~A.}\ \bibnamefont
  {Bandurin}},\ }\bibfield  {title} {\bibinfo {title} {{Non-Majorana origin of
  anomalous current-phase relation and Josephson diode effect in Bi2Se3/NbSe2
  Josephson junctions}},\ }\bibfield  {journal} {\bibinfo  {journal} {Sci.
  Adv.}\ }\textbf {\bibinfo {volume} {11}},\ \href
  {https://doi.org/10.1126/sciadv.adw6925} {10.1126/sciadv.adw6925} (\bibinfo
  {year} {2025})\BibitemShut {NoStop}%
\bibitem [{\citenamefont {Cao}\ \emph {et~al.}(2018{\natexlab{a}})\citenamefont
  {Cao}, \citenamefont {Fatemi}, \citenamefont {Fang}, \citenamefont
  {Watanabe}, \citenamefont {Taniguchi}, \citenamefont {Kaxiras},\ and\
  \citenamefont {Jarillo-Herrero}}]{Cao2018Apr2}%
  \BibitemOpen
  \bibfield  {author} {\bibinfo {author} {\bibfnamefont {Y.}~\bibnamefont
  {Cao}}, \bibinfo {author} {\bibfnamefont {V.}~\bibnamefont {Fatemi}},
  \bibinfo {author} {\bibfnamefont {S.}~\bibnamefont {Fang}}, \bibinfo {author}
  {\bibfnamefont {K.}~\bibnamefont {Watanabe}}, \bibinfo {author}
  {\bibfnamefont {T.}~\bibnamefont {Taniguchi}}, \bibinfo {author}
  {\bibfnamefont {E.}~\bibnamefont {Kaxiras}},\ and\ \bibinfo {author}
  {\bibfnamefont {P.}~\bibnamefont {Jarillo-Herrero}},\ }\bibfield  {title}
  {\bibinfo {title} {{Unconventional superconductivity in magic-angle graphene
  superlattices}},\ }\href {https://doi.org/10.1038/nature26160} {\bibfield
  {journal} {\bibinfo  {journal} {Nature}\ }\textbf {\bibinfo {volume} {556}},\
  \bibinfo {pages} {43} (\bibinfo {year} {2018}{\natexlab{a}})}\BibitemShut
  {NoStop}%
\bibitem [{\citenamefont {Khalaf}\ \emph {et~al.}(2019)\citenamefont {Khalaf},
  \citenamefont {Kruchkov}, \citenamefont {Tarnopolsky},\ and\ \citenamefont
  {Vishwanath}}]{Khalaf2019Aug}%
  \BibitemOpen
  \bibfield  {author} {\bibinfo {author} {\bibfnamefont {E.}~\bibnamefont
  {Khalaf}}, \bibinfo {author} {\bibfnamefont {A.~J.}\ \bibnamefont
  {Kruchkov}}, \bibinfo {author} {\bibfnamefont {G.}~\bibnamefont
  {Tarnopolsky}},\ and\ \bibinfo {author} {\bibfnamefont {A.}~\bibnamefont
  {Vishwanath}},\ }\bibfield  {title} {\bibinfo {title} {{Magic angle hierarchy
  in twisted graphene multilayers}},\ }\href
  {https://doi.org/10.1103/PhysRevB.100.085109} {\bibfield  {journal} {\bibinfo
   {journal} {Phys. Rev. B}\ }\textbf {\bibinfo {volume} {100}},\ \bibinfo
  {pages} {085109} (\bibinfo {year} {2019})}\BibitemShut {NoStop}%
\bibitem [{\citenamefont {Park}\ \emph
  {et~al.}(2021{\natexlab{a}})\citenamefont {Park}, \citenamefont {Cao},
  \citenamefont {Watanabe}, \citenamefont {Taniguchi},\ and\ \citenamefont
  {Jarillo-Herrero}}]{Park2021Feb}%
  \BibitemOpen
  \bibfield  {author} {\bibinfo {author} {\bibfnamefont {J.~M.}\ \bibnamefont
  {Park}}, \bibinfo {author} {\bibfnamefont {Y.}~\bibnamefont {Cao}}, \bibinfo
  {author} {\bibfnamefont {K.}~\bibnamefont {Watanabe}}, \bibinfo {author}
  {\bibfnamefont {T.}~\bibnamefont {Taniguchi}},\ and\ \bibinfo {author}
  {\bibfnamefont {P.}~\bibnamefont {Jarillo-Herrero}},\ }\bibfield  {title}
  {\bibinfo {title} {{Tunable strongly coupled superconductivity in magic-angle
  twisted trilayer graphene}},\ }\href
  {https://doi.org/10.1038/s41586-021-03192-0} {\bibfield  {journal} {\bibinfo
  {journal} {Nature}\ }\textbf {\bibinfo {volume} {590}},\ \bibinfo {pages}
  {249} (\bibinfo {year} {2021}{\natexlab{a}})}\BibitemShut {NoStop}%
\bibitem [{\citenamefont {Park}\ \emph {et~al.}(2022)\citenamefont {Park},
  \citenamefont {Cao}, \citenamefont {Xia}, \citenamefont {Sun}, \citenamefont
  {Watanabe}, \citenamefont {Taniguchi},\ and\ \citenamefont
  {Jarillo-Herrero}}]{Park2022Aug}%
  \BibitemOpen
  \bibfield  {author} {\bibinfo {author} {\bibfnamefont {J.~M.}\ \bibnamefont
  {Park}}, \bibinfo {author} {\bibfnamefont {Y.}~\bibnamefont {Cao}}, \bibinfo
  {author} {\bibfnamefont {L.-Q.}\ \bibnamefont {Xia}}, \bibinfo {author}
  {\bibfnamefont {S.}~\bibnamefont {Sun}}, \bibinfo {author} {\bibfnamefont
  {K.}~\bibnamefont {Watanabe}}, \bibinfo {author} {\bibfnamefont
  {T.}~\bibnamefont {Taniguchi}},\ and\ \bibinfo {author} {\bibfnamefont
  {P.}~\bibnamefont {Jarillo-Herrero}},\ }\bibfield  {title} {\bibinfo {title}
  {{Robust superconductivity in magic-angle multilayer graphene family}},\
  }\href {https://doi.org/10.1038/s41563-022-01287-1} {\bibfield  {journal}
  {\bibinfo  {journal} {Nat. Mater.}\ }\textbf {\bibinfo {volume} {21}},\
  \bibinfo {pages} {877} (\bibinfo {year} {2022})}\BibitemShut {NoStop}%
\bibitem [{\citenamefont {Zhang}\ \emph {et~al.}(2022)\citenamefont {Zhang},
  \citenamefont {Polski}, \citenamefont {Lewandowski}, \citenamefont {Thomson},
  \citenamefont {Peng}, \citenamefont {Choi}, \citenamefont {Kim},
  \citenamefont {Watanabe}, \citenamefont {Taniguchi}, \citenamefont {Alicea},
  \citenamefont {von Oppen}, \citenamefont {Refael},\ and\ \citenamefont
  {Nadj-Perge}}]{Zhang2022Sep}%
  \BibitemOpen
  \bibfield  {author} {\bibinfo {author} {\bibfnamefont {Y.}~\bibnamefont
  {Zhang}}, \bibinfo {author} {\bibfnamefont {R.}~\bibnamefont {Polski}},
  \bibinfo {author} {\bibfnamefont {C.}~\bibnamefont {Lewandowski}}, \bibinfo
  {author} {\bibfnamefont {A.}~\bibnamefont {Thomson}}, \bibinfo {author}
  {\bibfnamefont {Y.}~\bibnamefont {Peng}}, \bibinfo {author} {\bibfnamefont
  {Y.}~\bibnamefont {Choi}}, \bibinfo {author} {\bibfnamefont {H.}~\bibnamefont
  {Kim}}, \bibinfo {author} {\bibfnamefont {K.}~\bibnamefont {Watanabe}},
  \bibinfo {author} {\bibfnamefont {T.}~\bibnamefont {Taniguchi}}, \bibinfo
  {author} {\bibfnamefont {J.}~\bibnamefont {Alicea}}, \bibinfo {author}
  {\bibfnamefont {F.}~\bibnamefont {von Oppen}}, \bibinfo {author}
  {\bibfnamefont {G.}~\bibnamefont {Refael}},\ and\ \bibinfo {author}
  {\bibfnamefont {S.}~\bibnamefont {Nadj-Perge}},\ }\bibfield  {title}
  {\bibinfo {title} {{Promotion of superconductivity in magic-angle graphene
  multilayers}},\ }\href {https://doi.org/10.1126/science.abn8585} {\bibfield
  {journal} {\bibinfo  {journal} {Science}\ }\textbf {\bibinfo {volume}
  {377}},\ \bibinfo {pages} {1538} (\bibinfo {year} {2022})}\BibitemShut
  {NoStop}%
\bibitem [{\citenamefont {Hao}\ \emph {et~al.}(2021)\citenamefont {Hao},
  \citenamefont {Zimmerman}, \citenamefont {Ledwith}, \citenamefont {Khalaf},
  \citenamefont {Najafabadi}, \citenamefont {Watanabe}, \citenamefont
  {Taniguchi}, \citenamefont {Vishwanath},\ and\ \citenamefont
  {Kim}}]{Hao2021Mar}%
  \BibitemOpen
  \bibfield  {author} {\bibinfo {author} {\bibfnamefont {Z.}~\bibnamefont
  {Hao}}, \bibinfo {author} {\bibfnamefont {A.~M.}\ \bibnamefont {Zimmerman}},
  \bibinfo {author} {\bibfnamefont {P.}~\bibnamefont {Ledwith}}, \bibinfo
  {author} {\bibfnamefont {E.}~\bibnamefont {Khalaf}}, \bibinfo {author}
  {\bibfnamefont {D.~H.}\ \bibnamefont {Najafabadi}}, \bibinfo {author}
  {\bibfnamefont {K.}~\bibnamefont {Watanabe}}, \bibinfo {author}
  {\bibfnamefont {T.}~\bibnamefont {Taniguchi}}, \bibinfo {author}
  {\bibfnamefont {A.}~\bibnamefont {Vishwanath}},\ and\ \bibinfo {author}
  {\bibfnamefont {P.}~\bibnamefont {Kim}},\ }\bibfield  {title} {\bibinfo
  {title} {{Electric field{\textendash}tunable superconductivity in
  alternating-twist magic-angle trilayer graphene}},\ }\href
  {https://doi.org/10.1126/science.abg0399} {\bibfield  {journal} {\bibinfo
  {journal} {Science}\ }\textbf {\bibinfo {volume} {371}},\ \bibinfo {pages}
  {1133} (\bibinfo {year} {2021})}\BibitemShut {NoStop}%
\bibitem [{\citenamefont {Lu}\ \emph {et~al.}(2019)\citenamefont {Lu},
  \citenamefont {Stepanov}, \citenamefont {Yang}, \citenamefont {Xie},
  \citenamefont {Aamir}, \citenamefont {Das}, \citenamefont {Urgell},
  \citenamefont {Watanabe}, \citenamefont {Taniguchi}, \citenamefont {Zhang},
  \citenamefont {Bachtold}, \citenamefont {MacDonald},\ and\ \citenamefont
  {Efetov}}]{Lu2019Oct}%
  \BibitemOpen
  \bibfield  {author} {\bibinfo {author} {\bibfnamefont {X.}~\bibnamefont
  {Lu}}, \bibinfo {author} {\bibfnamefont {P.}~\bibnamefont {Stepanov}},
  \bibinfo {author} {\bibfnamefont {W.}~\bibnamefont {Yang}}, \bibinfo {author}
  {\bibfnamefont {M.}~\bibnamefont {Xie}}, \bibinfo {author} {\bibfnamefont
  {M.~A.}\ \bibnamefont {Aamir}}, \bibinfo {author} {\bibfnamefont
  {I.}~\bibnamefont {Das}}, \bibinfo {author} {\bibfnamefont {C.}~\bibnamefont
  {Urgell}}, \bibinfo {author} {\bibfnamefont {K.}~\bibnamefont {Watanabe}},
  \bibinfo {author} {\bibfnamefont {T.}~\bibnamefont {Taniguchi}}, \bibinfo
  {author} {\bibfnamefont {G.}~\bibnamefont {Zhang}}, \bibinfo {author}
  {\bibfnamefont {A.}~\bibnamefont {Bachtold}}, \bibinfo {author}
  {\bibfnamefont {A.~H.}\ \bibnamefont {MacDonald}},\ and\ \bibinfo {author}
  {\bibfnamefont {D.~K.}\ \bibnamefont {Efetov}},\ }\bibfield  {title}
  {\bibinfo {title} {{Superconductors, orbital magnets and correlated states in
  magic-angle bilayer graphene}},\ }\href
  {https://doi.org/10.1038/s41586-019-1695-0} {\bibfield  {journal} {\bibinfo
  {journal} {Nature}\ }\textbf {\bibinfo {volume} {574}},\ \bibinfo {pages}
  {653} (\bibinfo {year} {2019})}\BibitemShut {NoStop}%
\bibitem [{\citenamefont {Yankowitz}\ \emph {et~al.}(2019)\citenamefont
  {Yankowitz}, \citenamefont {Chen}, \citenamefont {Polshyn}, \citenamefont
  {Zhang}, \citenamefont {Watanabe}, \citenamefont {Taniguchi}, \citenamefont
  {Graf}, \citenamefont {Young},\ and\ \citenamefont
  {Dean}}]{Yankowitz2019Mar}%
  \BibitemOpen
  \bibfield  {author} {\bibinfo {author} {\bibfnamefont {M.}~\bibnamefont
  {Yankowitz}}, \bibinfo {author} {\bibfnamefont {S.}~\bibnamefont {Chen}},
  \bibinfo {author} {\bibfnamefont {H.}~\bibnamefont {Polshyn}}, \bibinfo
  {author} {\bibfnamefont {Y.}~\bibnamefont {Zhang}}, \bibinfo {author}
  {\bibfnamefont {K.}~\bibnamefont {Watanabe}}, \bibinfo {author}
  {\bibfnamefont {T.}~\bibnamefont {Taniguchi}}, \bibinfo {author}
  {\bibfnamefont {D.}~\bibnamefont {Graf}}, \bibinfo {author} {\bibfnamefont
  {A.~F.}\ \bibnamefont {Young}},\ and\ \bibinfo {author} {\bibfnamefont
  {C.~R.}\ \bibnamefont {Dean}},\ }\bibfield  {title} {\bibinfo {title}
  {{Tuning superconductivity in twisted bilayer graphene}},\ }\href
  {https://doi.org/10.1126/science.aav1910} {\bibfield  {journal} {\bibinfo
  {journal} {Science}\ }\textbf {\bibinfo {volume} {363}},\ \bibinfo {pages}
  {1059} (\bibinfo {year} {2019})}\BibitemShut {NoStop}%
\bibitem [{\citenamefont {Oh}\ \emph {et~al.}(2021)\citenamefont {Oh},
  \citenamefont {Nuckolls}, \citenamefont {Wong}, \citenamefont {Lee},
  \citenamefont {Liu}, \citenamefont {Watanabe}, \citenamefont {Taniguchi},\
  and\ \citenamefont {Yazdani}}]{Oh2021Dec}%
  \BibitemOpen
  \bibfield  {author} {\bibinfo {author} {\bibfnamefont {M.}~\bibnamefont
  {Oh}}, \bibinfo {author} {\bibfnamefont {K.~P.}\ \bibnamefont {Nuckolls}},
  \bibinfo {author} {\bibfnamefont {D.}~\bibnamefont {Wong}}, \bibinfo {author}
  {\bibfnamefont {R.~L.}\ \bibnamefont {Lee}}, \bibinfo {author} {\bibfnamefont
  {X.}~\bibnamefont {Liu}}, \bibinfo {author} {\bibfnamefont {K.}~\bibnamefont
  {Watanabe}}, \bibinfo {author} {\bibfnamefont {T.}~\bibnamefont
  {Taniguchi}},\ and\ \bibinfo {author} {\bibfnamefont {A.}~\bibnamefont
  {Yazdani}},\ }\bibfield  {title} {\bibinfo {title} {{Evidence for
  unconventional superconductivity in twisted bilayer graphene}},\ }\href
  {https://doi.org/10.1038/s41586-021-04121-x} {\bibfield  {journal} {\bibinfo
  {journal} {Nature}\ }\textbf {\bibinfo {volume} {600}},\ \bibinfo {pages}
  {240} (\bibinfo {year} {2021})}\BibitemShut {NoStop}%
\bibitem [{\citenamefont {Cao}\ \emph {et~al.}(2018{\natexlab{b}})\citenamefont
  {Cao}, \citenamefont {Fatemi}, \citenamefont {Demir}, \citenamefont {Fang},
  \citenamefont {Tomarken}, \citenamefont {Luo}, \citenamefont
  {Sanchez-Yamagishi}, \citenamefont {Watanabe}, \citenamefont {Taniguchi},
  \citenamefont {Kaxiras}, \citenamefont {Ashoori},\ and\ \citenamefont
  {Jarillo-Herrero}}]{Cao2018Apr}%
  \BibitemOpen
  \bibfield  {author} {\bibinfo {author} {\bibfnamefont {Y.}~\bibnamefont
  {Cao}}, \bibinfo {author} {\bibfnamefont {V.}~\bibnamefont {Fatemi}},
  \bibinfo {author} {\bibfnamefont {A.}~\bibnamefont {Demir}}, \bibinfo
  {author} {\bibfnamefont {S.}~\bibnamefont {Fang}}, \bibinfo {author}
  {\bibfnamefont {S.~L.}\ \bibnamefont {Tomarken}}, \bibinfo {author}
  {\bibfnamefont {J.~Y.}\ \bibnamefont {Luo}}, \bibinfo {author} {\bibfnamefont
  {J.~D.}\ \bibnamefont {Sanchez-Yamagishi}}, \bibinfo {author} {\bibfnamefont
  {K.}~\bibnamefont {Watanabe}}, \bibinfo {author} {\bibfnamefont
  {T.}~\bibnamefont {Taniguchi}}, \bibinfo {author} {\bibfnamefont
  {E.}~\bibnamefont {Kaxiras}}, \bibinfo {author} {\bibfnamefont {R.~C.}\
  \bibnamefont {Ashoori}},\ and\ \bibinfo {author} {\bibfnamefont
  {P.}~\bibnamefont {Jarillo-Herrero}},\ }\bibfield  {title} {\bibinfo {title}
  {{Correlated insulator behaviour at half-filling in magic-angle graphene
  superlattices}},\ }\href {https://doi.org/10.1038/nature26154} {\bibfield
  {journal} {\bibinfo  {journal} {Nature}\ }\textbf {\bibinfo {volume} {556}},\
  \bibinfo {pages} {80} (\bibinfo {year} {2018}{\natexlab{b}})}\BibitemShut
  {NoStop}%
\bibitem [{\citenamefont {Stepanov}\ \emph {et~al.}(2020)\citenamefont
  {Stepanov}, \citenamefont {Das}, \citenamefont {Lu}, \citenamefont
  {Fahimniya}, \citenamefont {Watanabe}, \citenamefont {Taniguchi},
  \citenamefont {Koppens}, \citenamefont {Lischner}, \citenamefont {Levitov},\
  and\ \citenamefont {Efetov}}]{Stepanov2020Jul}%
  \BibitemOpen
  \bibfield  {author} {\bibinfo {author} {\bibfnamefont {P.}~\bibnamefont
  {Stepanov}}, \bibinfo {author} {\bibfnamefont {I.}~\bibnamefont {Das}},
  \bibinfo {author} {\bibfnamefont {X.}~\bibnamefont {Lu}}, \bibinfo {author}
  {\bibfnamefont {A.}~\bibnamefont {Fahimniya}}, \bibinfo {author}
  {\bibfnamefont {K.}~\bibnamefont {Watanabe}}, \bibinfo {author}
  {\bibfnamefont {T.}~\bibnamefont {Taniguchi}}, \bibinfo {author}
  {\bibfnamefont {F.~H.~L.}\ \bibnamefont {Koppens}}, \bibinfo {author}
  {\bibfnamefont {J.}~\bibnamefont {Lischner}}, \bibinfo {author}
  {\bibfnamefont {L.}~\bibnamefont {Levitov}},\ and\ \bibinfo {author}
  {\bibfnamefont {D.~K.}\ \bibnamefont {Efetov}},\ }\bibfield  {title}
  {\bibinfo {title} {{Untying the insulating and superconducting orders in
  magic-angle graphene}},\ }\href {https://doi.org/10.1038/s41586-020-2459-6}
  {\bibfield  {journal} {\bibinfo  {journal} {Nature}\ }\textbf {\bibinfo
  {volume} {583}},\ \bibinfo {pages} {375} (\bibinfo {year} {2020})},\ \Eprint
  {https://arxiv.org/abs/32632215} {32632215} \BibitemShut {NoStop}%
\bibitem [{\citenamefont {Sharpe}\ \emph {et~al.}(2019)\citenamefont {Sharpe},
  \citenamefont {Fox}, \citenamefont {Barnard}, \citenamefont {Finney},
  \citenamefont {Watanabe}, \citenamefont {Taniguchi}, \citenamefont
  {Kastner},\ and\ \citenamefont {Goldhaber-Gordon}}]{Sharpe2019Jul}%
  \BibitemOpen
  \bibfield  {author} {\bibinfo {author} {\bibfnamefont {A.~L.}\ \bibnamefont
  {Sharpe}}, \bibinfo {author} {\bibfnamefont {E.~J.}\ \bibnamefont {Fox}},
  \bibinfo {author} {\bibfnamefont {A.~W.}\ \bibnamefont {Barnard}}, \bibinfo
  {author} {\bibfnamefont {J.}~\bibnamefont {Finney}}, \bibinfo {author}
  {\bibfnamefont {K.}~\bibnamefont {Watanabe}}, \bibinfo {author}
  {\bibfnamefont {T.}~\bibnamefont {Taniguchi}}, \bibinfo {author}
  {\bibfnamefont {M.~A.}\ \bibnamefont {Kastner}},\ and\ \bibinfo {author}
  {\bibfnamefont {D.}~\bibnamefont {Goldhaber-Gordon}},\ }\bibfield  {title}
  {\bibinfo {title} {{Emergent ferromagnetism near three-quarters filling in
  twisted bilayer graphene}},\ }\href {https://doi.org/10.1126/science.aaw3780}
  {\bibfield  {journal} {\bibinfo  {journal} {Science}\ }\textbf {\bibinfo
  {volume} {365}},\ \bibinfo {pages} {605} (\bibinfo {year}
  {2019})}\BibitemShut {NoStop}%
\bibitem [{\citenamefont {Serlin}\ \emph {et~al.}(2020)\citenamefont {Serlin},
  \citenamefont {Tschirhart}, \citenamefont {Polshyn}, \citenamefont {Zhang},
  \citenamefont {Zhu}, \citenamefont {Watanabe}, \citenamefont {Taniguchi},
  \citenamefont {Balents},\ and\ \citenamefont {Young}}]{Serlin2020Feb}%
  \BibitemOpen
  \bibfield  {author} {\bibinfo {author} {\bibfnamefont {M.}~\bibnamefont
  {Serlin}}, \bibinfo {author} {\bibfnamefont {C.~L.}\ \bibnamefont
  {Tschirhart}}, \bibinfo {author} {\bibfnamefont {H.}~\bibnamefont {Polshyn}},
  \bibinfo {author} {\bibfnamefont {Y.}~\bibnamefont {Zhang}}, \bibinfo
  {author} {\bibfnamefont {J.}~\bibnamefont {Zhu}}, \bibinfo {author}
  {\bibfnamefont {K.}~\bibnamefont {Watanabe}}, \bibinfo {author}
  {\bibfnamefont {T.}~\bibnamefont {Taniguchi}}, \bibinfo {author}
  {\bibfnamefont {L.}~\bibnamefont {Balents}},\ and\ \bibinfo {author}
  {\bibfnamefont {A.~F.}\ \bibnamefont {Young}},\ }\bibfield  {title} {\bibinfo
  {title} {{Intrinsic quantized anomalous Hall effect in a
  moir{\ifmmode\acute{e}\else\'{e}\fi} heterostructure}},\ }\href
  {https://doi.org/10.1126/science.aay5533} {\bibfield  {journal} {\bibinfo
  {journal} {Science}\ }\textbf {\bibinfo {volume} {367}},\ \bibinfo {pages}
  {900} (\bibinfo {year} {2020})}\BibitemShut {NoStop}%
\bibitem [{\citenamefont {Stepanov}\ \emph {et~al.}(2021)\citenamefont
  {Stepanov}, \citenamefont {Xie}, \citenamefont {Taniguchi}, \citenamefont
  {Watanabe}, \citenamefont {Lu}, \citenamefont {MacDonald}, \citenamefont
  {Bernevig},\ and\ \citenamefont {Efetov}}]{Stepanov2021Nov}%
  \BibitemOpen
  \bibfield  {author} {\bibinfo {author} {\bibfnamefont {P.}~\bibnamefont
  {Stepanov}}, \bibinfo {author} {\bibfnamefont {M.}~\bibnamefont {Xie}},
  \bibinfo {author} {\bibfnamefont {T.}~\bibnamefont {Taniguchi}}, \bibinfo
  {author} {\bibfnamefont {K.}~\bibnamefont {Watanabe}}, \bibinfo {author}
  {\bibfnamefont {X.}~\bibnamefont {Lu}}, \bibinfo {author} {\bibfnamefont
  {A.~H.}\ \bibnamefont {MacDonald}}, \bibinfo {author} {\bibfnamefont {B.~A.}\
  \bibnamefont {Bernevig}},\ and\ \bibinfo {author} {\bibfnamefont {D.~K.}\
  \bibnamefont {Efetov}},\ }\bibfield  {title} {\bibinfo {title} {{Competing
  Zero-Field Chern Insulators in Superconducting Twisted Bilayer Graphene}},\
  }\href {https://doi.org/10.1103/PhysRevLett.127.197701} {\bibfield  {journal}
  {\bibinfo  {journal} {Phys. Rev. Lett.}\ }\textbf {\bibinfo {volume} {127}},\
  \bibinfo {pages} {197701} (\bibinfo {year} {2021})}\BibitemShut {NoStop}%
\bibitem [{\citenamefont {Das}\ \emph {et~al.}(2021)\citenamefont {Das},
  \citenamefont {Lu}, \citenamefont {Herzog-Arbeitman}, \citenamefont {Song},
  \citenamefont {Watanabe}, \citenamefont {Taniguchi}, \citenamefont
  {Bernevig},\ and\ \citenamefont {Efetov}}]{Das2021Jun}%
  \BibitemOpen
  \bibfield  {author} {\bibinfo {author} {\bibfnamefont {I.}~\bibnamefont
  {Das}}, \bibinfo {author} {\bibfnamefont {X.}~\bibnamefont {Lu}}, \bibinfo
  {author} {\bibfnamefont {J.}~\bibnamefont {Herzog-Arbeitman}}, \bibinfo
  {author} {\bibfnamefont {Z.-D.}\ \bibnamefont {Song}}, \bibinfo {author}
  {\bibfnamefont {K.}~\bibnamefont {Watanabe}}, \bibinfo {author}
  {\bibfnamefont {T.}~\bibnamefont {Taniguchi}}, \bibinfo {author}
  {\bibfnamefont {B.~A.}\ \bibnamefont {Bernevig}},\ and\ \bibinfo {author}
  {\bibfnamefont {D.~K.}\ \bibnamefont {Efetov}},\ }\bibfield  {title}
  {\bibinfo {title} {{Symmetry-broken Chern insulators and Rashba-like
  Landau-level crossings in magic-angle bilayer graphene}},\ }\href
  {https://doi.org/10.1038/s41567-021-01186-3} {\bibfield  {journal} {\bibinfo
  {journal} {Nat. Phys.}\ }\textbf {\bibinfo {volume} {17}},\ \bibinfo {pages}
  {710} (\bibinfo {year} {2021})}\BibitemShut {NoStop}%
\bibitem [{\citenamefont {Rodan-Legrain}\ \emph {et~al.}(2021)\citenamefont
  {Rodan-Legrain}, \citenamefont {Cao}, \citenamefont {Park}, \citenamefont
  {de~la Barrera}, \citenamefont {Randeria}, \citenamefont {Watanabe},
  \citenamefont {Taniguchi},\ and\ \citenamefont
  {Jarillo-Herrero}}]{Rodan-Legrain2021Jul}%
  \BibitemOpen
  \bibfield  {author} {\bibinfo {author} {\bibfnamefont {D.}~\bibnamefont
  {Rodan-Legrain}}, \bibinfo {author} {\bibfnamefont {Y.}~\bibnamefont {Cao}},
  \bibinfo {author} {\bibfnamefont {J.~M.}\ \bibnamefont {Park}}, \bibinfo
  {author} {\bibfnamefont {S.~C.}\ \bibnamefont {de~la Barrera}}, \bibinfo
  {author} {\bibfnamefont {M.~T.}\ \bibnamefont {Randeria}}, \bibinfo {author}
  {\bibfnamefont {K.}~\bibnamefont {Watanabe}}, \bibinfo {author}
  {\bibfnamefont {T.}~\bibnamefont {Taniguchi}},\ and\ \bibinfo {author}
  {\bibfnamefont {P.}~\bibnamefont {Jarillo-Herrero}},\ }\bibfield  {title}
  {\bibinfo {title} {{Highly tunable junctions and non-local Josephson effect
  in magic-angle graphene tunnelling devices}},\ }\href
  {https://doi.org/10.1038/s41565-021-00894-4} {\bibfield  {journal} {\bibinfo
  {journal} {Nat. Nanotechnol.}\ }\textbf {\bibinfo {volume} {16}},\ \bibinfo
  {pages} {769} (\bibinfo {year} {2021})}\BibitemShut {NoStop}%
\bibitem [{\citenamefont {de~Vries}\ \emph {et~al.}(2021)\citenamefont
  {de~Vries}, \citenamefont {Portol{\ifmmode\acute{e}\else\'{e}\fi}s},
  \citenamefont {Zheng}, \citenamefont {Taniguchi}, \citenamefont {Watanabe},
  \citenamefont {Ihn}, \citenamefont {Ensslin},\ and\ \citenamefont
  {Rickhaus}}]{deVries2021Jul}%
  \BibitemOpen
  \bibfield  {author} {\bibinfo {author} {\bibfnamefont {F.~K.}\ \bibnamefont
  {de~Vries}}, \bibinfo {author} {\bibfnamefont {E.}~\bibnamefont
  {Portol{\ifmmode\acute{e}\else\'{e}\fi}s}}, \bibinfo {author} {\bibfnamefont
  {G.}~\bibnamefont {Zheng}}, \bibinfo {author} {\bibfnamefont
  {T.}~\bibnamefont {Taniguchi}}, \bibinfo {author} {\bibfnamefont
  {K.}~\bibnamefont {Watanabe}}, \bibinfo {author} {\bibfnamefont
  {T.}~\bibnamefont {Ihn}}, \bibinfo {author} {\bibfnamefont {K.}~\bibnamefont
  {Ensslin}},\ and\ \bibinfo {author} {\bibfnamefont {P.}~\bibnamefont
  {Rickhaus}},\ }\bibfield  {title} {\bibinfo {title} {{Gate-defined Josephson
  junctions in magic-angle twisted bilayer graphene}},\ }\href
  {https://doi.org/10.1038/s41565-021-00896-2} {\bibfield  {journal} {\bibinfo
  {journal} {Nat. Nanotechnol.}\ }\textbf {\bibinfo {volume} {16}},\ \bibinfo
  {pages} {760} (\bibinfo {year} {2021})}\BibitemShut {NoStop}%
\bibitem [{\citenamefont {Portol{\ifmmode\acute{e}\else\'{e}\fi}s}\ \emph
  {et~al.}(2022)\citenamefont {Portol{\ifmmode\acute{e}\else\'{e}\fi}s},
  \citenamefont {Iwakiri}, \citenamefont {Zheng}, \citenamefont {Rickhaus},
  \citenamefont {Taniguchi}, \citenamefont {Watanabe}, \citenamefont {Ihn},
  \citenamefont {Ensslin},\ and\ \citenamefont {de~Vries}}]{Portoles2022Nov}%
  \BibitemOpen
  \bibfield  {author} {\bibinfo {author} {\bibfnamefont {E.}~\bibnamefont
  {Portol{\ifmmode\acute{e}\else\'{e}\fi}s}}, \bibinfo {author} {\bibfnamefont
  {S.}~\bibnamefont {Iwakiri}}, \bibinfo {author} {\bibfnamefont
  {G.}~\bibnamefont {Zheng}}, \bibinfo {author} {\bibfnamefont
  {P.}~\bibnamefont {Rickhaus}}, \bibinfo {author} {\bibfnamefont
  {T.}~\bibnamefont {Taniguchi}}, \bibinfo {author} {\bibfnamefont
  {K.}~\bibnamefont {Watanabe}}, \bibinfo {author} {\bibfnamefont
  {T.}~\bibnamefont {Ihn}}, \bibinfo {author} {\bibfnamefont {K.}~\bibnamefont
  {Ensslin}},\ and\ \bibinfo {author} {\bibfnamefont {F.~K.}\ \bibnamefont
  {de~Vries}},\ }\bibfield  {title} {\bibinfo {title} {{A tunable monolithic
  SQUID in twisted bilayer graphene}},\ }\href
  {https://doi.org/10.1038/s41565-022-01222-0} {\bibfield  {journal} {\bibinfo
  {journal} {Nat. Nanotechnol.}\ }\textbf {\bibinfo {volume} {17}},\ \bibinfo
  {pages} {1159} (\bibinfo {year} {2022})}\BibitemShut {NoStop}%
\bibitem [{\citenamefont {Perego}\ \emph {et~al.}(2024)\citenamefont {Perego},
  \citenamefont {Agero}, \citenamefont {Tor{\ifmmode\grave{a}\else\`{a}\fi}},
  \citenamefont {Portol{\ifmmode\acute{e}\else\'{e}\fi}s}, \citenamefont
  {Denisov}, \citenamefont {Taniguchi}, \citenamefont {Watanabe}, \citenamefont
  {Gaggioli}, \citenamefont {Geshkenbein}, \citenamefont {Blatter},
  \citenamefont {Ihn},\ and\ \citenamefont {Ensslin}}]{Perego2024Oct}%
  \BibitemOpen
  \bibfield  {author} {\bibinfo {author} {\bibfnamefont {M.}~\bibnamefont
  {Perego}}, \bibinfo {author} {\bibfnamefont {C.~G.}\ \bibnamefont {Agero}},
  \bibinfo {author} {\bibfnamefont {A.~M.}\ \bibnamefont
  {Tor{\ifmmode\grave{a}\else\`{a}\fi}}}, \bibinfo {author} {\bibfnamefont
  {E.}~\bibnamefont {Portol{\ifmmode\acute{e}\else\'{e}\fi}s}}, \bibinfo
  {author} {\bibfnamefont {A.~O.}\ \bibnamefont {Denisov}}, \bibinfo {author}
  {\bibfnamefont {T.}~\bibnamefont {Taniguchi}}, \bibinfo {author}
  {\bibfnamefont {K.}~\bibnamefont {Watanabe}}, \bibinfo {author}
  {\bibfnamefont {F.}~\bibnamefont {Gaggioli}}, \bibinfo {author}
  {\bibfnamefont {V.}~\bibnamefont {Geshkenbein}}, \bibinfo {author}
  {\bibfnamefont {G.}~\bibnamefont {Blatter}}, \bibinfo {author} {\bibfnamefont
  {T.}~\bibnamefont {Ihn}},\ and\ \bibinfo {author} {\bibfnamefont
  {K.}~\bibnamefont {Ensslin}},\ }\bibfield  {title} {\bibinfo {title}
  {{Experimental detection of vortices in magic-angle graphene}},\ }\bibfield
  {journal} {\bibinfo  {journal} {arXiv}\ }\href
  {https://doi.org/10.48550/arXiv.2410.03508} {10.48550/arXiv.2410.03508}
  (\bibinfo {year} {2024}),\ \Eprint {https://arxiv.org/abs/2410.03508}
  {2410.03508} \BibitemShut {NoStop}%
\bibitem [{\citenamefont {Portol{\ifmmode\acute{e}\else\'{e}\fi}s}\ \emph
  {et~al.}(2025)\citenamefont {Portol{\ifmmode\acute{e}\else\'{e}\fi}s},
  \citenamefont {Perego}, \citenamefont {Volkov}, \citenamefont {Toschini},
  \citenamefont {Kemna}, \citenamefont
  {Mestre-Tor{\ifmmode\grave{a}\else\`{a}\fi}}, \citenamefont {Zheng},
  \citenamefont {Denisov}, \citenamefont {Vries}, \citenamefont {Rickhaus},
  \citenamefont {Taniguchi}, \citenamefont {Watanabe}, \citenamefont {Pixley},
  \citenamefont {Ihn},\ and\ \citenamefont {Ensslin}}]{Portoles2025May}%
  \BibitemOpen
  \bibfield  {author} {\bibinfo {author} {\bibfnamefont {E.}~\bibnamefont
  {Portol{\ifmmode\acute{e}\else\'{e}\fi}s}}, \bibinfo {author} {\bibfnamefont
  {M.}~\bibnamefont {Perego}}, \bibinfo {author} {\bibfnamefont {P.~A.}\
  \bibnamefont {Volkov}}, \bibinfo {author} {\bibfnamefont {M.}~\bibnamefont
  {Toschini}}, \bibinfo {author} {\bibfnamefont {Y.}~\bibnamefont {Kemna}},
  \bibinfo {author} {\bibfnamefont {A.}~\bibnamefont
  {Mestre-Tor{\ifmmode\grave{a}\else\`{a}\fi}}}, \bibinfo {author}
  {\bibfnamefont {G.}~\bibnamefont {Zheng}}, \bibinfo {author} {\bibfnamefont
  {A.~O.}\ \bibnamefont {Denisov}}, \bibinfo {author} {\bibfnamefont
  {F.~K.~d.}\ \bibnamefont {Vries}}, \bibinfo {author} {\bibfnamefont
  {P.}~\bibnamefont {Rickhaus}}, \bibinfo {author} {\bibfnamefont
  {T.}~\bibnamefont {Taniguchi}}, \bibinfo {author} {\bibfnamefont
  {K.}~\bibnamefont {Watanabe}}, \bibinfo {author} {\bibfnamefont {J.~H.}\
  \bibnamefont {Pixley}}, \bibinfo {author} {\bibfnamefont {T.}~\bibnamefont
  {Ihn}},\ and\ \bibinfo {author} {\bibfnamefont {K.}~\bibnamefont {Ensslin}},\
  }\bibfield  {title} {\bibinfo {title} {{Quasiparticle and superfluid dynamics
  in Magic-Angle Graphene}},\ }\href
  {https://doi.org/10.1038/s41467-025-58325-0} {\bibfield  {journal} {\bibinfo
  {journal} {Nat. Commun.}\ }\textbf {\bibinfo {volume} {16}},\ \bibinfo
  {pages} {4273} (\bibinfo {year} {2025})}\BibitemShut {NoStop}%
\bibitem [{\citenamefont {Davydova}\ \emph {et~al.}(2022)\citenamefont
  {Davydova}, \citenamefont {Prembabu},\ and\ \citenamefont
  {Fu}}]{Davydova2022Jun}%
  \BibitemOpen
  \bibfield  {author} {\bibinfo {author} {\bibfnamefont {M.}~\bibnamefont
  {Davydova}}, \bibinfo {author} {\bibfnamefont {S.}~\bibnamefont {Prembabu}},\
  and\ \bibinfo {author} {\bibfnamefont {L.}~\bibnamefont {Fu}},\ }\bibfield
  {title} {\bibinfo {title} {{Universal Josephson diode effect}},\ }\href
  {https://doi.org/10.1126/sciadv.abo0309} {\bibfield  {journal} {\bibinfo
  {journal} {Sci. Adv.}\ }\textbf {\bibinfo {volume} {8}},\ \bibinfo {pages}
  {eabo0309} (\bibinfo {year} {2022})}\BibitemShut {NoStop}%
\bibitem [{\citenamefont {Wei}\ \emph {et~al.}(2022)\citenamefont {Wei},
  \citenamefont {Liu}, \citenamefont {Wang},\ and\ \citenamefont
  {Liu}}]{Wei2022Oct}%
  \BibitemOpen
  \bibfield  {author} {\bibinfo {author} {\bibfnamefont {Y.-J.}\ \bibnamefont
  {Wei}}, \bibinfo {author} {\bibfnamefont {H.-L.}\ \bibnamefont {Liu}},
  \bibinfo {author} {\bibfnamefont {J.}~\bibnamefont {Wang}},\ and\ \bibinfo
  {author} {\bibfnamefont {J.-F.}\ \bibnamefont {Liu}},\ }\bibfield  {title}
  {\bibinfo {title} {{Supercurrent rectification effect in graphene-based
  Josephson junctions}},\ }\href {https://doi.org/10.1103/PhysRevB.106.165419}
  {\bibfield  {journal} {\bibinfo  {journal} {Phys. Rev. B}\ }\textbf {\bibinfo
  {volume} {106}},\ \bibinfo {pages} {165419} (\bibinfo {year}
  {2022})}\BibitemShut {NoStop}%
\bibitem [{\citenamefont {Hu}\ \emph {et~al.}(2023)\citenamefont {Hu},
  \citenamefont {Sun}, \citenamefont {Xie},\ and\ \citenamefont
  {Law}}]{Hu2023Jun}%
  \BibitemOpen
  \bibfield  {author} {\bibinfo {author} {\bibfnamefont {J.-X.}\ \bibnamefont
  {Hu}}, \bibinfo {author} {\bibfnamefont {Z.-T.}\ \bibnamefont {Sun}},
  \bibinfo {author} {\bibfnamefont {Y.-M.}\ \bibnamefont {Xie}},\ and\ \bibinfo
  {author} {\bibfnamefont {K.~T.}\ \bibnamefont {Law}},\ }\bibfield  {title}
  {\bibinfo {title} {{Josephson Diode Effect Induced by Valley Polarization in
  Twisted Bilayer Graphene}},\ }\href
  {https://doi.org/10.1103/PhysRevLett.130.266003} {\bibfield  {journal}
  {\bibinfo  {journal} {Phys. Rev. Lett.}\ }\textbf {\bibinfo {volume} {130}},\
  \bibinfo {pages} {266003} (\bibinfo {year} {2023})}\BibitemShut {NoStop}%
\bibitem [{\citenamefont {Xie}\ \emph {et~al.}(2023)\citenamefont {Xie},
  \citenamefont {Efetov},\ and\ \citenamefont {Law}}]{Xie2023Apr}%
  \BibitemOpen
  \bibfield  {author} {\bibinfo {author} {\bibfnamefont {Y.-M.}\ \bibnamefont
  {Xie}}, \bibinfo {author} {\bibfnamefont {D.~K.}\ \bibnamefont {Efetov}},\
  and\ \bibinfo {author} {\bibfnamefont {K.~T.}\ \bibnamefont {Law}},\
  }\bibfield  {title} {\bibinfo {title}
  {{${\ensuremath{\varphi}}_{0}$-Josephson junction in twisted bilayer graphene
  induced by a valley-polarized state}},\ }\href
  {https://doi.org/10.1103/PhysRevResearch.5.023029} {\bibfield  {journal}
  {\bibinfo  {journal} {Phys. Rev. Res.}\ }\textbf {\bibinfo {volume} {5}},\
  \bibinfo {pages} {023029} (\bibinfo {year} {2023})}\BibitemShut {NoStop}%
\bibitem [{\citenamefont {Alvarado}\ \emph {et~al.}(2023)\citenamefont
  {Alvarado}, \citenamefont {Burset},\ and\ \citenamefont
  {Yeyati}}]{Alvarado2023Sep}%
  \BibitemOpen
  \bibfield  {author} {\bibinfo {author} {\bibfnamefont {M.}~\bibnamefont
  {Alvarado}}, \bibinfo {author} {\bibfnamefont {P.}~\bibnamefont {Burset}},\
  and\ \bibinfo {author} {\bibfnamefont {A.~L.}\ \bibnamefont {Yeyati}},\
  }\bibfield  {title} {\bibinfo {title} {{Intrinsic nonmagnetic
  ${\ensuremath{\phi}}_{0}$ Josephson junctions in twisted bilayer graphene}},\
  }\href {https://doi.org/10.1103/PhysRevResearch.5.L032033} {\bibfield
  {journal} {\bibinfo  {journal} {Phys. Rev. Res.}\ }\textbf {\bibinfo {volume}
  {5}},\ \bibinfo {pages} {L032033} (\bibinfo {year} {2023})}\BibitemShut
  {NoStop}%
\bibitem [{\citenamefont {Annunziata}\ \emph {et~al.}(2010)\citenamefont
  {Annunziata}, \citenamefont {Santavicca}, \citenamefont {Frunzio},
  \citenamefont {Catelani}, \citenamefont {Rooks}, \citenamefont {Frydman},\
  and\ \citenamefont {Prober}}]{Annunziata2010Oct}%
  \BibitemOpen
  \bibfield  {author} {\bibinfo {author} {\bibfnamefont {A.~J.}\ \bibnamefont
  {Annunziata}}, \bibinfo {author} {\bibfnamefont {D.~F.}\ \bibnamefont
  {Santavicca}}, \bibinfo {author} {\bibfnamefont {L.}~\bibnamefont {Frunzio}},
  \bibinfo {author} {\bibfnamefont {G.}~\bibnamefont {Catelani}}, \bibinfo
  {author} {\bibfnamefont {M.~J.}\ \bibnamefont {Rooks}}, \bibinfo {author}
  {\bibfnamefont {A.}~\bibnamefont {Frydman}},\ and\ \bibinfo {author}
  {\bibfnamefont {D.~E.}\ \bibnamefont {Prober}},\ }\bibfield  {title}
  {\bibinfo {title} {{Tunable superconducting nanoinductors}},\ }\href
  {https://doi.org/10.1088/0957-4484/21/44/445202} {\bibfield  {journal}
  {\bibinfo  {journal} {Nanotechnology}\ }\textbf {\bibinfo {volume} {21}},\
  \bibinfo {pages} {445202} (\bibinfo {year} {2010})}\BibitemShut {NoStop}%
\bibitem [{\citenamefont {Barone}\ and\ \citenamefont
  {Patern{\ifmmode\grave{o}\else\`{o}\fi}}(1982)}]{Barone1982Jul}%
  \BibitemOpen
  \bibfield  {author} {\bibinfo {author} {\bibfnamefont {A.}~\bibnamefont
  {Barone}}\ and\ \bibinfo {author} {\bibfnamefont {G.}~\bibnamefont
  {Patern{\ifmmode\grave{o}\else\`{o}\fi}}},\ }\href
  {https://doi.org/10.1002/352760278X} {\emph {\bibinfo {title} {{Physics and
  Applications of the Josephson Effect}}}}\ (\bibinfo  {publisher} {John Wiley
  \& Sons, Ltd},\ \bibinfo {year} {1982})\ Chap.~\bibinfo {chapter}
  {5}\BibitemShut {NoStop}%
\bibitem [{\citenamefont
  {L{\ifmmode\acute{o}\else\'{o}\fi}pez-N{\ifmmode\acute{u}\else\'{u}\fi}{\ifmmode\tilde{n}\else\~{n}\fi}ez}\
  \emph {et~al.}(2023)\citenamefont
  {L{\ifmmode\acute{o}\else\'{o}\fi}pez-N{\ifmmode\acute{u}\else\'{u}\fi}{\ifmmode\tilde{n}\else\~{n}\fi}ez},
  \citenamefont {Torras-Coloma}, \citenamefont {Montserrat}, \citenamefont
  {Bertoldo}, \citenamefont {Cozzolino}, \citenamefont {Rius}, \citenamefont
  {Mart{\ifmmode\acute{\imath}\else\'{\i}\fi}nez},\ and\ \citenamefont
  {Forn-D{\ifmmode\acute{\imath}\else\'{\i}\fi}az}}]{Lopez-Nunez2023Nov}%
  \BibitemOpen
  \bibfield  {author} {\bibinfo {author} {\bibfnamefont {D.}~\bibnamefont
  {L{\ifmmode\acute{o}\else\'{o}\fi}pez-N{\ifmmode\acute{u}\else\'{u}\fi}{\ifmmode\tilde{n}\else\~{n}\fi}ez}},
  \bibinfo {author} {\bibfnamefont {A.}~\bibnamefont {Torras-Coloma}}, \bibinfo
  {author} {\bibfnamefont {Q.~P.}\ \bibnamefont {Montserrat}}, \bibinfo
  {author} {\bibfnamefont {E.}~\bibnamefont {Bertoldo}}, \bibinfo {author}
  {\bibfnamefont {L.}~\bibnamefont {Cozzolino}}, \bibinfo {author}
  {\bibfnamefont {G.}~\bibnamefont {Rius}}, \bibinfo {author} {\bibfnamefont
  {M.}~\bibnamefont {Mart{\ifmmode\acute{\imath}\else\'{\i}\fi}nez}},\ and\
  \bibinfo {author} {\bibfnamefont {P.}~\bibnamefont
  {Forn-D{\ifmmode\acute{\imath}\else\'{\i}\fi}az}},\ }\bibfield  {title}
  {\bibinfo {title} {{Magnetic penetration depth of Aluminum thin films}},\
  }\bibfield  {journal} {\bibinfo  {journal} {arXiv}\ }\href
  {https://doi.org/10.48550/arXiv.2311.14119} {10.48550/arXiv.2311.14119}
  (\bibinfo {year} {2023}),\ \Eprint {https://arxiv.org/abs/2311.14119}
  {2311.14119} \BibitemShut {NoStop}%
\bibitem [{\citenamefont {Banszerus}\ \emph {et~al.}(2025)\citenamefont
  {Banszerus}, \citenamefont {Andersson}, \citenamefont {Marshall},
  \citenamefont {Lindemann}, \citenamefont {Manfra}, \citenamefont {Marcus},\
  and\ \citenamefont
  {Vaitiek{\ifmmode\dot{e}\else\.{e}\fi}nas}}]{Banszerus2025Feb}%
  \BibitemOpen
  \bibfield  {author} {\bibinfo {author} {\bibfnamefont {L.}~\bibnamefont
  {Banszerus}}, \bibinfo {author} {\bibfnamefont {C.~W.}\ \bibnamefont
  {Andersson}}, \bibinfo {author} {\bibfnamefont {W.}~\bibnamefont {Marshall}},
  \bibinfo {author} {\bibfnamefont {T.}~\bibnamefont {Lindemann}}, \bibinfo
  {author} {\bibfnamefont {M.~J.}\ \bibnamefont {Manfra}}, \bibinfo {author}
  {\bibfnamefont {C.~M.}\ \bibnamefont {Marcus}},\ and\ \bibinfo {author}
  {\bibfnamefont {S.}~\bibnamefont
  {Vaitiek{\ifmmode\dot{e}\else\.{e}\fi}nas}},\ }\bibfield  {title} {\bibinfo
  {title} {{Hybrid Josephson Rhombus: A Superconducting Element with Tailored
  Current-Phase Relation}},\ }\href
  {https://doi.org/10.1103/PhysRevX.15.011021} {\bibfield  {journal} {\bibinfo
  {journal} {Phys. Rev. X}\ }\textbf {\bibinfo {volume} {15}},\ \bibinfo
  {pages} {011021} (\bibinfo {year} {2025})}\BibitemShut {NoStop}%
\bibitem [{\citenamefont {Jha}\ \emph {et~al.}(2025)\citenamefont {Jha},
  \citenamefont {Endres}, \citenamefont {Watanabe}, \citenamefont {Taniguchi},
  \citenamefont {Banerjee}, \citenamefont
  {Sch{\ifmmode\ddot{o}\else\"{o}\fi}nenberger},\ and\ \citenamefont
  {Karnatak}}]{Jha2024Mar}%
  \BibitemOpen
  \bibfield  {author} {\bibinfo {author} {\bibfnamefont {R.}~\bibnamefont
  {Jha}}, \bibinfo {author} {\bibfnamefont {M.}~\bibnamefont {Endres}},
  \bibinfo {author} {\bibfnamefont {K.}~\bibnamefont {Watanabe}}, \bibinfo
  {author} {\bibfnamefont {T.}~\bibnamefont {Taniguchi}}, \bibinfo {author}
  {\bibfnamefont {M.}~\bibnamefont {Banerjee}}, \bibinfo {author}
  {\bibfnamefont {C.}~\bibnamefont
  {Sch{\ifmmode\ddot{o}\else\"{o}\fi}nenberger}},\ and\ \bibinfo {author}
  {\bibfnamefont {P.}~\bibnamefont {Karnatak}},\ }\bibfield  {title} {\bibinfo
  {title} {{Large Tunable Kinetic Inductance in a Twisted Graphene
  Superconductor}},\ }\href {https://doi.org/10.1103/PhysRevLett.134.216001}
  {\bibfield  {journal} {\bibinfo  {journal} {Phys. Rev. Lett.}\ }\textbf
  {\bibinfo {volume} {134}},\ \bibinfo {pages} {216001} (\bibinfo {year}
  {2025})}\BibitemShut {NoStop}%
\bibitem [{\citenamefont {Tinkham}(2004)}]{tinkham1975}%
  \BibitemOpen
  \bibfield  {author} {\bibinfo {author} {\bibfnamefont {M.}~\bibnamefont
  {Tinkham}},\ }\href@noop {} {\emph {\bibinfo {title} {{Introduction to
  Superconductivity}}}},\ Vol.~\bibinfo {volume} {2}\ (\bibinfo  {publisher}
  {Dover Publications},\ \bibinfo {year} {2004})\BibitemShut {NoStop}%
\bibitem [{\citenamefont
  {Sac{\ifmmode\acute{e}\else\'{e}\fi}p{\ifmmode\acute{e}\else\'{e}\fi}}\ \emph
  {et~al.}(2020)\citenamefont
  {Sac{\ifmmode\acute{e}\else\'{e}\fi}p{\ifmmode\acute{e}\else\'{e}\fi}},
  \citenamefont {Feigel{'}man},\ and\ \citenamefont
  {Klapwijk}}]{Sacepe2020Jul}%
  \BibitemOpen
  \bibfield  {author} {\bibinfo {author} {\bibfnamefont {B.}~\bibnamefont
  {Sac{\ifmmode\acute{e}\else\'{e}\fi}p{\ifmmode\acute{e}\else\'{e}\fi}}},
  \bibinfo {author} {\bibfnamefont {M.}~\bibnamefont {Feigel{'}man}},\ and\
  \bibinfo {author} {\bibfnamefont {T.~M.}\ \bibnamefont {Klapwijk}},\
  }\bibfield  {title} {\bibinfo {title} {{Quantum breakdown of
  superconductivity in low-dimensional materials}},\ }\href
  {https://doi.org/10.1038/s41567-020-0905-x} {\bibfield  {journal} {\bibinfo
  {journal} {Nat. Phys.}\ }\textbf {\bibinfo {volume} {16}},\ \bibinfo {pages}
  {734} (\bibinfo {year} {2020})}\BibitemShut {NoStop}%
\bibitem [{\citenamefont {Glick}\ \emph {et~al.}(2017)\citenamefont {Glick},
  \citenamefont {Khasawneh}, \citenamefont {Niedzielski}, \citenamefont
  {Loloee}, \citenamefont {Pratt}, \citenamefont {Birge}, \citenamefont
  {Gingrich}, \citenamefont {Kotula},\ and\ \citenamefont
  {Missert}}]{Glick2017Oct}%
  \BibitemOpen
  \bibfield  {author} {\bibinfo {author} {\bibfnamefont {J.~A.}\ \bibnamefont
  {Glick}}, \bibinfo {author} {\bibfnamefont {M.~A.}\ \bibnamefont
  {Khasawneh}}, \bibinfo {author} {\bibfnamefont {B.~M.}\ \bibnamefont
  {Niedzielski}}, \bibinfo {author} {\bibfnamefont {R.}~\bibnamefont {Loloee}},
  \bibinfo {author} {\bibfnamefont {W.~P.}\ \bibnamefont {Pratt}}, \bibinfo
  {author} {\bibfnamefont {N.~O.}\ \bibnamefont {Birge}}, \bibinfo {author}
  {\bibfnamefont {E.~C.}\ \bibnamefont {Gingrich}}, \bibinfo {author}
  {\bibfnamefont {P.~G.}\ \bibnamefont {Kotula}},\ and\ \bibinfo {author}
  {\bibfnamefont {N.}~\bibnamefont {Missert}},\ }\bibfield  {title} {\bibinfo
  {title} {{Critical current oscillations of elliptical Josephson junctions
  with single-domain ferromagnetic layers}},\ }\href
  {https://doi.org/10.1063/1.4989392} {\bibfield  {journal} {\bibinfo
  {journal} {J. Appl. Phys.}\ }\textbf {\bibinfo {volume} {122}},\ \bibinfo
  {pages} {133906} (\bibinfo {year} {2017})}\BibitemShut {NoStop}%
\bibitem [{\citenamefont {Ivanchenko}\ and\ \citenamefont
  {Zil'berman}(1969)}]{Ivanchenko1969Jun}%
  \BibitemOpen
  \bibfield  {author} {\bibinfo {author} {\bibfnamefont {Y.~M.}\ \bibnamefont
  {Ivanchenko}}\ and\ \bibinfo {author} {\bibfnamefont {L.~A.}\ \bibnamefont
  {Zil'berman}},\ }\bibfield  {title} {\bibinfo {title} {{The Josephson Effect
  in Small Tunnel Contacts}},\ }\href@noop {} {\bibfield  {journal} {\bibinfo
  {journal} {Sov. Phys. J. Exp. Theor. Phys.}\ }\textbf {\bibinfo {volume}
  {28}},\ \bibinfo {pages} {1272} (\bibinfo {year} {1969})}\BibitemShut
  {NoStop}%
\bibitem [{\citenamefont {Ambegaokar}\ and\ \citenamefont
  {Halperin}(1969)}]{Ambegaokar1969Jun}%
  \BibitemOpen
  \bibfield  {author} {\bibinfo {author} {\bibfnamefont {V.}~\bibnamefont
  {Ambegaokar}}\ and\ \bibinfo {author} {\bibfnamefont {B.~I.}\ \bibnamefont
  {Halperin}},\ }\bibfield  {title} {\bibinfo {title} {{Voltage Due to Thermal
  Noise in the dc Josephson Effect}},\ }\href
  {https://doi.org/10.1103/PhysRevLett.22.1364} {\bibfield  {journal} {\bibinfo
   {journal} {Phys. Rev. Lett.}\ }\textbf {\bibinfo {volume} {22}},\ \bibinfo
  {pages} {1364} (\bibinfo {year} {1969})}\BibitemShut {NoStop}%
\bibitem [{\citenamefont {Schiela}\ \emph {et~al.}(2025)\citenamefont
  {Schiela}, \citenamefont {Mikalsen}, \citenamefont {Crawford}, \citenamefont
  {Ili{\ifmmode\acute{c}\else\'{c}\fi}}, \citenamefont {Strickland},
  \citenamefont {Bergeret},\ and\ \citenamefont {Shabani}}]{Schiela2025Apr}%
  \BibitemOpen
  \bibfield  {author} {\bibinfo {author} {\bibfnamefont {W.~F.}\ \bibnamefont
  {Schiela}}, \bibinfo {author} {\bibfnamefont {M.}~\bibnamefont {Mikalsen}},
  \bibinfo {author} {\bibfnamefont {D.}~\bibnamefont {Crawford}}, \bibinfo
  {author} {\bibfnamefont {S.}~\bibnamefont
  {Ili{\ifmmode\acute{c}\else\'{c}\fi}}}, \bibinfo {author} {\bibfnamefont
  {W.~M.}\ \bibnamefont {Strickland}}, \bibinfo {author} {\bibfnamefont
  {F.~S.}\ \bibnamefont {Bergeret}},\ and\ \bibinfo {author} {\bibfnamefont
  {J.}~\bibnamefont {Shabani}},\ }\bibfield  {title} {\bibinfo {title}
  {{Gate-tunable polarity inversions and three-fold rotation symmetry of the
  superconducting diode effect}},\ }\bibfield  {journal} {\bibinfo  {journal}
  {arXiv}\ }\href {https://doi.org/10.48550/arXiv.2504.21470}
  {10.48550/arXiv.2504.21470} (\bibinfo {year} {2025}),\ \Eprint
  {https://arxiv.org/abs/2504.21470} {2504.21470} \BibitemShut {NoStop}%
\bibitem [{\citenamefont {McGilly}\ \emph {et~al.}(2020)\citenamefont
  {McGilly}, \citenamefont {Kerelsky}, \citenamefont {Finney}, \citenamefont
  {Shapovalov}, \citenamefont {Shih}, \citenamefont {Ghiotto}, \citenamefont
  {Zeng}, \citenamefont {Moore}, \citenamefont {Wu}, \citenamefont {Bai},
  \citenamefont {Watanabe}, \citenamefont {Taniguchi}, \citenamefont {Stengel},
  \citenamefont {Zhou}, \citenamefont {Hone}, \citenamefont {Zhu},
  \citenamefont {Basov}, \citenamefont {Dean}, \citenamefont {Dreyer},\ and\
  \citenamefont {Pasupathy}}]{McGilly2020Jul}%
  \BibitemOpen
  \bibfield  {author} {\bibinfo {author} {\bibfnamefont {L.~J.}\ \bibnamefont
  {McGilly}}, \bibinfo {author} {\bibfnamefont {A.}~\bibnamefont {Kerelsky}},
  \bibinfo {author} {\bibfnamefont {N.~R.}\ \bibnamefont {Finney}}, \bibinfo
  {author} {\bibfnamefont {K.}~\bibnamefont {Shapovalov}}, \bibinfo {author}
  {\bibfnamefont {E.-M.}\ \bibnamefont {Shih}}, \bibinfo {author}
  {\bibfnamefont {A.}~\bibnamefont {Ghiotto}}, \bibinfo {author} {\bibfnamefont
  {Y.}~\bibnamefont {Zeng}}, \bibinfo {author} {\bibfnamefont {S.~L.}\
  \bibnamefont {Moore}}, \bibinfo {author} {\bibfnamefont {W.}~\bibnamefont
  {Wu}}, \bibinfo {author} {\bibfnamefont {Y.}~\bibnamefont {Bai}}, \bibinfo
  {author} {\bibfnamefont {K.}~\bibnamefont {Watanabe}}, \bibinfo {author}
  {\bibfnamefont {T.}~\bibnamefont {Taniguchi}}, \bibinfo {author}
  {\bibfnamefont {M.}~\bibnamefont {Stengel}}, \bibinfo {author} {\bibfnamefont
  {L.}~\bibnamefont {Zhou}}, \bibinfo {author} {\bibfnamefont {J.}~\bibnamefont
  {Hone}}, \bibinfo {author} {\bibfnamefont {X.}~\bibnamefont {Zhu}}, \bibinfo
  {author} {\bibfnamefont {D.~N.}\ \bibnamefont {Basov}}, \bibinfo {author}
  {\bibfnamefont {C.}~\bibnamefont {Dean}}, \bibinfo {author} {\bibfnamefont
  {C.~E.}\ \bibnamefont {Dreyer}},\ and\ \bibinfo {author} {\bibfnamefont
  {A.~N.}\ \bibnamefont {Pasupathy}},\ }\bibfield  {title} {\bibinfo {title}
  {{Visualization of moir{\ifmmode\acute{e}\else\'{e}\fi} superlattices}},\
  }\href {https://doi.org/10.1038/s41565-020-0708-3} {\bibfield  {journal}
  {\bibinfo  {journal} {Nat. Nanotechnol.}\ }\textbf {\bibinfo {volume} {15}},\
  \bibinfo {pages} {580} (\bibinfo {year} {2020})}\BibitemShut {NoStop}%
\bibitem [{\citenamefont {Sch{\ifmmode\ddot{a}\else\"{a}\fi}pers}\ \emph
  {et~al.}(2022)\citenamefont {Sch{\ifmmode\ddot{a}\else\"{a}\fi}pers},
  \citenamefont {Sonntag}, \citenamefont {Valerius}, \citenamefont {Pestka},
  \citenamefont {Strasdas}, \citenamefont {Watanabe}, \citenamefont
  {Taniguchi}, \citenamefont {Wirtz}, \citenamefont {Morgenstern},
  \citenamefont {Beschoten}, \citenamefont {Dolleman},\ and\ \citenamefont
  {Stampfer}}]{Schapers2022Jul}%
  \BibitemOpen
  \bibfield  {author} {\bibinfo {author} {\bibfnamefont {A.}~\bibnamefont
  {Sch{\ifmmode\ddot{a}\else\"{a}\fi}pers}}, \bibinfo {author} {\bibfnamefont
  {J.}~\bibnamefont {Sonntag}}, \bibinfo {author} {\bibfnamefont
  {L.}~\bibnamefont {Valerius}}, \bibinfo {author} {\bibfnamefont
  {B.}~\bibnamefont {Pestka}}, \bibinfo {author} {\bibfnamefont
  {J.}~\bibnamefont {Strasdas}}, \bibinfo {author} {\bibfnamefont
  {K.}~\bibnamefont {Watanabe}}, \bibinfo {author} {\bibfnamefont
  {T.}~\bibnamefont {Taniguchi}}, \bibinfo {author} {\bibfnamefont
  {L.}~\bibnamefont {Wirtz}}, \bibinfo {author} {\bibfnamefont
  {M.}~\bibnamefont {Morgenstern}}, \bibinfo {author} {\bibfnamefont
  {B.}~\bibnamefont {Beschoten}}, \bibinfo {author} {\bibfnamefont {R.~J.}\
  \bibnamefont {Dolleman}},\ and\ \bibinfo {author} {\bibfnamefont
  {C.}~\bibnamefont {Stampfer}},\ }\bibfield  {title} {\bibinfo {title} {{Raman
  imaging of twist angle variations in twisted bilayer graphene at intermediate
  angles}},\ }\href {https://doi.org/10.1088/2053-1583/ac7e59} {\bibfield
  {journal} {\bibinfo  {journal} {2D Mater.}\ }\textbf {\bibinfo {volume}
  {9}},\ \bibinfo {pages} {045009} (\bibinfo {year} {2022})}\BibitemShut
  {NoStop}%
\bibitem [{\citenamefont {Uri}\ \emph {et~al.}(2020)\citenamefont {Uri},
  \citenamefont {Grover}, \citenamefont {Cao}, \citenamefont {Crosse},
  \citenamefont {Bagani}, \citenamefont {Rodan-Legrain}, \citenamefont
  {Myasoedov}, \citenamefont {Watanabe}, \citenamefont {Taniguchi},
  \citenamefont {Moon}, \citenamefont {Koshino}, \citenamefont
  {Jarillo-Herrero},\ and\ \citenamefont {Zeldov}}]{Uri2020May}%
  \BibitemOpen
  \bibfield  {author} {\bibinfo {author} {\bibfnamefont {A.}~\bibnamefont
  {Uri}}, \bibinfo {author} {\bibfnamefont {S.}~\bibnamefont {Grover}},
  \bibinfo {author} {\bibfnamefont {Y.}~\bibnamefont {Cao}}, \bibinfo {author}
  {\bibfnamefont {J.~A.}\ \bibnamefont {Crosse}}, \bibinfo {author}
  {\bibfnamefont {K.}~\bibnamefont {Bagani}}, \bibinfo {author} {\bibfnamefont
  {D.}~\bibnamefont {Rodan-Legrain}}, \bibinfo {author} {\bibfnamefont
  {Y.}~\bibnamefont {Myasoedov}}, \bibinfo {author} {\bibfnamefont
  {K.}~\bibnamefont {Watanabe}}, \bibinfo {author} {\bibfnamefont
  {T.}~\bibnamefont {Taniguchi}}, \bibinfo {author} {\bibfnamefont
  {P.}~\bibnamefont {Moon}}, \bibinfo {author} {\bibfnamefont {M.}~\bibnamefont
  {Koshino}}, \bibinfo {author} {\bibfnamefont {P.}~\bibnamefont
  {Jarillo-Herrero}},\ and\ \bibinfo {author} {\bibfnamefont {E.}~\bibnamefont
  {Zeldov}},\ }\bibfield  {title} {\bibinfo {title} {{Mapping the twist-angle
  disorder and Landau levels in magic-angle graphene}},\ }\href
  {https://doi.org/10.1038/s41586-020-2255-3} {\bibfield  {journal} {\bibinfo
  {journal} {Nature}\ }\textbf {\bibinfo {volume} {581}},\ \bibinfo {pages}
  {47} (\bibinfo {year} {2020})}\BibitemShut {NoStop}%
\bibitem [{\citenamefont {Choi}\ \emph {et~al.}(2021)\citenamefont {Choi},
  \citenamefont {Kim}, \citenamefont {Peng}, \citenamefont {Thomson},
  \citenamefont {Lewandowski}, \citenamefont {Polski}, \citenamefont {Zhang},
  \citenamefont {Arora}, \citenamefont {Watanabe}, \citenamefont {Taniguchi},
  \citenamefont {Alicea},\ and\ \citenamefont {Nadj-Perge}}]{Choi2021Jan}%
  \BibitemOpen
  \bibfield  {author} {\bibinfo {author} {\bibfnamefont {Y.}~\bibnamefont
  {Choi}}, \bibinfo {author} {\bibfnamefont {H.}~\bibnamefont {Kim}}, \bibinfo
  {author} {\bibfnamefont {Y.}~\bibnamefont {Peng}}, \bibinfo {author}
  {\bibfnamefont {A.}~\bibnamefont {Thomson}}, \bibinfo {author} {\bibfnamefont
  {C.}~\bibnamefont {Lewandowski}}, \bibinfo {author} {\bibfnamefont
  {R.}~\bibnamefont {Polski}}, \bibinfo {author} {\bibfnamefont
  {Y.}~\bibnamefont {Zhang}}, \bibinfo {author} {\bibfnamefont {H.~S.}\
  \bibnamefont {Arora}}, \bibinfo {author} {\bibfnamefont {K.}~\bibnamefont
  {Watanabe}}, \bibinfo {author} {\bibfnamefont {T.}~\bibnamefont {Taniguchi}},
  \bibinfo {author} {\bibfnamefont {J.}~\bibnamefont {Alicea}},\ and\ \bibinfo
  {author} {\bibfnamefont {S.}~\bibnamefont {Nadj-Perge}},\ }\bibfield  {title}
  {\bibinfo {title} {{Correlation-driven topological phases in magic-angle
  twisted bilayer graphene}},\ }\href
  {https://doi.org/10.1038/s41586-020-03159-7} {\bibfield  {journal} {\bibinfo
  {journal} {Nature}\ }\textbf {\bibinfo {volume} {589}},\ \bibinfo {pages}
  {536} (\bibinfo {year} {2021})}\BibitemShut {NoStop}%
\bibitem [{\citenamefont {Dolleman}\ \emph {et~al.}(2024)\citenamefont
  {Dolleman}, \citenamefont {Rothstein}, \citenamefont {Fischer}, \citenamefont
  {Klebl}, \citenamefont {Waldecker}, \citenamefont {Watanabe}, \citenamefont
  {Taniguchi}, \citenamefont {Kennes}, \citenamefont {Libisch}, \citenamefont
  {Beschoten},\ and\ \citenamefont {Stampfer}}]{Dolleman2024Apr}%
  \BibitemOpen
  \bibfield  {author} {\bibinfo {author} {\bibfnamefont {R.~J.}\ \bibnamefont
  {Dolleman}}, \bibinfo {author} {\bibfnamefont {A.}~\bibnamefont {Rothstein}},
  \bibinfo {author} {\bibfnamefont {A.}~\bibnamefont {Fischer}}, \bibinfo
  {author} {\bibfnamefont {L.}~\bibnamefont {Klebl}}, \bibinfo {author}
  {\bibfnamefont {L.}~\bibnamefont {Waldecker}}, \bibinfo {author}
  {\bibfnamefont {K.}~\bibnamefont {Watanabe}}, \bibinfo {author}
  {\bibfnamefont {T.}~\bibnamefont {Taniguchi}}, \bibinfo {author}
  {\bibfnamefont {D.~M.}\ \bibnamefont {Kennes}}, \bibinfo {author}
  {\bibfnamefont {F.}~\bibnamefont {Libisch}}, \bibinfo {author} {\bibfnamefont
  {B.}~\bibnamefont {Beschoten}},\ and\ \bibinfo {author} {\bibfnamefont
  {C.}~\bibnamefont {Stampfer}},\ }\bibfield  {title} {\bibinfo {title}
  {{Negative electronic compressibility in charge islands in twisted bilayer
  graphene}},\ }\href {https://doi.org/10.1103/PhysRevB.109.155430} {\bibfield
  {journal} {\bibinfo  {journal} {Phys. Rev. B}\ }\textbf {\bibinfo {volume}
  {109}},\ \bibinfo {pages} {155430} (\bibinfo {year} {2024})}\BibitemShut
  {NoStop}%
\bibitem [{\citenamefont {Lau}\ \emph {et~al.}(2022)\citenamefont {Lau},
  \citenamefont {Bockrath}, \citenamefont {Mak},\ and\ \citenamefont
  {Zhang}}]{Lau2022Feb}%
  \BibitemOpen
  \bibfield  {author} {\bibinfo {author} {\bibfnamefont {C.~N.}\ \bibnamefont
  {Lau}}, \bibinfo {author} {\bibfnamefont {M.~W.}\ \bibnamefont {Bockrath}},
  \bibinfo {author} {\bibfnamefont {K.~F.}\ \bibnamefont {Mak}},\ and\ \bibinfo
  {author} {\bibfnamefont {F.}~\bibnamefont {Zhang}},\ }\bibfield  {title}
  {\bibinfo {title} {{Reproducibility in the fabrication and physics of
  moir{\ifmmode\acute{e}\else\'{e}\fi} materials}},\ }\href
  {https://doi.org/10.1038/s41586-021-04173-z} {\bibfield  {journal} {\bibinfo
  {journal} {Nature}\ }\textbf {\bibinfo {volume} {602}},\ \bibinfo {pages}
  {41} (\bibinfo {year} {2022})}\BibitemShut {NoStop}%
\bibitem [{\citenamefont {Reinhardt}\ \emph {et~al.}(2024)\citenamefont
  {Reinhardt}, \citenamefont {Ascherl}, \citenamefont {Costa}, \citenamefont
  {Berger}, \citenamefont {Gronin}, \citenamefont {Gardner}, \citenamefont
  {Lindemann}, \citenamefont {Manfra}, \citenamefont {Fabian}, \citenamefont
  {Kochan}, \citenamefont {Strunk},\ and\ \citenamefont
  {Paradiso}}]{Reinhardt2024May}%
  \BibitemOpen
  \bibfield  {author} {\bibinfo {author} {\bibfnamefont {S.}~\bibnamefont
  {Reinhardt}}, \bibinfo {author} {\bibfnamefont {T.}~\bibnamefont {Ascherl}},
  \bibinfo {author} {\bibfnamefont {A.}~\bibnamefont {Costa}}, \bibinfo
  {author} {\bibfnamefont {J.}~\bibnamefont {Berger}}, \bibinfo {author}
  {\bibfnamefont {S.}~\bibnamefont {Gronin}}, \bibinfo {author} {\bibfnamefont
  {G.~C.}\ \bibnamefont {Gardner}}, \bibinfo {author} {\bibfnamefont
  {T.}~\bibnamefont {Lindemann}}, \bibinfo {author} {\bibfnamefont {M.~J.}\
  \bibnamefont {Manfra}}, \bibinfo {author} {\bibfnamefont {J.}~\bibnamefont
  {Fabian}}, \bibinfo {author} {\bibfnamefont {D.}~\bibnamefont {Kochan}},
  \bibinfo {author} {\bibfnamefont {C.}~\bibnamefont {Strunk}},\ and\ \bibinfo
  {author} {\bibfnamefont {N.}~\bibnamefont {Paradiso}},\ }\bibfield  {title}
  {\bibinfo {title} {{Link between supercurrent diode and anomalous Josephson
  effect revealed by gate-controlled interferometry}},\ }\href
  {https://doi.org/10.1038/s41467-024-48741-z} {\bibfield  {journal} {\bibinfo
  {journal} {Nat. Commun.}\ }\textbf {\bibinfo {volume} {15}},\ \bibinfo
  {pages} {1} (\bibinfo {year} {2024})}\BibitemShut {NoStop}%
\bibitem [{\citenamefont {Randle}\ \emph {et~al.}(2023)\citenamefont {Randle},
  \citenamefont {Hosoda}, \citenamefont {Deacon}, \citenamefont {Ohtomo},
  \citenamefont {Zellekens}, \citenamefont {Watanabe}, \citenamefont
  {Taniguchi}, \citenamefont {Okazaki}, \citenamefont {Sasagawa}, \citenamefont
  {Kawaguchi}, \citenamefont {Sato},\ and\ \citenamefont
  {Ishibashi}}]{Randle2023Sep}%
  \BibitemOpen
  \bibfield  {author} {\bibinfo {author} {\bibfnamefont {M.~D.}\ \bibnamefont
  {Randle}}, \bibinfo {author} {\bibfnamefont {M.}~\bibnamefont {Hosoda}},
  \bibinfo {author} {\bibfnamefont {R.~S.}\ \bibnamefont {Deacon}}, \bibinfo
  {author} {\bibfnamefont {M.}~\bibnamefont {Ohtomo}}, \bibinfo {author}
  {\bibfnamefont {P.}~\bibnamefont {Zellekens}}, \bibinfo {author}
  {\bibfnamefont {K.}~\bibnamefont {Watanabe}}, \bibinfo {author}
  {\bibfnamefont {T.}~\bibnamefont {Taniguchi}}, \bibinfo {author}
  {\bibfnamefont {S.}~\bibnamefont {Okazaki}}, \bibinfo {author} {\bibfnamefont
  {T.}~\bibnamefont {Sasagawa}}, \bibinfo {author} {\bibfnamefont
  {K.}~\bibnamefont {Kawaguchi}}, \bibinfo {author} {\bibfnamefont
  {S.}~\bibnamefont {Sato}},\ and\ \bibinfo {author} {\bibfnamefont
  {K.}~\bibnamefont {Ishibashi}},\ }\bibfield  {title} {\bibinfo {title}
  {{Gate-Defined Josephson Weak-Links in Monolayer WTe$_2$}},\ }\href
  {https://doi.org/10.1002/adma.202301683} {\bibfield  {journal} {\bibinfo
  {journal} {Adv. Mater.}\ }\textbf {\bibinfo {volume} {35}},\ \bibinfo {pages}
  {2301683} (\bibinfo {year} {2023})}\BibitemShut {NoStop}%
\bibitem [{\citenamefont {Hooper}\ \emph {et~al.}(2004)\citenamefont {Hooper},
  \citenamefont {Mao}, \citenamefont {Nelson}, \citenamefont {Liu},
  \citenamefont {Wada},\ and\ \citenamefont {Maeno}}]{Hooper2004Jul}%
  \BibitemOpen
  \bibfield  {author} {\bibinfo {author} {\bibfnamefont {J.}~\bibnamefont
  {Hooper}}, \bibinfo {author} {\bibfnamefont {Z.~Q.}\ \bibnamefont {Mao}},
  \bibinfo {author} {\bibfnamefont {K.~D.}\ \bibnamefont {Nelson}}, \bibinfo
  {author} {\bibfnamefont {Y.}~\bibnamefont {Liu}}, \bibinfo {author}
  {\bibfnamefont {M.}~\bibnamefont {Wada}},\ and\ \bibinfo {author}
  {\bibfnamefont {Y.}~\bibnamefont {Maeno}},\ }\bibfield  {title} {\bibinfo
  {title} {{Anomalous Josephson network in the
  $\mathrm{Ru}\text{\penalty1000-\hskip0pt}{\mathrm{Sr}}_{2}{\mathrm{Ru}\mathrm{O}}_{4}$
  eutectic system}},\ }\href {https://doi.org/10.1103/PhysRevB.70.014510}
  {\bibfield  {journal} {\bibinfo  {journal} {Phys. Rev. B}\ }\textbf {\bibinfo
  {volume} {70}},\ \bibinfo {pages} {014510} (\bibinfo {year}
  {2004})}\BibitemShut {NoStop}%
\bibitem [{\citenamefont {Gupta}\ \emph {et~al.}(2023)\citenamefont {Gupta},
  \citenamefont {Graziano}, \citenamefont {Pendharkar}, \citenamefont {Dong},
  \citenamefont {Dempsey}, \citenamefont {Palmstr{\o}m},\ and\ \citenamefont
  {Pribiag}}]{Gupta2023May}%
  \BibitemOpen
  \bibfield  {author} {\bibinfo {author} {\bibfnamefont {M.}~\bibnamefont
  {Gupta}}, \bibinfo {author} {\bibfnamefont {G.~V.}\ \bibnamefont {Graziano}},
  \bibinfo {author} {\bibfnamefont {M.}~\bibnamefont {Pendharkar}}, \bibinfo
  {author} {\bibfnamefont {J.~T.}\ \bibnamefont {Dong}}, \bibinfo {author}
  {\bibfnamefont {C.~P.}\ \bibnamefont {Dempsey}}, \bibinfo {author}
  {\bibfnamefont {C.}~\bibnamefont {Palmstr{\o}m}},\ and\ \bibinfo {author}
  {\bibfnamefont {V.~S.}\ \bibnamefont {Pribiag}},\ }\bibfield  {title}
  {\bibinfo {title} {{Gate-tunable superconducting diode effect in a
  three-terminal Josephson device}},\ }\href
  {https://doi.org/10.1038/s41467-023-38856-0} {\bibfield  {journal} {\bibinfo
  {journal} {Nat. Commun.}\ }\textbf {\bibinfo {volume} {14}},\ \bibinfo
  {pages} {1} (\bibinfo {year} {2023})}\BibitemShut {NoStop}%
\bibitem [{\citenamefont {Albrecht}\ \emph {et~al.}(2017)\citenamefont
  {Albrecht}, \citenamefont {Moers},\ and\ \citenamefont
  {Hermanns}}]{Albrecht2017May}%
  \BibitemOpen
  \bibfield  {author} {\bibinfo {author} {\bibfnamefont {W.}~\bibnamefont
  {Albrecht}}, \bibinfo {author} {\bibfnamefont {J.}~\bibnamefont {Moers}},\
  and\ \bibinfo {author} {\bibfnamefont {B.}~\bibnamefont {Hermanns}},\
  }\bibfield  {title} {\bibinfo {title} {{HNF - Helmholtz Nano Facility}},\
  }\href {https://doi.org/10.17815/jlsrf-3-158} {\bibfield  {journal} {\bibinfo
   {journal} {Journal of Large-Scale Research Facilities}\ }\textbf {\bibinfo
  {volume} {3}},\ \bibinfo {pages} {112} (\bibinfo {year} {2017})}\BibitemShut
  {NoStop}%
\bibitem [{\citenamefont {Park}\ \emph
  {et~al.}(2021{\natexlab{b}})\citenamefont {Park}, \citenamefont {Cao},
  \citenamefont {Watanabe}, \citenamefont {Taniguchi},\ and\ \citenamefont
  {Jarillo-Herrero}}]{Park2021Apr}%
  \BibitemOpen
  \bibfield  {author} {\bibinfo {author} {\bibfnamefont {J.~M.}\ \bibnamefont
  {Park}}, \bibinfo {author} {\bibfnamefont {Y.}~\bibnamefont {Cao}}, \bibinfo
  {author} {\bibfnamefont {K.}~\bibnamefont {Watanabe}}, \bibinfo {author}
  {\bibfnamefont {T.}~\bibnamefont {Taniguchi}},\ and\ \bibinfo {author}
  {\bibfnamefont {P.}~\bibnamefont {Jarillo-Herrero}},\ }\bibfield  {title}
  {\bibinfo {title} {{Flavour Hund{'}s coupling, Chern gaps and charge
  diffusivity in moir{\ifmmode\acute{e}\else\'{e}\fi} graphene}},\ }\href
  {https://doi.org/10.1038/s41586-021-03366-w} {\bibfield  {journal} {\bibinfo
  {journal} {Nature}\ }\textbf {\bibinfo {volume} {592}},\ \bibinfo {pages}
  {43} (\bibinfo {year} {2021}{\natexlab{b}})}\BibitemShut {NoStop}%
\bibitem [{\citenamefont {Kim}\ \emph {et~al.}(2016)\citenamefont {Kim},
  \citenamefont {Yankowitz}, \citenamefont {Fallahazad}, \citenamefont {Kang},
  \citenamefont {Movva}, \citenamefont {Huang}, \citenamefont {Larentis},
  \citenamefont {Corbet}, \citenamefont {Taniguchi}, \citenamefont {Watanabe},
  \citenamefont {Banerjee}, \citenamefont {LeRoy},\ and\ \citenamefont
  {Tutuc}}]{Kim2016Mar}%
  \BibitemOpen
  \bibfield  {author} {\bibinfo {author} {\bibfnamefont {K.}~\bibnamefont
  {Kim}}, \bibinfo {author} {\bibfnamefont {M.}~\bibnamefont {Yankowitz}},
  \bibinfo {author} {\bibfnamefont {B.}~\bibnamefont {Fallahazad}}, \bibinfo
  {author} {\bibfnamefont {S.}~\bibnamefont {Kang}}, \bibinfo {author}
  {\bibfnamefont {H.~C.~P.}\ \bibnamefont {Movva}}, \bibinfo {author}
  {\bibfnamefont {S.}~\bibnamefont {Huang}}, \bibinfo {author} {\bibfnamefont
  {S.}~\bibnamefont {Larentis}}, \bibinfo {author} {\bibfnamefont {C.~M.}\
  \bibnamefont {Corbet}}, \bibinfo {author} {\bibfnamefont {T.}~\bibnamefont
  {Taniguchi}}, \bibinfo {author} {\bibfnamefont {K.}~\bibnamefont {Watanabe}},
  \bibinfo {author} {\bibfnamefont {S.~K.}\ \bibnamefont {Banerjee}}, \bibinfo
  {author} {\bibfnamefont {B.~J.}\ \bibnamefont {LeRoy}},\ and\ \bibinfo
  {author} {\bibfnamefont {E.}~\bibnamefont {Tutuc}},\ }\bibfield  {title}
  {\bibinfo {title} {{van der Waals Heterostructures with High Accuracy
  Rotational Alignment}},\ }\href
  {https://doi.org/10.1021/acs.nanolett.5b05263} {\bibfield  {journal}
  {\bibinfo  {journal} {Nano Lett.}\ }\textbf {\bibinfo {volume} {16}},\
  \bibinfo {pages} {1989} (\bibinfo {year} {2016})}\BibitemShut {NoStop}%
\bibitem [{\citenamefont {Bisswanger}\ \emph {et~al.}(2022)\citenamefont
  {Bisswanger}, \citenamefont {Winter}, \citenamefont {Schmidt}, \citenamefont
  {Volmer}, \citenamefont {Watanabe}, \citenamefont {Taniguchi}, \citenamefont
  {Stampfer},\ and\ \citenamefont {Beschoten}}]{Bisswanger2022Jun}%
  \BibitemOpen
  \bibfield  {author} {\bibinfo {author} {\bibfnamefont {T.}~\bibnamefont
  {Bisswanger}}, \bibinfo {author} {\bibfnamefont {Z.}~\bibnamefont {Winter}},
  \bibinfo {author} {\bibfnamefont {A.}~\bibnamefont {Schmidt}}, \bibinfo
  {author} {\bibfnamefont {F.}~\bibnamefont {Volmer}}, \bibinfo {author}
  {\bibfnamefont {K.}~\bibnamefont {Watanabe}}, \bibinfo {author}
  {\bibfnamefont {T.}~\bibnamefont {Taniguchi}}, \bibinfo {author}
  {\bibfnamefont {C.}~\bibnamefont {Stampfer}},\ and\ \bibinfo {author}
  {\bibfnamefont {B.}~\bibnamefont {Beschoten}},\ }\bibfield  {title} {\bibinfo
  {title} {{CVD Bilayer Graphene Spin Valves with 26 {$\mu$}m Spin Diffusion
  Length at Room Temperature}},\ }\href
  {https://doi.org/10.1021/acs.nanolett.2c01119} {\bibfield  {journal}
  {\bibinfo  {journal} {Nano Lett.}\ }\textbf {\bibinfo {volume} {2022}},\
  \bibinfo {pages} {,22} (\bibinfo {year} {2022})}\BibitemShut {NoStop}%
\bibitem [{\citenamefont {Wang}\ \emph {et~al.}(2013)\citenamefont {Wang},
  \citenamefont {Meric}, \citenamefont {Huang}, \citenamefont {Gao},
  \citenamefont {Gao}, \citenamefont {Tran}, \citenamefont {Taniguchi},
  \citenamefont {Watanabe}, \citenamefont {Campos}, \citenamefont {Muller},
  \citenamefont {Guo}, \citenamefont {Kim}, \citenamefont {Hone}, \citenamefont
  {Shepard},\ and\ \citenamefont {Dean}}]{Wang2013Nov}%
  \BibitemOpen
  \bibfield  {author} {\bibinfo {author} {\bibfnamefont {L.}~\bibnamefont
  {Wang}}, \bibinfo {author} {\bibfnamefont {I.}~\bibnamefont {Meric}},
  \bibinfo {author} {\bibfnamefont {P.~Y.}\ \bibnamefont {Huang}}, \bibinfo
  {author} {\bibfnamefont {Q.}~\bibnamefont {Gao}}, \bibinfo {author}
  {\bibfnamefont {Y.}~\bibnamefont {Gao}}, \bibinfo {author} {\bibfnamefont
  {H.}~\bibnamefont {Tran}}, \bibinfo {author} {\bibfnamefont {T.}~\bibnamefont
  {Taniguchi}}, \bibinfo {author} {\bibfnamefont {K.}~\bibnamefont {Watanabe}},
  \bibinfo {author} {\bibfnamefont {L.~M.}\ \bibnamefont {Campos}}, \bibinfo
  {author} {\bibfnamefont {D.~A.}\ \bibnamefont {Muller}}, \bibinfo {author}
  {\bibfnamefont {J.}~\bibnamefont {Guo}}, \bibinfo {author} {\bibfnamefont
  {P.}~\bibnamefont {Kim}}, \bibinfo {author} {\bibfnamefont {J.}~\bibnamefont
  {Hone}}, \bibinfo {author} {\bibfnamefont {K.~L.}\ \bibnamefont {Shepard}},\
  and\ \bibinfo {author} {\bibfnamefont {C.~R.}\ \bibnamefont {Dean}},\
  }\bibfield  {title} {\bibinfo {title} {{One-Dimensional Electrical Contact to
  a Two-Dimensional Material}},\ }\href
  {https://doi.org/10.1126/science.1244358} {\bibfield  {journal} {\bibinfo
  {journal} {Science}\ }\textbf {\bibinfo {volume} {342}},\ \bibinfo {pages}
  {614} (\bibinfo {year} {2013})}\BibitemShut {NoStop}%
\bibitem [{\citenamefont {Uwanno}\ \emph {et~al.}(2018)\citenamefont {Uwanno},
  \citenamefont {Taniguchi}, \citenamefont {Watanabe},\ and\ \citenamefont
  {Nagashio}}]{Uwanno2018Aug}%
  \BibitemOpen
  \bibfield  {author} {\bibinfo {author} {\bibfnamefont {T.}~\bibnamefont
  {Uwanno}}, \bibinfo {author} {\bibfnamefont {T.}~\bibnamefont {Taniguchi}},
  \bibinfo {author} {\bibfnamefont {K.}~\bibnamefont {Watanabe}},\ and\
  \bibinfo {author} {\bibfnamefont {K.}~\bibnamefont {Nagashio}},\ }\bibfield
  {title} {\bibinfo {title} {{Electrically Inert h-BN/Bilayer Graphene
  Interface in All-Two-Dimensional Heterostructure Field Effect Transistors}},\
  }\href {https://doi.org/10.1021/acsami.8b08959} {\bibfield  {journal}
  {\bibinfo  {journal} {ACS Appl. Mater. Interfaces}\ }\textbf {\bibinfo
  {volume} {10}},\ \bibinfo {pages} {28780} (\bibinfo {year}
  {2018})}\BibitemShut {NoStop}%
\end{thebibliography}
\end{document}


\author{A.~Rothstein}
\email{alexander.rothstein@rwth-aachen.de}
\affiliation{JARA-FIT and 2nd Institute of Physics, RWTH Aachen University, 52074 Aachen, Germany,~EU}%
\affiliation{Peter Gr\"unberg Institute  (PGI-9), Forschungszentrum J\"ulich, 52425 J\"ulich,~Germany,~EU}

\author{R. J.~Dolleman}
\affiliation{JARA-FIT and 2nd Institute of Physics, RWTH Aachen University, 52074 Aachen, Germany,~EU}%

\author{L. Klebl}
\affiliation{I. Institute for Theoretical Physics, University of Hamburg, Notkestraße 9-11, 22607 Hamburg, Germany, EU}
\affiliation{Institute for Theoretical Physics and Astrophysics, University of Würzburg, Am Hubland, 97074 Würzburg, Germany,~EU}

\author{A. Achtermann}
\affiliation{JARA-FIT and 2nd Institute of Physics, RWTH Aachen University, 52074 Aachen, Germany,~EU}%

\author{F. Volmer}
\affiliation{JARA-FIT and 2nd Institute of Physics, RWTH Aachen University, 52074 Aachen, Germany,~EU}%

\author{K.~Watanabe}
\affiliation{Research Center for Electronic and Optical Materials, National Institute for Materials Science, 1-1 Namiki, Tsukuba 305-0044, Japan}

\author{T.~Taniguchi}
\affiliation{Research Center for Materials Nanoarchitectonics, National Institute for Materials Science,  1-1 Namiki, Tsukuba 305-0044, Japan}%

\author{F.~Hassler}
\affiliation{Institute for Quantum Information, RWTH Aachen University, 52056 Aachen, Germany, EU}

\author{L. Banszerus}
\affiliation{Faculty of Physics, University of Vienna, Boltzmanngasse 5, 1090 Vienna, Austria, EU}

\author{B.~Beschoten}
\affiliation{JARA-FIT and 2nd Institute of Physics, RWTH Aachen University, 52074 Aachen, Germany,~EU}%

\author{C.~Stampfer}
\email{stampfer@physik.rwth-aachen.de}
\affiliation{JARA-FIT and 2nd Institute of Physics, RWTH Aachen University, 52074 Aachen, Germany,~EU}%
\affiliation{Peter Gr\"unberg Institute  (PGI-9), Forschungszentrum J\"ulich, 52425 J\"ulich,~Germany,~EU}%

\title{Supporting Information - Gate-tunable Josephson diodes in magic-angle twisted bilayer graphene}

\date{\today}

\maketitle

\renewcommand{\theequation}{S\arabic{equation}}
\renewcommand{\thesection}{S\arabic{section}}
\renewcommand{\thefigure}{S\arabic{figure}}
\renewcommand{\thetable}{S\arabic{table}}
\setcounter{figure}{0}
\setcounter{table}{0}


\tableofcontents
\newpage
\section{Optical image of the sample}
\begin{figure*}[!htbp]
\centering
\includegraphics[draft=false,keepaspectratio=true,clip,width=1\linewidth]{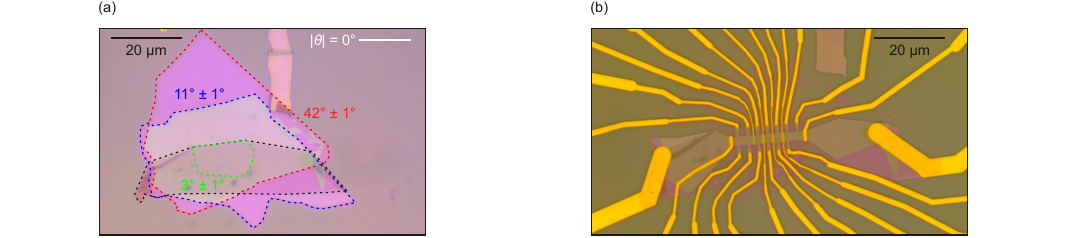}
\caption{ 
(a) 
Optical image of the finished van-der-Waals heterostructure. The dashed lines show the outlines of the top hBN (red), MATBG (green), bottom hBN (blue) and graphite (black) flake. 
The angles are given with respect to the horizontal reference (white) showing the misalignment of the hBN flakes to the MATBG.  
(b)
Optical image of the finished Hall bar structure. 
}
\label{Sf1}
\end{figure*}

In \cref{Sf1}, we show an image of the heterostructure after stacking and of the finished Hall bar device. 

\section{Calculation of charge carrier density and twist-angle extraction}
\begin{figure*}[!htbp]
\centering
\includegraphics[draft=false,keepaspectratio=true,clip,width=1\linewidth]{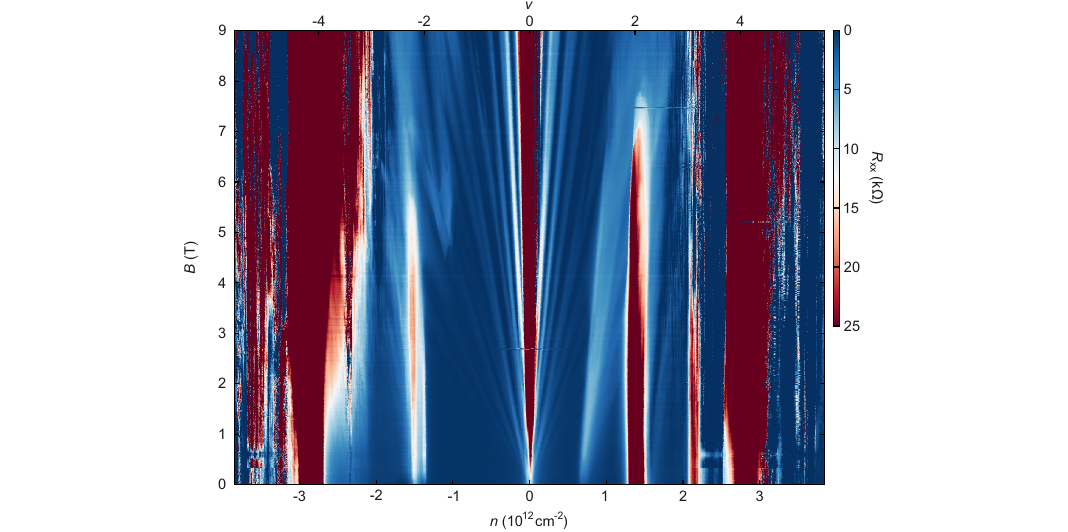}
\caption{Landau fan measurement showing the longitudinal resistance $R_\mathrm{xx}$ measured in the area of the device used to form the individual JJs as a function of out-of-plane magnetic field $B$ and charge carrier density $n$. 
Top gate voltages are chosen such that the graphite back gate tunes the charge carrier density homogeneously.
}
\label{Sf2}
\end{figure*}

The charge carrier density $n$ of the MATBG layer is electrostatically tuned by the global graphite back gate and (in the dual-gated regions) locally under the individual top gates. 
The local charge carrier density below the top gates is given by: 
\begin{align}
n_\mathrm{JJ} = \alpha_\mathrm{BG}\left[(V_\mathrm{BG} - V_\mathrm{BG,off}) + \frac{\alpha_\mathrm{TG}}{\alpha_\mathrm{BG}}(V_\mathrm{TG} - V_\mathrm{TG,off}) \right]. \label{nJJ}
\end{align} 
Here, $\alpha_\mathrm{BG/TG}$ is the lever arm of the graphite back gate and the used top gate, respectively. 
The lever arm $\alpha_\mathrm{BG}$ of the graphite back gate is extracted from the slopes of the visible Landau levels emerging from the charge neutrality point in the magneto-transport measurement shown in \cref{Sf2}. 

We perform least-square fitting to the minima of the resistance and calculate the lever arm via:
\begin{align}
B_\mathrm{LL} = \frac{h}{\nu_\mathrm{LL}e}\alpha_\mathrm{BG}V_\mathrm{BG} + \mathrm{const}.,
\end{align}
where $\nu_\mathrm{LL}$ denotes the observed Landau level filling sequence of $\nu_\mathrm{LL} = \pm 1, \pm 2, \pm 3, \pm 4$.
This results in a value of $\alpha_\mathrm{BG} \approx (5.50 \pm 0.10) \times 10^{15}\, \mathrm{V^{-1}m^{-2}}$. 
This value is in reasonable agreement with the geometric lever arm expected from a simple plate capacitor model:
\begin{align}
\alpha_\mathrm{BG} = \varepsilon_0 \varepsilon_\mathrm{hBN} \frac{1}{ed_\mathrm{b}},
\end{align}
which yields a value of $\alpha_\mathrm{BG} \approx 5.37 \times 10^{15} \, \mathrm{V^{-1}m^{-2}}$.
Here we used $\varepsilon_\mathrm{hBN} \approx 3.4$ for the hBN permittivity \cite{Pierret2022Jun, Laturia2018Mar} and $d_\mathrm{b} \approx 35$~nm for the bottom hBN thickness, which was extracted via atomic force microscopy.
Additionally, we extract from the magnetotransport measurement an offset value of $V_\mathrm{BG,off} \approx -130$~mV.
To extract the gate lever arm of the relevant top gates, we use the dual-gated maps as shown in Fig.~1 of the main manuscript.
The diagonal features correspond to the areas in the sample which are affected by both, the corresponding top gate and the graphite back gate. 
The slope of the lines describe the relative lever arm of the top gate with respect to the back gate. 
For both top gates we find $\alpha_\mathrm{L}/\alpha_\mathrm{BG} \approx \alpha_\mathrm{R}/\alpha_\mathrm{BG} \approx 1$.
We therefore conclude that $\alpha_\mathrm{L} = \alpha_\mathrm{R} \approx \alpha_\mathrm{BG} = (5.50 \pm 0.10) \times 10^{15}\, \mathrm{V^{-1}m^{-2}}$.
This value is also also reasonable agreement with the value predicted by the place capacitor model ($\alpha_\mathrm{L/R} \approx 5.37 \times 10^{15} \, \mathrm{V^{-1}m^{-2}}$, top hBN thickness $d_\mathrm{t} \approx 35$~nm). 
From the dual-gated maps we also extract the offset values to be $V_\mathrm{L, off} \approx 504$~mV and $V_\mathrm{R, off} \approx 511$~mV. 
For these voltage values applied to the top gates, the charge carrier density is homogeneously tuned only by the graphite back gate. 
To extract the twist angle of the device we estimated the superlattice density, i.e. the charge carrier density where the flat bands are fully filled, from the Landau fan diagram in \cref{Sf2} to be $n_\mathrm{s} = (2.75 \pm 0.05) \times 10^{12} \, \mathrm{cm^{-2}}$. 
With $\theta_\mathrm{MATBG} = [\sqrt{3}n_\mathrm{s}a^2/8]^{1/2}$ \cite{Cao2018Apr} this results in a twist angle of $\theta_\mathrm{MATBG} \approx 1.09^\circ \pm 0.01^\circ$.

\newpage
\section{Bulk superconducting properties}
\begin{figure*}[!htbp]
\centering
\includegraphics[draft=false,keepaspectratio=true,clip,width=1\linewidth]{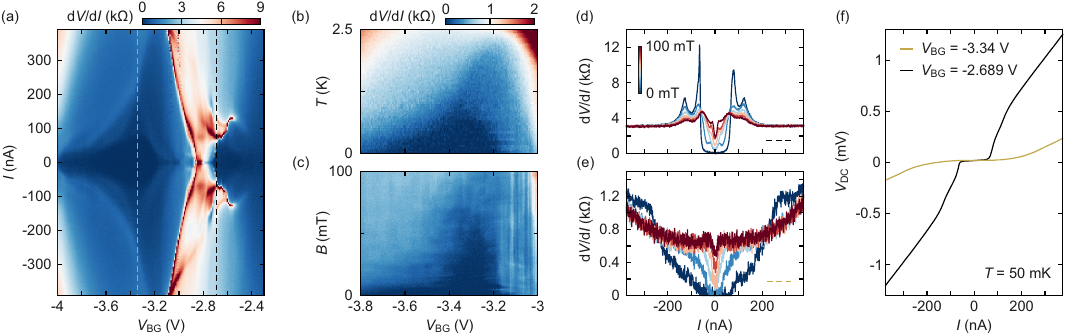}
\caption{
(a) Bias current dependence of the superconducting dome near to $\nu = -2$. 
(b) Temperature dependence around optimal doping at $V_\mathrm{BG} = -3.34 \, \mathrm{V}$. 
(c) Magnetic field dependence of the SC dome around optimal doping.
(d, e) Single linetraces of the differential resistance for different magnetic fields taken at $V_\mathrm{BG} = -3.34 \, \mathrm{V}$ and $V_\mathrm{BG} = -2.689 \, \mathrm{V}$.
(f) Current-voltage (DC) characteristic taken at $V_\mathrm{BG} = -3.34 \, \mathrm{V}$ and $V_\mathrm{BG} = -2.689 \, \mathrm{V}$.
All measurements are taken at $V_\mathrm{L} = 0.504\, \mathrm{V}$ and $V_\mathrm{R} = 0.511\, \mathrm{V}$.}
\label{Sf4}
\end{figure*}

We characterise the bulk superconductivity -- i.e. when homogeneously tuning the charge carrier density with the back gate -- of the device as shown in \cref{Sf4}.
The bias current dependent measurement in \cref{Sf4}(a) reveals that the superconducting dome is split into an asymmetric pair separated by the $\nu = -2$ correlated insulating state (compare with linecuts in \cref{Sf4}(d-f) taken at $V_\mathrm{BG} = -3.34 \, \mathrm{V}$ and $V_\mathrm{BG} = -2.689 \, \mathrm{V}$).
At optimal doping of $V_\mathrm{BG} = -3.34\, \mathrm{V}$ the critical current reaches its maximum of $I_\mathrm{c}^\mathrm{Bulk} \approx 75 \, \mathrm{nA}$. 
The temperature- and out-of-plane magnetic field dependent measurements around this optimal doping [\cref{Sf4}(b,c)] reveal critical values of $T_\mathrm{c}^\mathrm{Bulk} \approx 700 \, \mathrm{mK}$ and $B_\mathrm{c}^\mathrm{Bulk} \approx 50 \, \mathrm{mT}$.

\newpage
\section{Additional exemplary traces in the different junction regimes}
\begin{figure*}[!htbp]
\centering
\includegraphics[draft=false,keepaspectratio=true,clip,width=1\linewidth]{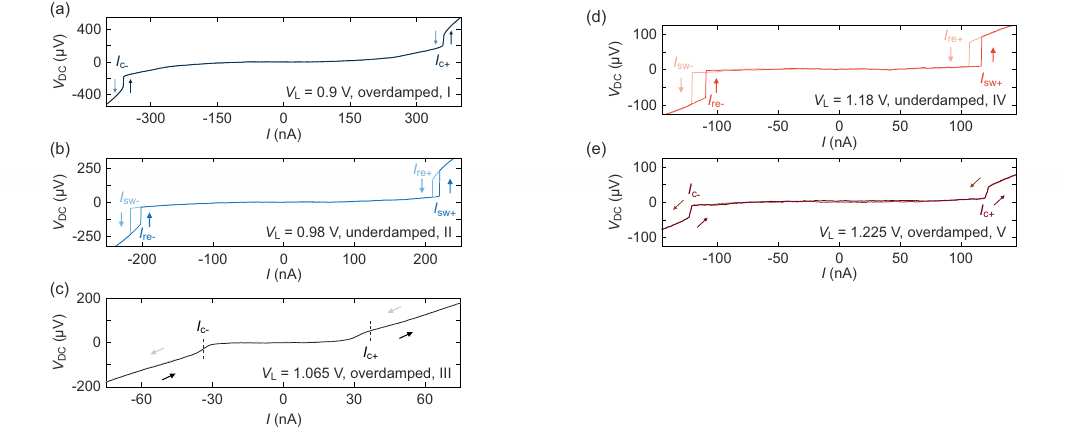}
\caption{Current-voltage characteristics for both bias current sweep directions measured exemplary in the five different junction dynamic regimes.
(a) region I, $V_\mathrm{L} = 0.9 \, \mathrm V$, (b) region II $V_\mathrm L = 0.98 \, \mathrm V$, (c) region III, $V_\mathrm L = 1.065 \, \mathrm V$, (d) region IV, $V_\mathrm L = 1.18 \, \mathrm V$ and (e) region V, $V_\mathrm L = 1.225 \, \mathrm V$.
}
\label{Sf-linetraces}
\end{figure*}

In \cref{Sf-linetraces}(a-e) we show additional current-voltage characteristics taken at exemplary positions in all five distinct junction dynamic regimes as indicated in Fig.~2 of the main text. 
We also want to highlight that especially in \cref{Sf-linetraces}(a) the residual resistance mentioned in the main text is visible before the characteristic voltage jump at the critical current is reached. 

\newpage
\section{Implementation of the Ivanchenko-Zil'berman model to extract the critical currents}
To quantitatively describe the superconducting diode effect in all regions of our device, a reliable technique to extract the critical and the retrapping currents is required.
For the hysteretic regions II and IV, this extraction is unproblematic, since the $I$-$V$ characteristic show a sharp increase when the current $I$ exceeds the switching current $I_\mathrm{sw}$.
These regimes follow the prediction of the resistively and capacitively shunted junction (RCSJ) model for a (underdamped) superconducting junction with $V \equiv V(R_\mathrm{N}, I_\mathrm{sw})$ \cite{Tinkham1975}
\begin{equation}
    V = R_\mathrm{N} \sqrt{I^2-I_\mathrm{sw}^2}, \label{IV1}
\end{equation}
where $R_\mathrm{N}$ is the normal state resistance.
However, the $I$-$V$ characteristics of the junction region III reveals a considerable deviation from this behavior, which makes it difficult to accurately determine the corresponding critical currents (we use the name critical current here, since the there is no hysteresis in region III).
Deviations from the square-root behavior in \cref{IV1} due to thermal fluctuations were studied by Ambegaokar and Halperin \cite{Ambegaokar1969Jun} as well as Ivanchenko and Zil'berman \cite{IZ1968}.
To extract the critical current in the non-hysteretic regime we adapt the Ivanchenko-Zil'berman model, ignoring macroscopic quantum tunneling effects and assuming classical fluctuations which result in the current voltage characteristic $V \equiv V(R_\mathrm{N}, I_\mathrm{c \pm}, I_\mathrm{f \pm}, V_\mathrm{0 \pm})$:
\begin{equation}
    V =  
\begin{cases}
    R_\mathrm{N} I_\mathrm{c+} \left( \frac{I}{I_\mathrm{c+}} +
    \mathfrak{I}_{1+\mathrm i\gamma^+}(\gamma^+_c) -
    \mathfrak{I}_{1-\mathrm i\gamma^+}(\gamma^+_c) \right) + V_{0+}, \quad
    \gamma^+ = I/I_\mathrm{f+},  \quad 
    \gamma^+_\mathrm{c}
    = I_\mathrm{c+}/I_\mathrm{f+}, &\quad \text{if } I\geq 0\\ \\
    R_\mathrm{N} I_\mathrm{c-} \left( \frac{I}{I_\mathrm{c-}} +
    \mathfrak{I}_{1+ \mathrm i\gamma^-}(\gamma^-_c) -
    \mathfrak{I}_{1- \mathrm i\gamma^-}(\gamma^-_c) \right) + V_{0-}, \quad 
    \gamma^- = I/I_\mathrm{f-},  \quad 
    \gamma^-_\mathrm{c}
    = I_\mathrm{c-}/I_\mathrm{f-}, &\quad \text{if } I< 0 
\end{cases} \label{IZ}
\end{equation}
where $I_\mathrm{f\pm}$ is an effective fluctuation current, $V_\mathrm{0\pm}$ is an additional offset that is required to obtain an accurate fit if the values of $I_\mathrm{c \pm}$ are comparable to the fluctuation-induced rounding. $\mathfrak{I}_\nu(z)$ is an expression formed from modified Bessel functions of the first kind~\cite{Glick2017Oct}, i.e., $\mathcal I_\nu(z)$:
\begin{equation}
\label{Bessel}
\mathfrak{I}_\nu(z) = \frac1{2 \mathrm i}\,\frac{\mathcal I_\nu(z)}{\mathcal I_{\nu-1}(z)} \,,
\end{equation}
with $\nu, z \in \mathbb{C}$ and $\nu$ being a non-integer~\footnote{
    We want to note here, that there is a typo in the definition of $\mathfrak{I_\pm}$ in Ref.~\cite{Glick2017Oct}, stating that $\mathfrak{I}_\pm = \frac{\mathcal I_{(1 \pm \mathrm i) \gamma}(z)}{2\mathrm i \mathcal I_{\pm \mathrm i \gamma}(z)}$ instead of the correct relation $\mathfrak{I}_\pm = \frac{\mathcal I_{(1 \pm \mathrm i \gamma) }(z)}{2\mathrm i \mathcal I_{\pm \mathrm i \gamma}(z)}$
}. The individual Bessel functions in \cref{Bessel} are difficult to evaluate numerically, and we therefore use a continued fraction representation~\cite{Gautschi1978, Abramowitz1972} to directly evaluate the ratio:
\begin{equation}\label{eq:gaussfraction}
2 \mathrm i\,\mathfrak I_\nu(z) = \frac{\mathcal I_{\nu}(z)}{\mathcal I_{\nu-1}(z)} =
\cfrac{1}{\frac{2}{z}\nu+\cfrac{1}{\frac{2}{z}(\nu+1) + \cfrac{1}{\frac{2}{z}(\nu+2) + \cdots \vphantom{\cfrac{1}{1}} }}} \,.
\end{equation}
We checked convergence of \cref{eq:gaussfraction} by comparing it to Fortran/C implementations of the individual Bessel functions of complex order and complex argument~\cite{Kodama2011, flint}, which are \emph{not} part of standard special function collections. Using $n=100$ iterations of the continued fraction results in an adequately converged value for most $\nu$ and $z$, with larger $n$ taken where necessary. In the data analysis, we evaluate \cref{eq:gaussfraction} with the following Python function:
\begin{lstlisting}[language=Python, basicstyle=\small]
def bessel_ratio(nu, z, n = 100)
    val = 0
    for k in range(-n, 1):
        i = -k
        val = 1./(val + 2*(nu + i)/z)
    return val    
\end{lstlisting}
Next, we use the SciPy's method \texttt{scipy.curve\_fit} to fit the model in \cref{IZ} to our data.

\newpage
\subsection{Examplary fits to the Ivanchenko-Zil'berman model}
In \ref{SI_Fits_IZ} we show exemplary fits of the Ivanchenko-Zil'berman model to the magnetotransport data from region III recorded for both junctions. 
The outline of the interference patterns are well reproduced when we not only take the critical currents $I_\mathrm{c\pm}$ (black curves) but also the fluctuation current $I_\mathrm{f \pm}$ (grey curves) into account by calculating the difference $I_\mathrm{c\pm} - I_\mathrm{f\pm}$. 

\begin{figure*}[!htbp]
\centering
\includegraphics[draft=false,keepaspectratio=true,clip,width=1\linewidth]{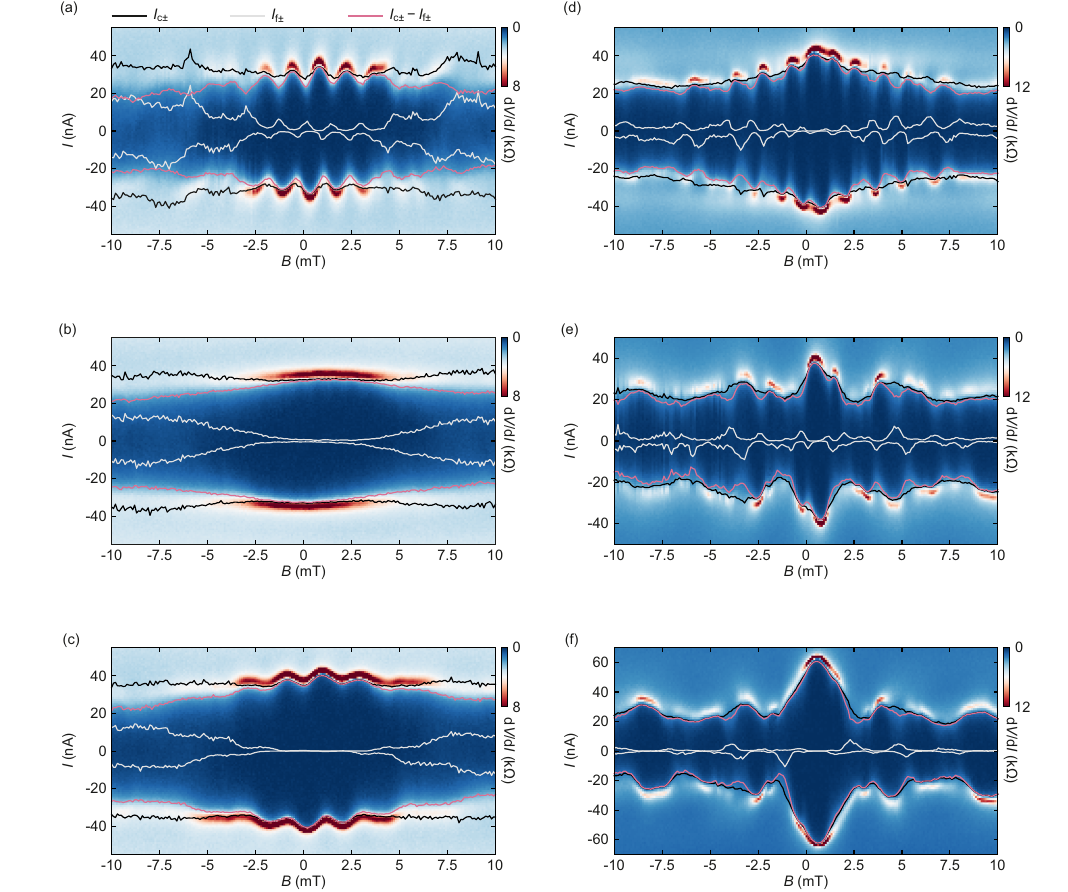}
\caption{
Exemplary fitting results of the Ivanchenko-Zil'berman model to the magnetospectroscopy in the non-hysteretic region III of the individual junctions. 
The fits are performed for the single line traces, i.e. at fixed magnetic field. 
Shown are the critical currents $I_\mathrm{c\pm}$ (black) and the fluctuation currents $I_\mathrm{f\pm}$ (grey) as well as the difference $I_\mathrm{c\pm} - I_\mathrm{f\pm}$ (pink).
(a) $V_\mathrm{L} = 1.06 \, \mathrm{V}$.
(b) $V_\mathrm{L} = 1.09 \, \mathrm{V}$.
(c) $V_\mathrm{L} = 1.12 \, \mathrm{V}$.
(d) $V_\mathrm{R} = 1.01 \, \mathrm{V}$.
(e) $V_\mathrm{R} = 1.04 \, \mathrm{V}$.
(f) $V_\mathrm{R} = 1.055 \, \mathrm{V}$.
}
\label{SI_Fits_IZ}
\end{figure*}

\newpage

\newpage
\section{Characterization of residual magnetic field}
\begin{figure*}[!htbp]
\centering
\includegraphics[draft=false,keepaspectratio=true,clip,width=1\linewidth]{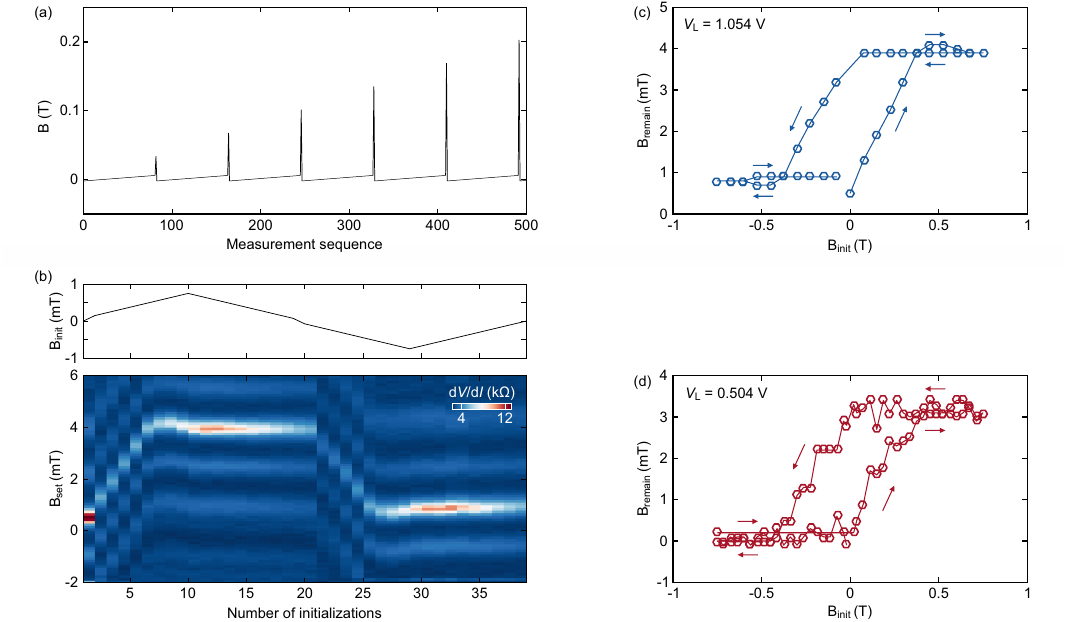}
\caption{(a) Magnetic field set to the instrument versus the number in the measurement sequence. (b) Shift in the superconducting oscillations during the progression of the measurement with different initializing fields. (c) Remaining magnetic field ($B_\mathrm{remain}$) versus the initializing field ($B_\mathrm{init}$) for $V_\mathrm{L} = 1.054~$V and $V_\mathrm{BG} = -3.34~$V. (d) Same as panel (c), but for $V_\mathrm{L} = 0.504~$V and $V_\mathrm{BG} = -2.69~$V.
} 
\label{SI_Magnetization}
\end{figure*}

In \cref{SI_Magnetization}, we characterize the residual magnetic field within our experimental setup.
We achieve this by measuring superconducting oscillations at a finite DC bias current and tracking their shifts with varying initializing magnetic fields.
\cref{SI_Magnetization}(a) shows the magnetic field set on our instrument ($B$) for each measurement point in the sequence. 
Initially, the magnetic field is set to an initializing value, followed by a smaller sweep to monitor the oscillations. 
In the example in \cref{SI_Magnetization}(b), we set $V_\mathrm{L} = 1.054~V$ and $V_\mathrm{BG} = -3.34~$V to create a junction around $\nu = -2$. 
We observe supercurrent oscillations between -2 and 6 mT, noting that their position shifts with different initializing fields $B_\mathrm{init}$. 
At high enough $B_\mathrm{init}$, no further shift in the oscillation position occurs, indicating a saturation in the remaining field $B_\mathrm{remain}$. 
We extract $B_\mathrm{remain}$ by tracking the position of the maximum in $\mathrm{d}V/\mathrm{d}I$, as shown in \cref{SI_Magnetization}(c) which visualizes the hysteresis loop. 
This resembles a typical ferromagnetic hysteresis loop that saturates at fields exceeding $|B|> 400~$mT. 
It is important to note that this loop is not centered around $B_\mathrm{remain} = 0~$mT because our definition makes it relative.
However, we presume that the actual magnetic field $B_\mathrm{eff} = 0$ in the center of the hysteresis loop.
To investigate whether the remaining field is intrinsic to either the superconducting or correlated insulating phase in our sample, we adjust the voltages to see if $B_\mathrm{remain}$ persists.
In \cref{SI_Magnetization}(d) we plot $B_\mathrm{remain}$ versus $B_\mathrm{init}$ for $V_\mathrm{L} = 0.504~V$ and $V_\mathrm{BG} = -2.69~$V.
In this regime, no junction forms, and the superconductor is tuned away from optimal doping; however, the hysteresis loop remains identical to that shown in \cref{SI_Magnetization}(c).  
This leads us to conclude that the remaining field is an artifact of our experimental setup.
To account for this effect, each measurement of the superconducting interference presented in the main text was initialized at -750 mT. 

\newpage
\def\brokenRef{\textbf{\color{red}REF BROKEN}}
\section{Simulation of the current-voltage characteristic}
To obtain the current-voltage (IV) characteristic, we simulate the dynamics of $N$ RCSJ junctions coupled in parallel, as shown in Fig.~5 of the main text. The dynamics for two coupled RSCJ junctions are well studied in the context of SQUIDs~\cite{Tesche1977Nov, Bruines1982Feb, deWaal1984Feb}, and we here extend the theory to $N$ parallel RCSJ junctions. Notably, the mutual inductance between each plaquette is treated in Ref.~\cite{deWaal1984Feb} without appearing explicitly, which generalizes to the present case.

\begin{figure*}[!b]
\centering
\includegraphics[draft=false,keepaspectratio=true,clip,width=1\linewidth]{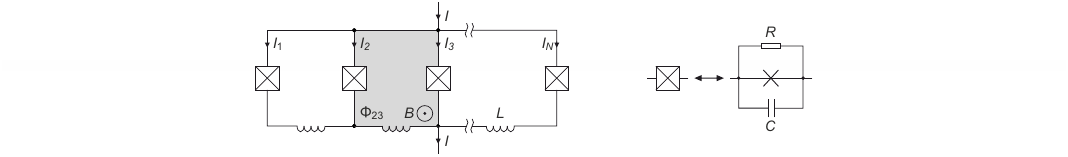}
\caption{Diagram for the circuit that is simulated numerically. The whole circuit is under the influence of a perpendicular magnetic field $B$. 
} 
\label{SI_Fits_IZ}
\end{figure*}

Following the derivation in Ref.~\cite{deWaal1984Feb}, the current through each junction is given by 
\begin{equation}
    I_i = I_0(1-\alpha_i) \sin(\delta_i) + \frac{V_i}{R} + C\frac{\dd V_i}{\dd t} \,.
    \label{SI_current_relation}
\end{equation}
In addition, the overall current must equal the sum over the currents in the individual paths, i.e. $I = \sum_i I_i$. For each plaquette, we can write a phase relation due to the induced current:
\begin{equation}
    \label{SI_phase_relation}
    \delta_i - \delta_{i+1} = \frac{2\pi}{\Phi_0}\big( \Phi_a + L(I_{i+1}-I_i) \big) \,,
\end{equation}
where $\Phi_a$ is the \emph{externally} applied magnetic flux. Through Josephson's relations, we can express the voltages in terms of the phases:
\begin{equation}
    \frac{\dd \delta_i}{\dd t} = \frac{4\pi e}{h}\, V_i \,.
\end{equation}
As we want to solve the $N$ coupled differential equations for $\delta_i$, we have to re-express \cref{SI_current_relation} as a function of $\delta_i$. The right hand side becomes
\begin{equation}
    \label{SI_diffeq_initial}
    I_i = I_0(1-\alpha_i) \sin(\delta_i) + \frac{h}{4\pi eR} \frac{\dd \delta_i}{\dd t} + \frac{Ch}{4\pi e}\frac{\dd^2 \delta_i}{\dd t^2} \,.
\end{equation}
In order to determine the dependence of the left hand side on $\delta_i$, we recast \cref{SI_phase_relation} together with the total current relation $I=\sum_i I_i$ into vector/matrix form:
\begin{equation}
    \underbrace{\begin{pmatrix}
        -1 & 1 \\
        & -1 & 1 \\
        && \ddots & \ddots \\
        1 && \cdots && 1
    \end{pmatrix}}_{\hat M}
    \begin{pmatrix}
        I_1 \\[0.7em] \vdots \\[0.7em] I_N
    \end{pmatrix} =
    \begin{pmatrix}
    L^{-1}[(\delta_1-\delta_2)\Phi_0/2\pi - \Phi_a] \\
    L^{-1}[(\delta_2-\delta_3)\Phi_0/2\pi - \Phi_a] \\
    \vdots \\
    I
    \end{pmatrix} \,.
\end{equation}
In simulations, we numerically invert the matrix $\hat M$ and apply it to the right hand column vector. This allows us to define a linear mapping $\bvec F(\bvec \delta) = \bvec I$:
\begin{equation}
    \begin{pmatrix}
        I_1 \\[0.7em] \vdots \\[0.7em] I_N
    \end{pmatrix} = \bvec F(\bvec \delta) = \hat M^{-1} \begin{pmatrix}
        L^{-1}[(\delta_1-\delta_2)\Phi_0/2\pi - \Phi_a] \\
        L^{-1}[(\delta_2-\delta_3)\Phi_0/2\pi - \Phi_a] \\
        \vdots \\
        I
    \end{pmatrix} \,.
\end{equation}
Here, we defined column vectors $\bvec \delta = (\delta_1, \dots, \delta_N)^T$ and $\bvec I = (I_1, \dots, I_N)^T$. Now, the reformulation of \cref{SI_diffeq_initial} in vector form becomes apparent:
\begin{equation}
    \label{SI_simeq_dim}
    \frac{C h}{4\pi e} \frac{\dd^2 \bvec \delta}{\dd t^2} = \bvec F(\bvec \delta) - I_0(1-\bvec \alpha)\odot\sin(\bvec \delta) - \frac{h}{4\pi e R} \frac{\dd \bvec \delta}{\dd t} \,,
\end{equation}
with ``$\odot$'' an element-wise multiplication of vectors, $1-\bvec \alpha = (1-\alpha_1, \dots, 1-\alpha_N)^T$, and $\sin(\bvec \delta) = (\sin(\delta_1), \dots, \sin(\delta_N))^T$.

For simulations, we employ the commonly used units $\Phi_0/(2\pi I_0 R)$ for time, $I_0$ for current, $I_0 R$ for voltage, and $\Phi_0$ for flux~\cite{deWaal1984Feb}. The dimensionless version of \cref{SI_simeq_dim} becomes
\begin{equation}
    \beta_c \frac{\dd^2 \bvec \delta}{\dd \vartheta^2} = \bvec f (\bvec \delta) - (1-\bvec \alpha) \odot \sin(\bvec \delta) - \frac{\dd \bvec \delta}{\dd \vartheta} \,,
    \label{SI_diffeq_final}
\end{equation}
with $\bvec f(\bvec \delta)$ given by
\begin{equation}
    \bvec f(\bvec \delta) = \hat M^{-1} \begin{pmatrix}
        2/(\pi\beta) \, [\delta_1-\delta_2 - 2\pi\phi_a] \\
        2/(\pi\beta) \, [\delta_2-\delta_3 - 2\pi\phi_a] \\
        \vdots \\
        i
    \end{pmatrix} \,,
\end{equation}
and $\beta = 4I_0L/\Phi_0$, $\beta_c = 2\pi I_0R^2 C/\Phi_0$, $i=I/I_0$, $\phi_a = \Phi_a/\Phi_0$, and $\vartheta$ corresponding to time measured in aforementioned units.

The total voltage must equal each of the individual voltages:
\begin{equation}
\begin{aligned}
    V
    &{}= V_1 - L/2 (\dot I_2 - \dot I_1) \\
    &{}= V_2 + L/2 (\dot I_2 - \dot I_1) - L/4 (\dot I_3 - \dot I_2) \\[-0.7em]
    &{}\hspace{0.5em}\vdots \\[-0.7em]
    &{}= V_i + L/2 (\dot I_i - \dot I_{i-1}) - L/4 (\dot I_{i+1} - \dot I_i) \\[-0.7em]
    &{}\hspace{0.5em}\vdots \\[-0.7em]
    &{}= V_N + L/2 (\dot I_N - \dot I_{N-1}) \,.
\end{aligned}
\end{equation}
We can therefore express $V$ as
\begin{equation}
    V = \frac1N\sum_i V_i = \frac{h}{4\pi e\, N} \sum_i \frac{\dd \delta_i}{\dd t} \,.
\end{equation}
In units of $I_0R$, we arrive at
\begin{equation}
    v = \frac{V}{I_0R} = \frac1N\sum_i \frac{\dd \delta_i}{\dd \vartheta} \,.
    \label{SI_eqvolt}
\end{equation}

\begin{figure*}[!htbp]
\centering
\includegraphics[draft=false,keepaspectratio=true,clip,width=0.9\linewidth]{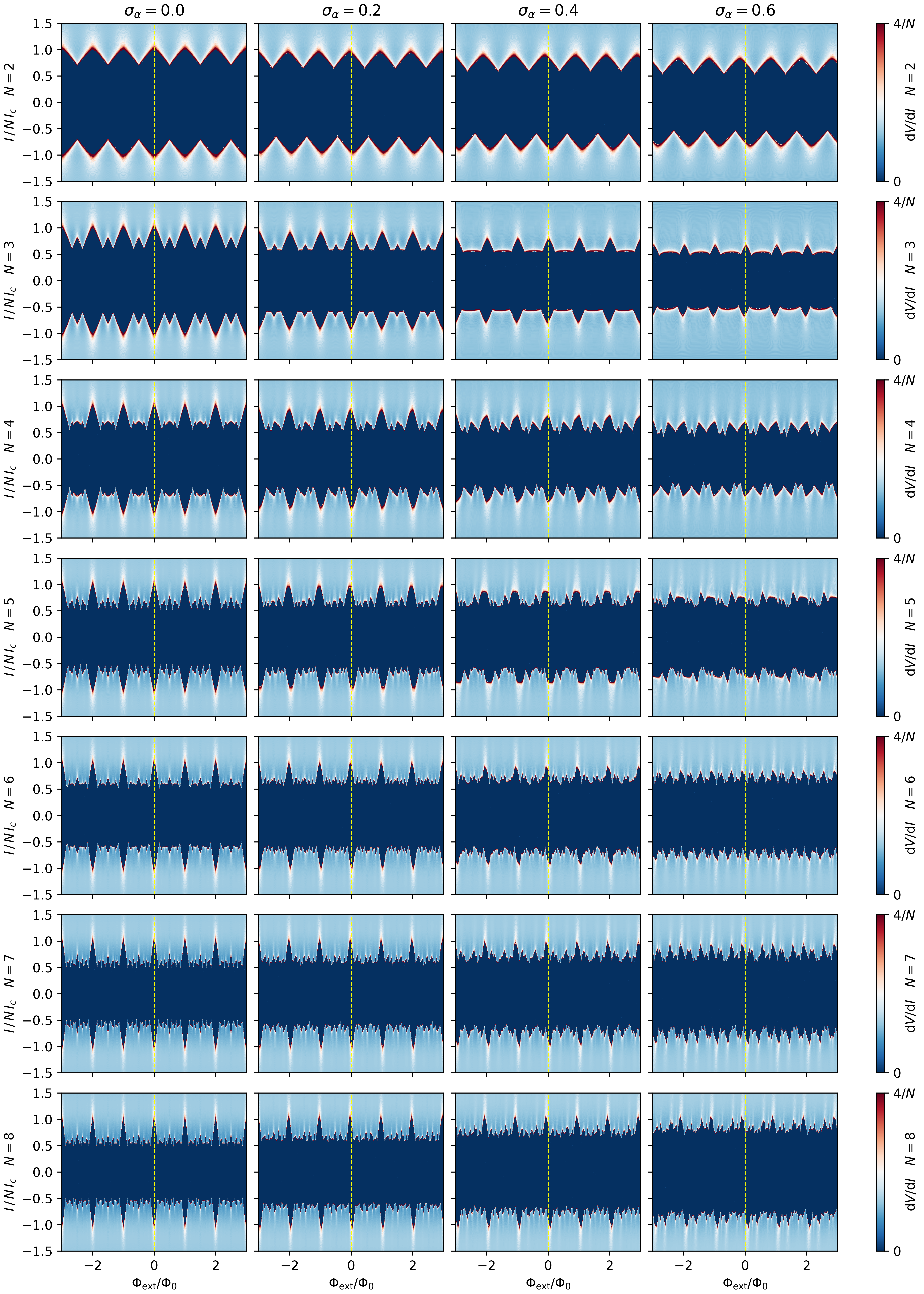}
\caption{Magnetotransport simulations for a different number of parallel junctions and different current imbalances.
} 
\label{simulation}
\end{figure*}

The results of Fig.~5 of the main text are obtained for RCSJ junctions in the overdamped regime, i.e., without hysteretic behavior in the $I$-$\dd V/\dd I$ characteristic.
We simulate \cref{SI_diffeq_final} using a Runge-Kutta 4 scheme implemented in the Boost-C++ libraries~\cite{Ahnert2011}. The characteristic parameters are given as $\beta = 2$ and $\beta_c = 0.1$. We sample currents in the range $i \in [-3N/2, 3N/2]$ (note that $N$ is the number of parallel junctions) and external fluxes in the range $\phi_a \in [-3,3]$. The imbalance vector $\bvec \alpha$ is sampled from a normal distribution centered around zero, with standard deviation $\sigma_\alpha\in [0, 0.2, 0.4, 0.6]$ (see \cref{SI_alpha_table} for values). Discrete Runge-Kutta time steps are taken at width $\dd\vartheta = 0.02$, with a maximum simulation time of $\vartheta_\mathrm{max} = 2000$. Fast oscillations in the resulting voltage are eliminated by averaging over the last $N_\mathrm{avg}=80,\!000$ time steps. The derivative $\dd v/\dd i$ is taken as a numerical forward-difference.
In \cref{simulation} we show the simulation results for $N = 2$ to $N = 8$ parallel junctions.

\begin{table}
    \caption{Numerical values for imbalance vector $\bvec \alpha$ used in simulations for Fig.~5, sampled from a Mersenne-Twister ``MT 19937'' engine with seed 12345. Depending on the number of parallel junctions, the first $N$ values for $\alpha_i$ are taken.}
    \label{SI_alpha_table}
    \centering
    \begin{ruledtabular}
    \def\arraystretch{1.1}
    \begin{tabular}{ccccccccc}
        $\sigma_\alpha$ &
        $\alpha_1$ & $\alpha_2$ & $\alpha_3$ & $\alpha_4$ &
        $\alpha_5$ & $\alpha_6$ & $\alpha_7$ & $\alpha_8$ \\[0.2em]
        $0$ &
        $0$ & $0$ & $0$ & $0$ & $0$ & $0$ & $0$ & $0$ \\
        $0.2$ &
        $0.119018$ & $0.00836681$ & $0.160089$ & $-0.0143843$ & $-0.208614$ & $-0.0762312$ & $-0.153216$ & $-0.315489$ \\
        $0.4$ &
        $0.238036$ & $0.0167336$ &  $0.320179$ & $-0.0287687$  & $-0.417228$ & $-0.152462$ & $-0.306431$ & $-0.630977$ \\
        $0.6$ &
        $0.357055$ & $0.0251004$ & $0.480268$ & $-0.043153$ & $-0.625841$ & $-0.228693$ & $-0.459647$ & $-0.946466$
    \end{tabular}
    \end{ruledtabular}
\end{table}

